\documentclass[longauth]{aa}

\usepackage{xspace}
\usepackage{txfonts}
\usepackage{graphicx}
\usepackage{xcolor}
\usepackage[colorlinks=true,linkcolor=blue,citecolor=blue,urlcolor=blue]{hyperref}
\usepackage{longtable}
\usepackage{booktabs}
\usepackage{supertabular}
\usepackage{lscape}
\usepackage{comment}

\usepackage{multirow}

\usepackage[switch,columnwise]{lineno}

\newcommand{\gwtrig}{S190814bv\xspace}
\newcommand{\UTGW}{2019 Aug 14 21:10:39.01}
\newcommand{\MJDGW}{58709.88240}  
\newcommand{\alertUTGW}{2019 Aug 14 21:31:40}  
\newcommand\alertMJDGW{58709.89699}  
\newcommand\GWDist{$267\pm52$\,Mpc} 
\newcommand{\sect}{\S\,}
\newcommand{\gps}{\ensuremath{g_{\rm P1}}}
\newcommand{\rps}{\ensuremath{r_{\rm P1}}}
\newcommand{\ips}{\ensuremath{i_{\rm P1}}}
\newcommand{\zps}{\ensuremath{z_{\rm P1}}}

\newcommand{\grizy}{\ensuremath{grizy_{\rm P1}}}

\newcommand{\sqd}{\ensuremath{\mathrm{deg}^2}}
\newcommand{\PS}{Pan-STARRS1}

\newcommand{\sjs}[1]{{\textcolor{magenta}{\texttt{Smartt: #1}} }}

\begin{document}

   \title{Observational constraints on the optical and near-infrared emission from the neutron star--black hole binary merger candidate S190814bv}

\author{K. Ackley\inst{1}\and
L. Amati\inst{2}\and
C. Barbieri\inst{3,4,5}\and
F. E. Bauer\inst{6,7,8}\and
S. Benetti\inst{9}\and
M. G. Bernardini\inst{3}\and
K. Bhirombhakdi\inst{10}\and
M. T. Botticella\inst{11}\and
M. Branchesi\inst{12,13}\and
E. Brocato\inst{14,15}
% ...
}

\institute{Affiliations at end of paper
}

\date{Received, accepted}

\abstract 
{Gravitational wave (GW) astronomy has rapidly reached maturity becoming a fundamental observing window for modern astrophysics. The coalescences of a few tens of black hole (BH) binaries have been detected, while the number of events possibly including a neutron star (NS) is still limited to a few. On 2019 August 14, the LIGO and Virgo interferometers detected a high-significance event labelled S190814bv. Preliminary analysis of the GW data suggests that the event was likely due to the merger of a compact binary system formed by a BH and a NS.} 
{In this paper, we present our extensive search campaign aimed at uncovering the potential optical/near infrared electromagnetic counterpart of S190814bv. We found no convincing electromagnetic counterpart in our data. We therefore use our non-detection to place limits on the properties of the putative outflows that could have been produced by the binary during and after the merger.}
{Thanks to the three-detector observation of S190814bv, and given the characteristics of the signal, the LIGO and Virgo Collaborations delivered a relatively narrow localisation in low latency -- a 50\% (90\%) credible area of 5\,\sqd\ (23\,\sqd) -- despite the relatively large distance of $267\pm52$\,Mpc. ElectromagNetic counterparts of GRAvitational wave sources at the VEry Large Telescope (ENGRAVE) collaboration members carried out an intensive multi-epoch, multi-instrument observational campaign to identify the possible optical/near infrared counterpart of the event. In addition, the ATLAS, GOTO, GRAWITA-VST, Pan-STARRS and VINROUGE projects also carried out a search on this event. In this paper, we describe the combined observational campaign of these groups.} 
{Our observations allow us to place limits on the presence of any counterpart and discuss the implications for the kilonova (KN) possibly generated by this NS-BH merger, and for the strategy of future searches. The typical depth of our wide-field observations, which cover most of the projected sky localisation probability (up to 99.8\%, depending on the night and filter considered), is $r\sim 22$ (resp.~$K\sim 21$) in the optical (resp.~near infrared). We reach deeper limits in a subset of our galaxy-targeted observations, which cover a total $\sim 50\%$ of the galaxy-mass-weighted localisation probability. Altogether, our observations allow us to exclude a KN with large ejecta mass $M\gtrsim 0.1\,\mathrm{M_\odot}$ to a high ($>90\%$) confidence,
and we can exclude much smaller masses in a subsample of our observations.
This disfavours the tidal disruption of the neutron star during the merger.}{Despite the sensitive instruments involved in the campaign, given the distance of S190814bv we could not reach sufficiently deep limits to constrain a KN comparable in luminosity to AT\,2017gfo on a large fraction of the localisation probability. This suggests that future (likely common) events at a few hundreds Mpc will be detected only by large facilities with both high sensitivity and large field of view. Galaxy-targeted observations can reach the needed depth over a relevant portion of the localisation probability with a smaller investment of resources, but the number of galaxies to be targeted in order to get a fairly complete coverage is large, even in the case of a localisation as good as that of this event.}


\titlerunning{Optical/near-infrared constraints on a NS-BH merger candidate}
\authorrunning{ENGRAVE collaboration}

\maketitle

\graphicspath{{./}{figures/}}

\section{Introduction}

The discovery of the binary black hole (BH) merger event GW150914
\citep{lvc_150914} was a major landmark in the history of physics.
It was the first detection of gravitational waves (GWs) and the beginning
of GW astronomy. The detection of the first confirmed binary neutron star (NS) merger, GW170817 \citep{lvc_170817}, and
the subsequent discovery of its electromagnetic (EM) counterparts -- the short GRB 170817A \citep{lvc17grb,Goldsteinetal2017, savchenko17} and the UV/optical/IR transient AT2017gfo 
\citep{coulter17,lipunov17,tanvir17,soares-santo17,valenti17} --
was a second major breakthrough, and marked the beginning of multi-messenger astrophysics with GWs \citep{lvc_mma}. 

The subsequent investigation of GW170817 convincingly linked NS-NS mergers with short duration gamma-ray bursts 
\citep[e.g.][]{lyman18,Dobieetal2018,mooley18b,lazzati18,resmi18,ghirlanda19,lamb19,Margutti2018ApJL,Nynka2018ApJL,Troja2018MNRAS,Davanzoetal2018} -- a
link for which the evidence had been accumulating for some time
\citep{fong13b,tanvir13,berger13,Jin16}. In addition, the
identification of its host galaxy, NGC\,4993 \citep[see][]{levan17},
and an assessment of its cosmological recession velocity \citep{hjorth17} 
permitted the first measurement of a cosmological parameter (the Hubble constant) using the GW distance measurement (thanks to the ``standard siren'' nature of compact object binaries) \citep{lvc_h0}.
The optical/near-infrared (NIR) monitoring campaigns of the transient also 
unveiled for the first time the developing kilonova (KN) emission
\citep{arcavi17b,chornock17,covino17,cowperthwaite17,drout17,evans17,kasliwal17,mccully17,nicholl17,pian17,shappee17,smartt17,tanvir17}
 due to the production and decay of \emph{r}-process elements
\citep[e.g.][]{metzger10,kasen17}, demonstrating that NS-NS mergers are
indeed a major source of these elements \citep{Galletal2017,watson19}, as previously suggested
\citep{lattimer77,eichler89,li98,freiburghaus99}.

Following the success of the GW170817 follow-up campaign, considerable effort has been expended in mounting similar campaigns with the aim of discovering and characterising the counterparts of new GW events. 
To optimize the science return of the demanding observations of GW counterparts, a large fraction of the GW/EM community in member states of the European Southern Observatory (ESO) has gathered together to form the ENGRAVE (``\textbf{E}lectromag\textbf{N}etic counterparts of \textbf{GRA}vitational wave sources at the \textbf{VE}ry Large Telescope'') consortium.\footnote{\url{http://www.engrave-eso.org/}} 

This paper introduces the collaboration and our first major campaign to search for an EM counterpart to a GW source, the NS-BH event merger candidate S190814bv, reported during the O3 run of the LIGO Scientific Collaboration and the Virgo Collaboration (LVC).\footnote{\url{https://gracedb.ligo.org/superevents/public/O3/}}

\subsection{The NS-BH merger candidate S190814bv}

\gwtrig was detected by the LVC on \UTGW\ UT (MJD \MJDGW) \citep{2019GCN.25324....1L}, and an alert was issued on \alertUTGW\ UT (MJD \alertMJDGW), approximately  $21$~min after the merger. 
The source was localised to a $50$\% credible region of 133\,\sqd\ in the initial report ($90$\% credible region of 772\,\sqd), which was reduced to 5\,\sqd\ at 50\% and 23\,\sqd\ at $90$\% half a day later \citep{2019GCN.25333....1T}
making this the best-localised candidate GW event so far. 
The source distance (as inferred directly from the GW observations) is $267\pm52$\,Mpc. The estimated False-Alarm Rate (FAR)
 is extremely low, at $2.033\times10^{-33}$\,Hz (1 per $1.559\times10^{25}$ years).

Preliminary parameter estimation \citep{veitch15} indicated that the lighter object had a mass $M_2<3$M$_{\odot}$, while the heavier object had a mass  $M_1>5$M$_{\odot}$, making this a NS-BH candidate according to the LVC classification criteria. The same preliminary analysis pointed to a negligible probability of any disrupted material remaining outside the final compact object (given by the parameter \texttt{HasRemnant} $<1$\%), implying that in this case an EM counterpart was unlikely. We note that the classification of one of the components as a NS is based solely on the mass being $<3$M$_{\odot}$ and that a low mass BH is not ruled out by such low-latency classification. 

No  $\gamma$-ray, X-ray, or neutrino signal could be connected to the event \citep{2019GCN.25323....1M,2019GCN.25326....1K, 2019GCN.25385....1O,2019GCN.25327....1P, 2019GCN.25335....1P,2019GCN.25329....1S,2019GCN.25341....1P, 2019GCN.25400....1E,2019GCN.25365....1C,2019GCN.25369....1S,2019GCN.25321....1I,2019GCN.25330....1A,2019GCN.25409....1A}.

The relatively small localisation region led to a world-wide follow-up effort with optical/NIR telescopes \citep[e.g. ][]{Gomezetal2019,Andreonietal2019,Dobieetal2019,Watson2020arXiv,Antieretal2020,Vieiraetal2020}. The ENGRAVE collaboration activated its search programmes to try to discover, or set limits on, an EM counterpart to \gwtrig.
ENGRAVE members ran wide-field searches for EM counterparts and the ATLAS, GOTO, GRAWITA-VST, Pan-STARRS, and VINROUGE projects also triggered their searches on this event. No promising EM counterpart was detected. In this paper we combine our ENGRAVE ESO/Very Large Telescope (VLT) data, a number of other narrow-field facilities, and the wide-field programmes to 
report the combined results of our search for a counterpart (\sect\ref{sect:observations} and \ref{sec:galaxyobs}). We place limits 
on the presence of a counterpart (\sect\ref{sect:incompleteness}), and discuss the implications of these limits for NS-BH mergers and future searches (\sect\ref{sect:discussion}).
Unless otherwise specified, errors are given at 68\% confidence level ($1\sigma$), upper limits are given at $3\sigma$,
and magnitudes are in the AB system. 
When needed, we assume a flat FLRW cosmology with $H_0=70\,\mathrm{km\,s^{-1}\,Mpc^{-1}}$ and $\Omega_\mathrm{m}=0.3$.

\section{Wide-field survey observations and results}
\label{sect:observations}

We employed two different approaches to search for an optical or NIR counterpart to {\gwtrig}. A number of wide-field facilities (with Field of View, FoV, of 1 \sqd\ or more) were used to tile the LVC GW sky localization probability maps (skymap) with the aim of covering as much of the 2D probability of S190814bv localisation as possible. These telescopes (with apertures of 0.4m to 4m) were situated in La Palma, Chile and Hawaii, giving a spread of latitude and temporal coverage. The second approach was to target specific galaxies in the 3D sky region 
with larger aperture (2m to 8m) telescopes and smaller FoV cameras. In the next subsections we summarize the search for transients with the different facilities.

\subsection{The search for transients with GOTO}

\begin{figure*}
    \centering
    \includegraphics[width=\linewidth]{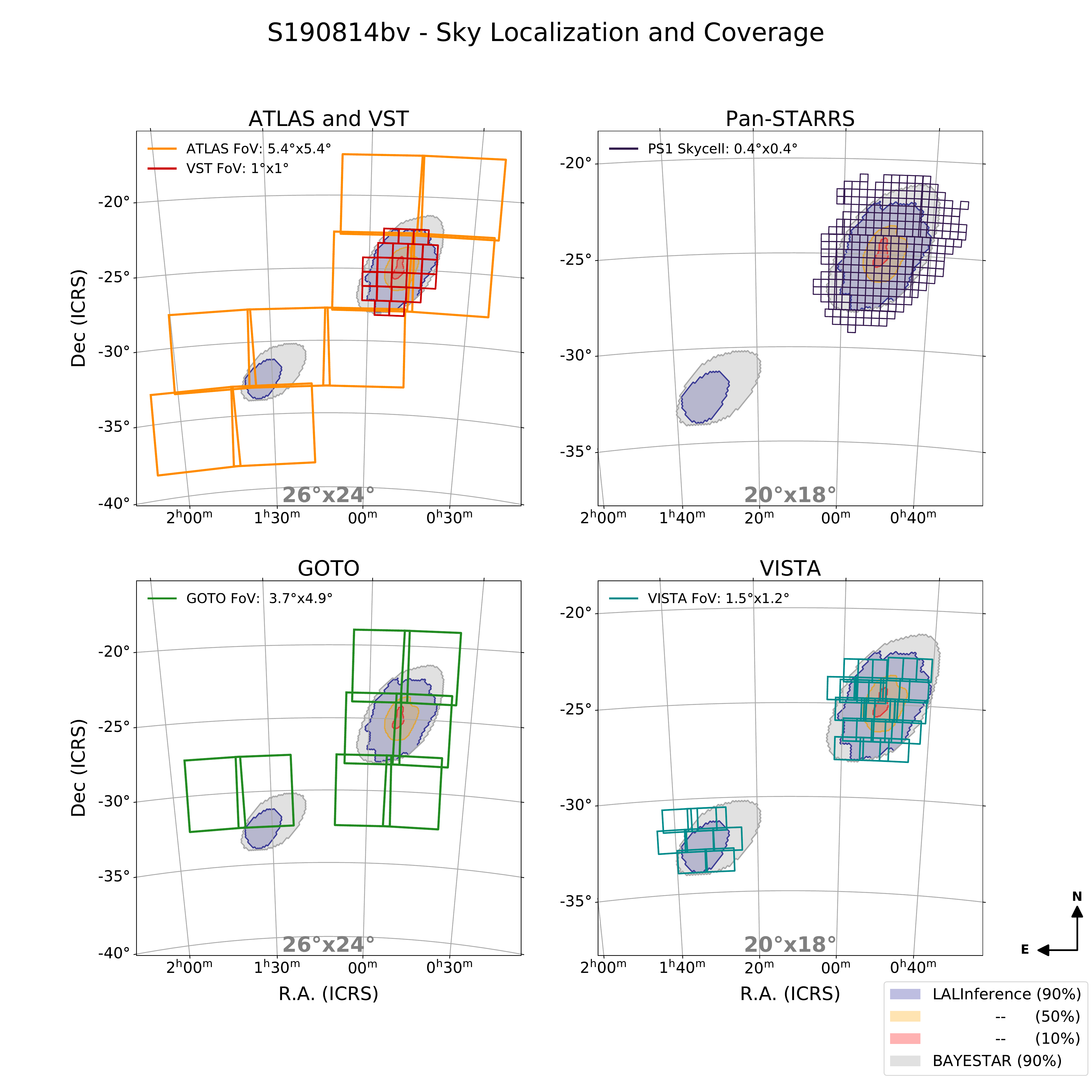}
    \caption{Coverage maps from the wide-field surveys as listed in Table\,\ref{tab:coverage_summary} with the probability contours of the initial skymap (\texttt{BAYESTAR}) and the refined skymap \texttt{LALInference}.}
    \label{fig:coveragemaps}
\end{figure*}

The Gravitational wave Optical Transient Observer (GOTO\footnote{\url{https://goto-observatory.org}}) is a robotic array of wide-field optical telescopes sited at the Roque de los Muchachos Observatory in La Palma. It is operated by the University of Warwick on behalf of an international collaboration. The hardware is modular in design and optimised to autonomously respond to GW events, being able to cover large areas of sky quickly. At the time of S190814bv, GOTO was equipped with 4 active unit telescopes, each having an aperture of 40\,cm at $f/2.5$ and featuring a 50-megapixel CCD detector. This corresponds to a plate-scale of 1\farcs25\,pix$^{-1}$ and a FoV of 5.9 \sqd\ per camera. 
Observations of the S190814bv error box were automatically scheduled on the basis of a ranked tiling pattern derived from the available skymaps \citep{dyer18, Gompertz20b}. Each tile was observed using sequences of 60\,s or 90\,s exposures with the GOTO-$L$ filter, which is a wide filter covering 400--700\,nm (slightly wider than Sloan $g$+$r$ combined).
The bulk of the final localisation probability, as given by the \texttt{LALInference} skymap (\texttt{LALInference.v1.fits}), was covered in 8 observable tiles (Fig.~\ref{fig:coveragemaps}), with 89.6\% probability covered over the timespan MJD$=$58710.091--58710.230 (5.09--8.34\,hr after the GW event).
Additional observations were obtained the following night, which covered 94.1\%. The Moon was closer to the relevant tiles during this second night, affecting the zeropoints achieved in the exposures sets.
Most tiles were observed multiple times (see Table~\ref{tab:widefield-obs-list}), though observing conditions were not optimal, given the presence of the Moon nearby and poor weather. The probability regions were also close to GOTO's lower declination limit, meaning some of the GW probability region could not be observed. Individual exposures are median combined in groups of 3--6 subsequent images and reached $5\sigma$ limiting magnitudes covering $V{=}17.4$--19.1 mag. These are derived by photometrically calibrating our photometry to AAVSO Photometric All-Sky Survey (APASS\footnote{\url{https://www.aavso.org/apass}}) stars using the $V$ band.

Raw GOTO images are transferred from La Palma to the Warwick data centre in real-time and processing begins minutes after acquisition using the GOTOphoto pipeline (GOTO collaboration, in prep). Image-level processing includes detector corrections, astrometry tied to {\it Gaia} and photometric zeropoints using a large number of field stars.
Difference imaging was performed on the median exposures using recent survey observations 
as reference. Source candidates were initially filtered using a trained classifier and cross-matched against a variety of catalogues, including the Minor Planet Center (MPC)\footnote{\url{https://minorplanetcenter.net/cgi-bin/checkmp.cgi}} and Pan-STARRS1 $3\pi$
survey \citep{chambers16}. The classifier employs a random forest algorithm based on image features, largely following the procedure of \citep{2012PASP..124.1175B}. It was trained using an injected source data-set. Human 
vetting was performed on 
the resulting candidates
using a web-based marshall interface. No viable optical counterpart candidates could be associated with S190814bv \citep{2019GCN.25337....1A}.

\subsection{The search for transients with the VST}

The \textit{VLT Survey Telescope} \citep[VST; ][]{Capaccioli11} is a 2.6-m facility located at Cerro Paranal, Chile, and managed by ESO.
The telescope is equipped with OmegaCam \citep{Kuijken11}, a 268-megapixel camera with a FoV of 1 $\times$ 1 \sqd\ and a resolution of 0\farcs21\,pix$^{-1}$.
The filter set includes Sloan $ugriz$ filters. The telescope is operated in service mode.
The GW trigger follow-up is performed using Guaranteed Time allocated to the Italian VST and OmegaCam Consortium.

The monitoring of the S190814bv sky area started on MJD 58710.36 \citep{2019GCN.25371....1G,2019GCN.25669....1G,2019GCN.25748....1Y}. For each pointing and epoch we obtained three dithered exposures for a total exposure time of 135\,s. 
The pointings were visited up to five times during a period of two weeks (see Table~\ref{tab:widefield-obs-list}). All exposures were obtained using the $r$ filter.

On the first night we imaged 15\,\sqd, covering 53.6\% of the localisation probability of the preliminary \texttt{BAYESTAR} skymap (which was the only map available at that time), and 60.7\% of the final \texttt{LALInference} skymap probability. 
Starting from the second epoch we revised the pointing list to optimize the sky coverage
for the updated \texttt{LALInference} skymap. 
The survey area increased to 23\,\sqd\ covering a maximum of 87.7\% of the localisation probability, as shown in Fig.~\ref{fig:coveragemaps}.

Details of the image processing and candidate detection are given in \citet{Brocato18} and \citet{Grado19}.
The area identified by the skymap is not fully covered by VST archive observations, hence we used both Pan-STARRS1
\citep{chambers16} and DECam \citep{Abbottetal2018ApJS}
archive images as templates for the comparison.
To select the candidates, we applied a random-forest machine learning algorithm trained on previous search instances \citep{grawitaml} and then visually inspected the candidates with the highest score. 
We detected a number of transients from which we removed objects detected only at one epoch and/or associated with stellar sources in the template images. The final list includes 27 transients (reported in Table \ref{tab:VSTtransients}).
Out of these, 21 objects were already discovered by other surveys and had been registered on the on the Transient Name Server (TNS)\footnote{\url{https://wis-tns.weizmann.ac.il/}}.
In Table~\ref{tab:VSTtransients} we include also three transients reported in TNS by other groups that are detected on our images but are below the detection threshold of our search and were therefore not independently detected in our search. 

All of these candidates show a slow evolution in the two weeks of the observing campaign ($\Delta m \textless 1$ mag between the first and last detection). Therefore we tentatively exclude that any of them are associated with the GW event (see \S\ref{sec:ruling_out_candidates} and Appendix~\ref{sect:candidates} for more details on the rejected candidates). The limiting magnitudes of the stacked images were estimated by means of artificial star experiments. 
The limiting magnitude for each pointing, defined as the magnitude at which $50\%$ of the artificial stars are recovered, is reported in Tab.~\ref{tab:widefield-obs-list}.
The limits are shallower in the first epoch because of a high background due to full Moon.

\begin{table*}[]
    \centering
    \caption{Summary of wide-field survey coverage and typical limiting magnitudes. The start MJD refers to the start of observations on that night (for reference, the GW trigger occurred at MJD\,58709.882). The given limiting magnitude is the median magnitude of the individual tiles that covered the probability listed. All times are in the observer frame.  }
    \begin{tabular}{llllll}\hline\hline
    Telescope    & Start MJD     &  Time after GW & Probability Coverage & Limiting mag & filter   \\\hline
   ATLAS         &    58709.52   &  -8.7\,hr   & 99.8\%         & 18.0       & $c$   \\
    GOTO         &    58710.09   &  +5.0\,hr    & 89.6\%    &  18.7        &   $L$ \\
    VST          &    58710.37   &  +11.5\,hr   & 60.7\%        & 20.9       & $r$ \\
    Pan-STARRS1  &    58710.528  &  +15.50\,hr  & 89.4\%          & 20.6, 20.3 & \ips, \zps\\
    ATLAS        &    58710.60   &  +17.23\,hr  & 99.8\%         & 18.0       & $o$   \\
    GOTO         &    58711.09   & +1.2\,d         & 94.1\%    &  18.1        & $L$  \\
    VISTA-wide        &    58711.17   & +1.3 -- 3.4\,d      & 94\%          &  21.0      & $K_{\rm s}$ \\
    VISTA-deep        &    58711.24   & +1.4\,d      & 21\%          &  22.0      & $K_{\rm s}$ \\ 
    VST          &    58711.2    & +1.5\,d      & 71.5\%        &  21.9      & $r$ \\
    ATLAS        &    58711.5    & +1.6\,d      & 99.8\%         & 17.6       & $o$   \\
    Pan-STARRS1  &    58713.5    & +3.6\,d      & 70.4\%          &  21.9      & \zps\\
        VST      &    58714.2    &  +4.3\,d     & 87.7\%        &  21.7      & $r$ \\
    Pan-STARRS1  &    58716.5    &  +6.6\,d     & 70.7\%          &  23.0      &  \zps\\
    VST          &    58717.1    &  +7.2\,d     & 87.7\%        &  21.8      & $r$ \\
 VISTA-wide        &    58719.05   & +9.2--10.5\,d      & 94\%          &  21.2      & $K_{\rm s}$ \\
 VISTA-deep        &    58720.15   & +10.3\,d      & 21\%          &  22.0      & $K_{\rm s}$ \\ 
 VST          &    58724.4    &  +14.5\,d    & 87.7\%        &  22.0      & $r$ \\
VISTA-wide        &    58750.1   & +40--41\,d      & 94\%          &  21.0      & $K_{\rm s}$ \\ 
VISTA-deep        &    58751.1   & +41\,d      & 21\%          &  22.0      & $K_{\rm s}$ \\ 
\hline
    \end{tabular}
    \label{tab:coverage_summary}
\end{table*}

\begin{table*}
\caption{Transients detected with the VST ($r$- band). }
\label{tab:VSTtransients}
\scriptsize
\centering
\begin{tabular}{cccccccc}
\hline\hline
 Name & TNS Name & RA (hms) & Dec (dms) & MJD & Mag (err)  & Note & tiling \\
\hline
VST J005109.17-221740.7 & AT2019qbu & 00:51:09.173 & -22:17:40.69 & 58715.16 & 21.19 +/- 0.09 & 1  & T13\\
VST J004414.33-250744.3 & AT2019qby & 00:44:14.334 & -25:07:44.32 & 58711.27 & 21.35 +/- 0.02 & 1  & T1\\
VST J005653.99-275921.4 & AT2019qbz & 00:56:53.987 & -27:59:21.37 & 58711.30 & 20.89 +/- 0.07  & 1  & T24\\
VST J004548.54-264939.0 & AT2019qca & 00:45:48.540 & -26:49:39.01 & 58725.33 & 21.51 +/- 0.05  & 1  & T9\\
VST J004619.06-260843.2 & AT2019qcb & 00:46:19.062 & -26:08:43.19 & 58714.24 & 21.55 +/- 0.04  & 1  & T8\\
VST J005349.82-244549.6 & AT2019qcc & 00:53:49.820 & -24:45:49.58 & 58725.35 & 22.14 +/- 0.09  & 1  & T21\\
VST J004656.70-252236.7 & AT2019npd & 00:46:56.711 & -25:22:36.43 & 58717.22 & 19.87 +/- 0.06  &  2 & T8\\
VST J004847.88-251823.5 & AT2019noq & 00:48:47.882 & -25:18:23.46 & 58711.23 & 19.96 +/- 0.03  & 2 & T16\\
VST J005605.55-243826.4 & AT2019nve & 00:56:05.510 & -24:38:26.40 & 58710.38 & 20.53 +/- 0.08  & 2  & T21\\
VST J004659.45-230559.5 & AT2019nyv & 00:46:59.451 & -23:05:59.50 & 58711.27 & 21.20 +/- 0.09  & 2  & T12\\
VST J005542.30-244149.9 & AT2019nvd & 00:55:42.301 & -24:41:49.93 & 58725.35 & 21.09 +/- 0.05 & 2  & T21\\
VST J005002.82-224118.8 & AT2019mwp & 00:50:02.820 & -22:41:18.78 & 58711.26 & 20.57 +/- 0.13 & 2  & T13\\
VST J004804.40-234750.9 & AT2019ntm & 00:48:04.398 & -23:47:50.94 & 58711.25 & 21.07 +/- 0.11 & 2  & T10\\
VST J010001.84-264251.3 & AT2019ntr & 01:00:01.843 & -26:42:51.32 & 58711.29 & 21.21 +/- 0.15 & 2  & T27\\
VST J005012.07-261152.6 & AT2019ntp & 00:50:12.072 & -26:11:52.56 & 58711.23 & 21.04 +/- 0.03 & 2  & T16\\
VST J005305.56-242138.7 & AT2019npz & 00:53:05.560 & -24:21:38.71 & 58714.25 & 20.87 +/- 0.08 & 2  & T15\\
VST J004616.81-242221.2 & AT2019nxe & 00:46:16.814 & -24:22:21.19 & 58714.25 & 20.87 +/- 0.07 & 2  & T7\\
VST J004320.49-255302.1 & SN2019mbq & 00:43:20.493 & -25:53:02.07 & 58715.17 & 18.83 +/- 0.04 & 2  & T2\\
VST J004901.74-231404.9 & AT2019nuj & 00:49:01.738 & -23:14:04.93 & 58715.16 & 21.72 +/- 0.18 & 2  & T13\\
VST J004133.33-234432.0 & AT2019npe & 00:41:33.330 & -23:44:31.95 & 58717.15 & 21.51 +/- 0.07 & 2  & T0\\
VST J005552.40-254659.8 & AT2019npw & 00:55:52.399 & -25:46:59.81 & 58711.24 & 21.34 +/- 0.05 & 2  & T23\\
VST J004330.16-224329.4 & AT2019nsm & 00:43:30.160 & -22:43:29.35 & 58717.15 & 21.39 +/- 0.09 & 2  & T5\\
VST J005531.60-225808.5 & AT2019num & 00:55:31.602 & -22:58:08.48 & 58717.14 & 21.39 +/- 0.09 & 2  & T19\\
VST J005243.34-233753.6 & AT2019nva & 00:52:43.339 & -23:37:53.64 & 58717.13 & 21.36 +/- 0.07 & 2  & T14\\
VST J005646.69-250933.3 & AT2019nqw & 00:56:46.693 & -25:09:33.29 & 58725.35 & 20.77 +/- 0.03 & 2  & T21\\
VST J005806.46-245014.3 & AT2019nzd & 00:58:06.456 & -24:50:14.28 & 58714.23 & 21.18 +/- 0.10 & 3  & T25\\
VST J005756.90-243400.5 & AT2019nys & 00:57:56.904 & -24:34:00.48 & 58714.23 & 21.31 +/- 0.10 & 3  & T25\\
VST J005332.32-234958.5 & SN2019npv & 00:53:32.316 & -23:49:58.50 & 58717.14 & 21.62 +/- 0.09 & 3  & T20\\
\hline
\end{tabular}
\begin{flushleft}
Notes: \\
1 - New transient candidates which we first reported in TNS.We verified that AT2019qcb and AT2019qcc are also visible in Pan-STARRS1 subtractions. AT2019qcb (VST J004619.06-260843.2) is detected at  $\zps = 21.1\pm 0.1$ mag, 
on MJD~58716.53. It is a nuclear transient coincident with a compact galaxy (Kron mag) $\rps=18.68$ mag. 
AT2019qcc (VST J005349.82-244549.6) is detected in two Pan-STARRS1 subtractions at 
$\zps = 21.6 \pm 0.2$ mag on both MJD 58713.54 and MJD 58716.54 and has a flat light curve in $\zps$. It is also coincident with a probable compact galaxy $\rps = 20.7$ (Kron mag).
\\
2 - Independent discoveries already reported in TNS by other groups. \\
3 - Candidates that are detected on our images but are below the threshold of our search criteria.
\end{flushleft}
\end{table*}

\subsection{The search for transients with VISTA}

The 4.1\,m \textit{Visible and Infrared Survey Telescope for Astronomy} (VISTA) is sited at Cerro Paranal and operated by the European Southern Observatory (ESO).
The VISTA InfraRed CAMera (VIRCAM) has 16 detectors arranged in a sparse array, and conventionally six pointings are combined with offsets to form a contiguous ``tile'' of $\sim$1.6\,\sqd \citep{sutherland2015}.

Observations were made with VISTA under the VINROUGE programme at three epochs, the first over several nights post-merger (beginning at MJD 58711.17), the second around a week later, and the final epoch roughly seven weeks post-merger which was used as our primary reference template.
We only observed in the $K_{\rm s}$ band ($2.15\,\mu$m), to optimise our search for a red KN component.
A large majority (>90\%, see Table~\ref{tab:coverage_summary}) of the \texttt{LALInference} localisation area (referred to here as `VISTA-wide') was covered at all three epochs, as shown in Fig.~\ref{fig:coveragemaps}. The single tile covering the highest likelihood region was re-imaged six times to provide deeper limits in that area (referred to as `VISTA-deep', enclosing $\sim 21\%$ of the sky localisation probability).
Full details of the area covered, timing and representative depth reached are given in Tables~\ref{tab:coverage_summary} and \ref{tab:widefield-obs-list}.

Initial processing of the data was performed using a pipeline based on
the VISTA Data Flow System \citep[VDFS;][]{ GonzalezFernandez18} modified for on-the-fly processing.
Subsequently, the VINROUGE in-house pipeline for the Identification of GW counterparts through NIR Image Subtraction (\texttt{IGNIS}) was used to aid the search for transient sources. 
Using object lists generated from the VDFS pipeline for both the science and template images, positions were cross-checked to create a list in which the majority of objects visible across multiple epochs were removed, along with objects associated with error flags.

Template and science images were paired based on area coverage, with templates resized to encompass the entire science area per image. Coordinates were aligned through a combination of the \texttt{astrometry.net} software \texttt{solve-field}, and \texttt{SExtractor} \citep{bertin1996} and \texttt{swarp} to match template image positions directly to their corresponding science images. Template images were then subtracted from science images using the \texttt{hotpants} tool\footnote{\url{http://www.astro.washington.edu/users/becker/v2.0/hotpants.htm}} \citep{becker2015}. 

We searched for potential counterparts in the subtracted images using two approaches: first by eyeballing the regions around obvious galaxies, particularly those thought to be in the distance range of interest, and secondly through an automated search for sources.

Candidate transients from the automated search were culled based on various criteria,
in particular, objects within low confidence regions (e.g.~tile edges), with $<5\,\sigma$ detection significance, coincident with foreground stars or the bright cores of  galaxies (for which the subtractions often left scars), or unusually sharp images suggestive of hot pixels rather than stellar sources. 
Moving sources were identified by reference to the MPC.
A final check involved human vetting of remaining candidates (typically 1--10 per science image).

This process was repeated for all available science data (including the VISTA-deep field) across the three epochs, cross-matching them over with as many template files as could be attributed to each. 

Of the sources found in the automated procedure, all were deemed to be image artefacts. Similarly, no convincing new sources were found in the eyeball search, with the exception of the known transient AT2019noq, which was found to have AB magnitudes $K=20.12\pm0.07$ (at MJD 58711.23) and $K=20.06\pm0.07$ (at MJD 58719.25). This source was marginally below the adopted significance threshold in the subtracted image.

\subsection{The search for transients with ATLAS}

The \textit{Asteroid Terrestrial-impact Last Alert System} (ATLAS) is a high cadence, near-Earth asteroid (NEA) survey with two telescopes located on two separate sites in Hawaii (Mauna Loa and Haleakala). The $f/2$ telescopes are 0.5\,m aperture  with 10.56k $\times$ 10.56k pixel CCD cameras \citep{tonry18}. The plate scale is 1\farcs86\,pix$^{-1}$, giving each camera a FoV of 29.2\,\sqd. Both units scan the sky between 
$-40^{\circ} < \delta < +80^{\circ}$ 
with a cadence of approximately two days, weather permitting. ATLAS survey mode uses  two composite filters - `cyan' and `orange' (\textit{c} and \textit{o}, respectively). Cyan covers the Sloan \textit{g} and \textit{r} filters and orange covers the Sloan \textit{r} and \textit{i} filters. 

A typical NEA survey observing cycle is comprised of a sequence of 4 slightly dithered exposures (which we call quads), each lasting 30\,s, with overheads and processing requiring an additional 10\,s.
The 4 exposures are typically separated by 15 minutes within a 1\,hr period to allow for detection and 
linking of fast-moving objects.  ATLAS frequently adjusts this NEA optimised schedule to carry out similar sequences of quads over the sky area of a GW sky map
\citep[e.g.][]{stalder17}. Observations are processed by an automatic pipeline to produce de-trended, sky-flattened images. These are corrected astrometrically with respect to the ICRS using {\it Gaia} stellar positions, and corrected photometrically with respect to a custom built reference catalogue \citep[Refcat2;][]{refcat2}. Difference imaging is employed to identify transients in the survey data and source extraction and measurement are carried out as described in \cite{tonry18}. All detections with $S/N \geq 5$ are read into a database at Queen's University Belfast and we require 3 or more detections at $S/N \geq 5$ to form an object detection. After such objects are defined, they are subject to various quality filters, machine
learning algorithms and cross-matching to known minor planet, star and galaxy catalogues. 

ATLAS was serendipitously observing the \gwtrig skymap region several hours {\em before} the GW detection during its normal survey mode. Hence any recent, young and bright transients would have been identified. Seven pointings of ATLAS covered the entirety of the \texttt{LALInference} skymap, and the first pre-discovery observation of the map started at MJD 58709.52 (8.7\,hr before S190814bv). 
The coverage continued until MJD 58709.635, or 2.8\,hr later (Figure \ref{fig:coveragemaps}). 
Only the Haleakala telescope observed, in the cyan ($c$) filter. Some of the earliest exposures were affected by cloud cover and moonlight. 
No new transient objects which are not cross-matched with stars or known AGN were found 
in our images. 
ATLAS re-observed the field on the next two subsequent nights, in the $o$ band (with the Mauna Loa unit).
The second night of observations began at MJD 58710.602 (17.26\,hr after S190814bv)
covering 99.8\%  of the  localisation area within a 1\,hr period. The third night of observations began at MJD 58711.6, again covering 99.8\% of the \texttt{LALInference} skymap probability. In none of the three post-event epochs did we find any new transients within the GW localisation area of ATLAS.

\subsection{The search for transients with Pan-STARRS1}

The Pan-STARRS system \citep{chambers16} comprises 2$\times$1.8\,m  telescopes on Haleakala, each with a 1.4-Gigapixel camera mounted at the Cassegrain $f/4.4$  focus of each unit. Here we describe observations with the Pan-STARRS1 telescope (PS1) and the camera GPC1. 
The GPC1 is composed of sixty Orthogonal Transfer Array devices (OTAs), each of which has a detector area of 4846 $\times$ 4868 pixels. The 10 micron pixels (0\farcs26) give a focal plane of 418.88\,mm in diameter or 3.0 degrees. This provides a
FoV area of 7.06 \sqd, and an active region of about 5\,\sqd\ 
\citep[see][for a description of the focal plane gaps]{chambers16}. 
The five filter system (generally denoted \grizy) is described in 
\cite{tonry12} and \cite{chambers16}. For filters in common, the PS1 
filters have similar transmission profiles as those from the Sloan Digital Sky Survey \citep[SDSS;][]{abazajian09}. 
Images from \PS\ are processed immediately with the Image Processing Pipeline \citep{magnier16a,waters16}. The existence of the PS1 3$\pi$ Survey data \citep{chambers16} provides a ready-made template image of the whole sky north of $\delta = -30^{\circ}$, and we furthermore have proprietary \ips\ data in a band between $-40^{\circ} < \delta < -30^{\circ}$, giving reference sky images in the \ips\ band down to this lower declination limit. All individual PS1 images have a reference sky subtracted from them and sources with at least two detections with $S/N \geq 5$ significance and spatially coincident to within 0\farcs5 are detected and measured. PS1 typically observes in a quad sequence similar to ATLAS, with a set of $4\times45$\,s exposures taken across a time span of 1\,hr to identify and link moving sources. The PS1 team can intervene at any moment and direct the telescope to observe a LVC GW sky map with a flexible choice of filter, exposure time, coverage, and dither and stack strategy. The difference images can be combined into deeper stacks or processed individually and the sources resulting from these are read into a large database at Queen's University Belfast. A series of quality control filters, machine learning algorithms and cross-matches against minor planet, stellar and galaxy catalogues are automatically run and human scanning occurs for all objects 
not coincident with known solar system objects, stars or catalogued AGN
\citep[see][for more details]{smartt16a,smartt16b}. 
At the detection time of S190814bv (\UTGW), Hawaii was in day time and PS1 began observing the field at 2019 Aug 15 12:40:37 UT, or 15.50\,hr after 
the LVC discovery time.

On the first night of observation 
the individual 45\,s exposures (called ``warps'') were processed individually to search for any fading transient over the 2\,hr\,33\,min period of observation. 
Image sensitivities are estimated by injecting 500 point sources per skycell across a range of magnitudes and  the limiting magnitude is defined when 50\%  of the sources are recovered \citep[described in the content of database table \texttt{DiffDetEffMeta} in ][]{Flewelling16}.  
Each chip exposure is warped onto a pre-defined tesselation \citep[called skycells, see][]{chambers16}, and the limits refer to these skycells. 
The 45\,s exposures were combined into a nightly stack on the first three nights of observing. The stacks are made by median combining the warps of each skycell.
On the first observing night 25 exposures were combined in \ips\ and 31 exposures in \zps\ in each skycell stack, giving a typical exposure time of 1125\,s and 1395\,s in \ips\ and \zps, respectively. We did not find any fading transient, but the true constraints are weak due to the dither strategy and fill factor.  

For the two subsequent nights, we did not process the individual images, rather we combined all the \zps band warps into a nightly stack. The effective exposure times were 12480\,s and 13440\,s on these respective nights. 
These were deeper than the 3$\pi$ reference stacks in \zps\ in this sky region, so over the next four weeks PS1 observed the region in \zps\ to make a deeper and more uniform reference stack.
The limiting magnitudes of the skycells on the three nights observing of \gwtrig\ were calculated using the new, custom-made deeper reference-stack for template subtraction. 
The final sky coverage is plotted in Figure \ref{fig:coveragemaps}. 

All images were processed through the Image Processing Pipeline described above.
Detections
coincident with known stellar objects from the {\it Gaia} DR1 \citep{2016A&A...595A...2G}, 
Guide Star Catalogue,\footnote{\url{https://archive.stsci.edu/gsc/}} 
Two Micron All Sky Survey \citep[2MASS][]{2006AJ....131.1163S}, 
SDSS DR12 \citep{2015ApJS..219...12A}
and PS1 \cite{chambers16}
catalogues were rejected. 
Additionally any objects coincident with known AGN 
were identified and excised from the transient search list. As discussed in \cite{smartt16a,smartt16b}, the AGN identification is based mostly on the \cite{veron01} and MILLIQUAS\footnote{\url{http://quasars.org/milliquas.htm}} catalogues \citep{MILLIQUAS}. The resultant objects are spatially cross-matched against known galaxies (mostly through the NASA Extragalactic Database, NED\footnote{\url{https://ned.ipac.caltech.edu/}})
and all are visually inspected. The objects discovered are listed in Table\,\ref{tab:PS1objects}, along with their likely classification
\citep{2019GCN.25356....1H,2019GCN.25417....1S,2019GCN.25455....1S}. None of these objects is a viable counterpart of S190814bv (see \S\ref{sec:ruling_out_candidates} for details on the candidate rejection).

\begin{table*}[]
\caption{Table of all PS1 objects discovered for S190814bv. The host galaxies are the primary names as catalogued now in NED. 
Spectroscopically classified events are noted, and ``Probable SN'' means the light curve points we have are consistent with it being an unrelated supernova. There are two objects in the nearby galaxy NGC 253, which are certain variable stars in the outskirts of the disk (labelled VS in the table). The final column gives the 2D skymap probability contour within which the transient position lies. Some of the sources are not in the redshift range of the GW event (i.e. $0.046 - 0.068$). A machine-readable file with all photometry for these candidates is available as online supplementary material.}
\scriptsize
\label{tab:PS1objects}
\begin{tabular}{lllccccccc}
\hline
PS1 ID & RA J2000      & Dec J2000           & Classification  Type & Disc. Epoch & Disc. Mag & Host                          & Redshift & IAU ID & Prob. Contour \\\hline
PS19ekf              & 00:46:57.39 & $-$24:21:42.6 & Probable SN    & 58710.547   & 19.66 (i)  & WISEA J004657.40-242142.6                & $-$      & AT2019nbp       & 40                \\
PS19epf              & 00:48:48.77 & $-$25:18:23.4 & Probable SN    & 58710.585   & 19.93 (i)  & WISEA J004847.51-251823.0                & $-$      & AT2019noq       & 20                \\
PS19eph              & 00:49:51.99 & $-$24:16:17.7 & Probable SN    & 58710.545   & 19.46 (i)  & 6dFJ0049520$-$241618           & 0.435622      & AT2019nor       & 10                \\
PS19epw$^{1}$              & 00:46:56.71 & $-$25:22:36.6 & VS in NGC 253       & 58710.586   & 20.28 (i)  &  NGC 253                           & 0.0008   & AT2019npd       & 50                \\
PS19epx              & 00:56:50.42 & $-$24:20:50.0 & Probable SN    & 58710.587   & 20.66 (z)  &  WISEA J005650.42-242050.3                 & $-$      & AT2019nqp       & 80                \\
PS19epz              & 00:50:26.34 & $-$25:52:57.8 & Probable SN        & 58713.541   & 21.85 (z)  &   faint, uncatalogued host                         & $-$      & AT2019nuw       & 20                \\
PS19eqa              & 00:50:21.01 & $-$23:42:46.7 & Probable SN        & 58713.541   & 21.75 (z)  & WISEA J005021.03-234246.0                            & $-$      & AT2019nux       & 50                \\
PS19eqb              & 00:50:50.39 & $-$25:29:29.5 & Probable SN        & 58713.541   & 21.01 (z)  & PSO J012.7099-25.4915                        & $-$      & AT2019nuy       & 20                \\
PS19eqc              & 00:49:52.26 & $-$25:31:25.6 & Probable SN        & 58713.541   & 21.89 (z)  &  PSO J012.4678-25.5238                          & $-$      & AT2019nuz       & 20                \\
PS19eqd              & 00:52:43.39 & $-$23:37:54.0 & Probable SN        & 58713.541   & 21.49 (z)  &  PSO J013.1807-23.6317                          & $-$      & AT2019nva       & 70                \\
PS19eqe              & 00:46:51.16 & $-$25:25:39.3 & VS in NGC 253         & 58713.541   & 21.72 (z)  &  NGC 253                           & $-$      & AT2019nvb       & 50                \\
PS19eqf$^{2}$              & 00:52:18.32 & $-$26:19:42.0 & SN II          & 58713.543   & 21.31 (z)  & WISEA J005218.36-261942.5  & 0.070     & AT2019nvc       & 50                \\
PS19eqg              & 00:55:42.39 & $-$24:41:50.2 & Probable SN         & 58713.541   & 21.47 (z)  & PSO J013.9262-24.6973                          & $-$      & AT2019nvd       & 60                \\
PS19eqh$^{3}$              & 00:56:05.51 & $-$24:38:26.3 & Probable SN   & 58713.541   & 21.30 (z)  &   PSO J014.0230-24.6407                          & $-$      & AT2019nve       & 60                \\
PS19eqi$^{4}$              & 00:53:32.30 & $-$23:49:58.6 & SN  Ib         & 58713.541   & 21.26 (z)  &  WISEA J005332.35-234955.8                          & 0.056    & SN2019npv       & 70                \\
PS19eqj$^{5}$              & 00:55:52.39 & $-$25:46:59.7 & SN  IIb        & 58713.544   & 21.35 (z)  &  PSO J013.9687-25.7831                          & 0.163    & AT2019npw       & 70                \\
PS19eqk$^{6}$              & 00:56:46.71 & $-$25:09:33.4 & Probable SN        & 58713.542   & 21.23 (z)  &  PSO J014.1947-25.1593                          & $-$      & AT2019nqw       & 80                \\
PS19eqo              & 00:48:16.08 & $-$25:28:14.9 & Probable SN         & 58713.539   & 20.89 (z)  & WISEA J004816.11-252814.8                          & $-$      & AT2019nvr       & 40                \\
PS19eqp              & 00:52:37.75 & $-$26:11:41.4 & Probable SN             & 58713.549   & 21.44 (z)  & WISEA J005237.72-261142.4                & $-$      & AT2019nvs       & 50                \\
PS19eqq$^{7}$              & 00:50:12.06 & $-$26:11:52.8 & SN Ic-BL       & 58713.541   & 21.31 (z)  & WISEA J005012.11-261154.7                            & $-$      & AT2019ntp       & 50                \\
PS19erd              & 00:55:19.23 & $-$26:11:50.7 & Probable SN     & 58716.542   & 21.41 (z)  &  WISEA J005519.14-261150.9                          & $-$      & AT2019ofb       & 70     \\\hline      
\end{tabular}
~\\
1. Discovered by DECam-Growth: DG19hqpgc.\\
2. Classified by \cite{2019GCN.25423....1R}. \\
3. Offset by 3\farcs8 from galaxy WISEA J005605.37-243830.5, but coincident with faint uncatalogued stellar source.\\
4. Discovered by DECam-Growth: DG19wxnjc. Classified by \cite{2019GCN.25483....1G}, \cite{2019GCN.25478....1D}, and \cite{2019GCN.25454....1J}.\\
5. Discovered by DECam-Growth: DG19wgmjc. Classified by \cite{2019GCN.25484....1T}.\\
6. Discovered by DECam-Growth: DG19xczjc.\\
7. Discovered by DECam-Growth: DG19gcwjc. Classification reported in \cite{2019GCN.25596....1W}, but no redshift given. \\
\end{table*}

\begin{figure}
\centering
\includegraphics[width=\linewidth]{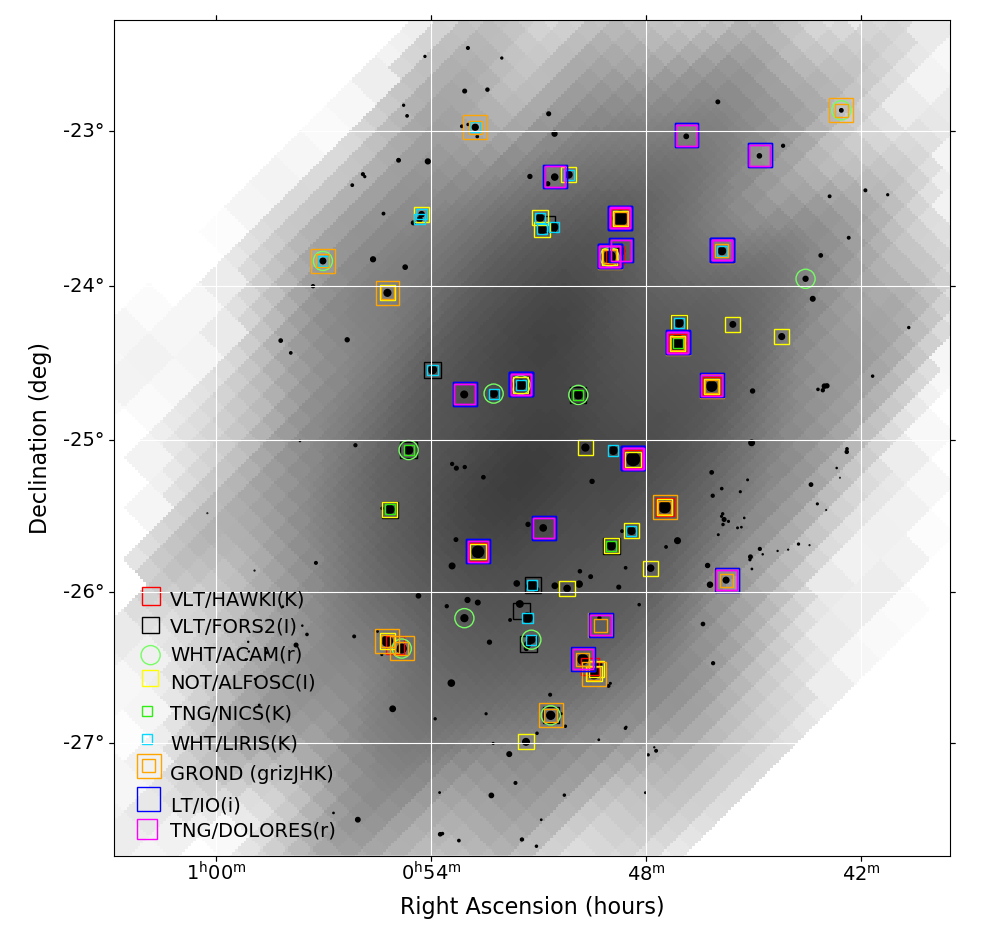}
\caption{The LVC skymap of S190814bv (\texttt{LALInference.v1.fits}) in greyscale, with 
galaxies selected through \texttt{HOGWARTS} over-plotted (black dots). Only the higher probability Northern region of the skymap is included, since the Southern region was not followed up by ENGRAVE. The size of the symbol of each galaxy (i.e. the black dots) is proportional to the probability of hosting the GW event (see table~\ref{tab:targeted-galaxies-LK-Pgal}) given the skymap and a weighing scheme following \citet{arcavi17b}. The various instruments are illustrated with different colours as in the figure legend, and the typical limiting magnitudes and filters used are given in table~\ref{tab:targeted-galaxies-LK-Pgal}.}
\label{fig:gal_target_coverage}
\end{figure}

\section{Galaxy targeted searches} 
\label{sec:galaxyobs}
In addition to the wide-field survey coverage, the 
unusually tight localisation map of \gwtrig\ (5 deg$^2$ at 50\%), and the distance estimate available from the GW signal ($267\pm52$ Mpc), allowed us to define a coordinated programme of multi-wavelength observations of galaxies within the localisation region \citep[e.g.][]{nissanke13, evans16a, gehrels16}. While these images cannot cover the whole 2D skymap, they can (often) significantly improve upon the depth of the wide-field surveys for a select number of high-luminosity galaxies (Fig.\,\ref{fig:gal_target_coverage}). 
To identify galaxies with the highest probability of hosting the GW event we utilized the \texttt{HOGWARTS} code\footnote{\url{https://gwtool.watchertelescope.ie/}} \citep{Salmonetal2019}, which ranks galaxies in the Galaxy List for the Advanced Detector Era \citep[GLADE;][]{dalya18} catalogue according to their probability of containing the corresponding merger given the 3D localization probability density\citep{singer16b}, and based on the expectation that NS-BH merger rates follow the galaxy mass distribution \citep{arcavi17b}.

Since the expected colours and luminosities of the counterparts of NS-BH mergers (see \S\ref{sec:discussKN}) still have significant uncertainties (largely due to the lack of observational constraints), our goal was to obtain multi-colour (optical and NIR) imaging, which we prioritised over observing a greater number of galaxies. This strategy enabled our observations to be sensitive to counterparts that were either blue (e.g.~disc-wind driven), or very red due to high lanthanide opacities in dynamical ejecta. 
While our observations targeted the most luminous galaxies weighted for the localization probability, individual telescope pointings were refined in order to capture additional (lower luminosity) galaxies within the localisation volume of the \texttt{LALInference} skymap.  
We obtained a series of coordinated observations using the Gamma-Ray burst Optical and Near-IR Detector (GROND), the Liverpool Telescope (LT), the Nordic Optical Telescope (NOT), the Telescopio Nazionale Galileo (TNG), the Very Large Telescope (VLT) and the William Herschel Telescope (WHT). In total, over 400 multi-wavelength 
($grizJHK$) images of the 67 most probable galaxies within the 3D volume 
were obtained in the ten days following the merger. When generating target lists for each telescope, we attempted to avoid unnecessary duplication of observations, while ensuring that the highest probability galaxies were observed to the greatest possible depth. In practice weather, seeing and other scheduling constraints meant that some duplication was unavoidable. Our global coverage
is shown in Figure~\ref{fig:gal_target_coverage} and a list of observed galaxies in order of decreasing probability (as defined in \S\ref{sec:probability}) 
is given in Table~\ref{tab:targeted-galaxies-LK-Pgal}. An example set
of observations with the VLT High Acuity Wide field $K_{\rm s}$ band Imager \citep[HAWK-I;][]{pirard04,casali06,kissler-patig08,siebenmorgen11} is shown in Figure~\ref{hawki}.

Our techniques for searching for transient objects depended on the nature of the data available. All images obtained were manually, and rapidly, compared against existing optical survey data, in particular the PS1 3$\pi$ survey
\citep[e.g. as was done in][when AT2017gfo was first discovered]{coulterGCN}.
Given the brightness and proximity of the Moon at the time of the observations, only the VLT data exceeded the depth of the PS1 3$\pi$ survey. For some observations these data remained the best comparison.

In most other cases, when reference images were subsequently obtained, we performed PSF-matched image subtraction using the \texttt{hotpants} code. The residual images were then manually inspected to identify any possible transient sources. We limited our search to the circle centered on the nucleus of the investigated galaxy, with radius  $1.5\times R_{25}$,  where $R_{25}$ is the galaxy isophotal radius 
at $B=25$\,mag\,arcsec$^{-2}$.\footnote{\url{http://leda.univ-lyon1.fr/}} In general, all galaxies are well subtracted except for some of the brighter nuclei that 
leave notable residuals and prevent the search for transients.
We confirmed that no transients are identified in the difference images.
To quantify the depth of these images we inserted artificial stars into the images in different positions within the galaxy search radius with a range of magnitudes and estimated their recovery in our difference images. 
The limiting magnitude is defined as the average magnitude of the faintest artificial stars that can be visually 
identified, where optical and NIR photometry is calibrated against the PS1 and 2MASS catalogues, respectively. In all cases these stars had S/N$\sim$3.
We found that the limiting magnitude is fairly constant at different positions within the galaxy search radius with the exception of the nuclei.

\subsection{VLT observations}
\label{sec:VLT}
\begin{figure}
\centering
\includegraphics[width=\linewidth]{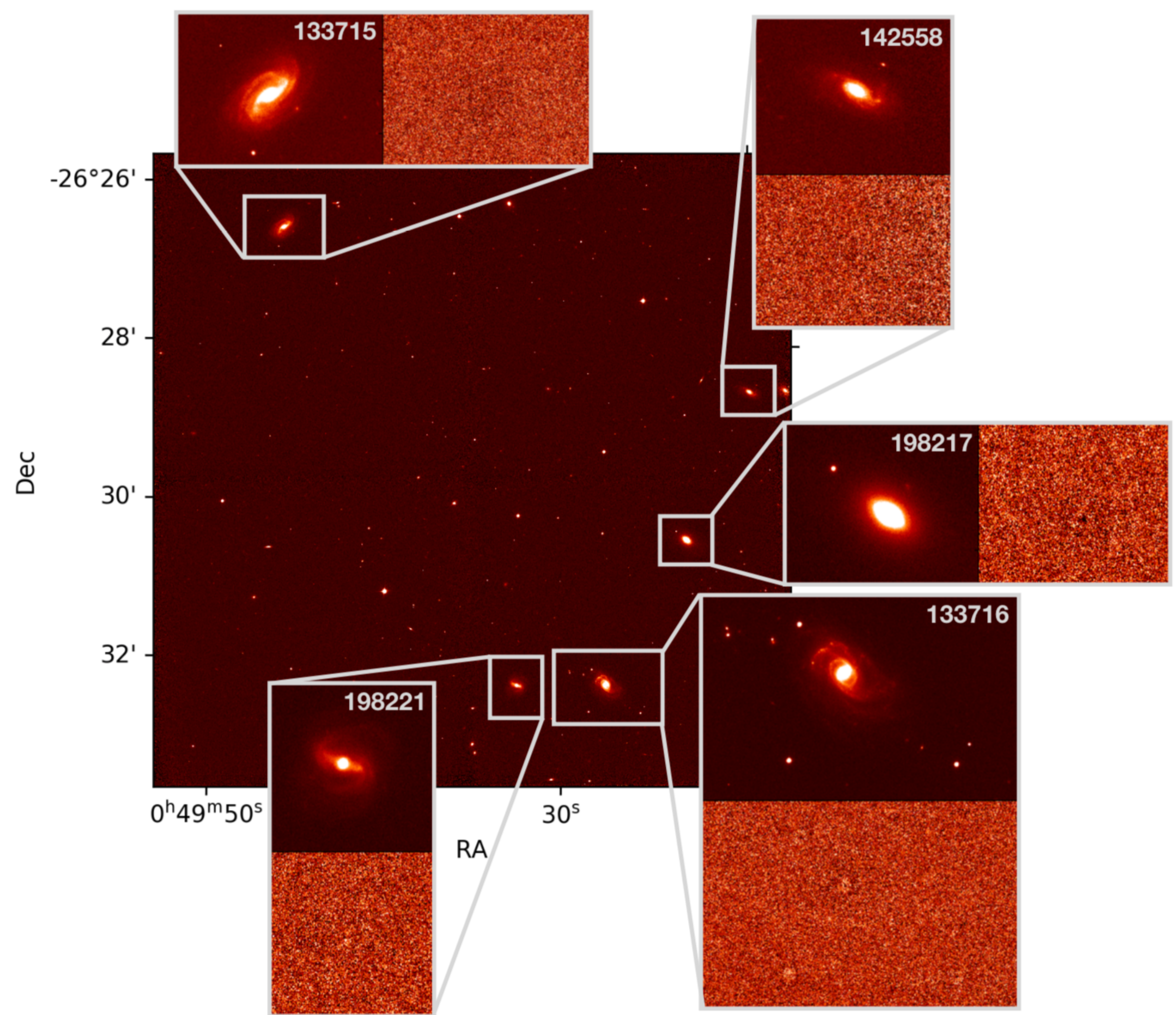}
\caption{A VLT/HAWK-I image of a galaxy targeted field. A number of catalogued galaxies (and at least one uncatalogued galaxy likely at the same redshift) are visible in the field. The insets show each individual galaxy as well as the resulting subtraction, demonstrating the absence of variable sources to the limits of the data in any of these possible host galaxies. Each galaxy inset is labelled by the HyperLEDA identifier from the GLADE catalogue, and the corresponding limiting magnitudes are listed in Table~\ref{tab:targeted-galaxies-LK-Pgal}.
}
\label{hawki}
\end{figure}

The Very Large Telescope (VLT) is a facility operated by ESO on Cerro Paranal in the Atacama Desert of northern Chile \citep{Arsenault2006}, which consists of four individual 8.2 m telescopes (UT 1 -- 4).
We obtained observations of 16 high-priority galaxies with the VLT using $i$ and $z$ imaging with the FOcal Reducer and low dispersion Spectrograph \citep[FORS --][mounted at UT1]{appenzeller98}, and NIR imaging for further 17 galaxies
with HAWK-I \citep[][mounted at UT4; see Fig.~\ref{hawki} for an example image]{kissler-patig08} in the $K_{\rm s}$ band (see Table~\ref{tab:targeted-galaxies-LK-Pgal}).
All VLT data were reduced using the standard EsoReflex graphical environment (v2.9.1) \citep{Freudling13}. 
The observations were performed in one epoch of FORS imaging on 2019 Aug 16 and three epochs of HAWK-I imaging on 2019 Aug 16, 22--23 and 2019 Sept 23--24.  
Given the likely slow rise time of KNe in the NIR bands, the first two HAWK-I $K_{\rm s}$-band epochs were intended to be sensitive to the peak of the KN a few days after merger time. This complemented the FORS optical imaging within the first 24--48\,hr, which is more sensitive to early emission. 

FORS observations consisted of $3\times100$ s observations in the $i$ band, although one field was erroneously observed in the $z$ band for the same exposure time. These images reached significantly deeper limiting magnitudes ($i \sim 23$--24.5 mag) across the field than those obtained by smaller aperture telescope searches. 
The cores of some galaxies were, on occasion, saturated, removing our ability to detect transients close to the nucleus. 

\subsection{WHT observations}
\label{sec:WHT}

A series of optical and NIR observations of 17 galaxies in the sample were taken with the William Herschel Telescope \citep[WHT, ][]{Boksenberg1985} from 2019 Aug 14--22. Optical observations of 17 galaxies 
were obtained in the $r$ band using the Auxiliary-port CAMera \citep[ACAM, ][]{Benn2008} instrument on 2019 Aug 15, while NIR observations of 12 galaxies were taken in the $K_{\rm s}$ band using the Long-slit Intermediate Resolution Infrared Spectrograph \citep[LIRIS, ][]{Acosta-Pulido2002} over the following nights. Both the LIRIS and ACAM images were reduced using standard \texttt{IRAF} procedures\footnote{\url{ https://iraf-community.github.io}} and the custom LIRIS package for LIRIS\footnote{\url{http://www.ing.iac.es/Astronomy/instruments/liris/liris\_ql.html}}.

\subsection{TNG observations}
\label{sec:TNG}

Optical and NIR images of a subset of 19 galaxies were carried out with the Italian 3.6-m Telescopio Nazionale Galileo \citep[TNG, ][]{Poretti2018}, situated on La Palma, using the optical DOLoRes \citep{Molinari1999} and near-infrared NICS \citep[][]{Oliva2001} instruments. Ten galaxies were observed in the $r$ band with relatively short (120 s) exposures obtained on 2019 Aug 15 between 02:13 and 02:54 UT \citep{2019GCN.25331....1D}.  Image reduction was carried out using standard \texttt{IRAF} procedures. Astrometry was performed using the USNO--B1.0\footnote{\url{http://www.nofs.navy.mil/data/fchpix/}}
catalogue. The typical upper limit is $r \sim 22.8$ ($3\sigma$ detection limit). 

NIR observations of five galaxies were undertaken on 2019 Aug 16, Aug 20 and Sept 5 (usually between 03:00 and 05:00 UT). In addition, the galaxy ESO474-026 was observed on 2019 Aug 17. Each galaxy was observed for 1200 s total exposure time with the $K_{\rm s}$ filter \citep{2019GCN.25361....1D}. Image reduction was carried out using the {\it jitter} task of the \texttt{ESO-eclipse} package.\footnote{\url{https://www.eso.org/sci/software/eclipse/}} Astrometry and photometric zeropoints were calculated using the 2MASS\footnote{\url{https://irsa.ipac.caltech.edu/Missions/2mass.html}} catalogue. 
The typical upper limit is $K_{\rm s} \sim 19.7 - 20.9$ ($3\sigma$ detection limit).

\subsection{GROND observations}
\label{sec:GROND}
We observed 36 galaxies simultaneously in $g^\prime, r^\prime, i^\prime, z^\prime, J, H, K_s$ with GROND 
\citep{greiner08}, 
mounted at the 2.2\,m MPG telescope at the ESO La Silla Observatory. 
For each galaxy we obtained an average exposure of 2.1 min in the optical bands and 3.9 min in the NIR bands, and the data were reduced using the GROND pipeline \citep{2008ApJ...685..376K}, which applies bias and flat-field corrections, stacks images and provides an astrometric calibration. 
The observations reached typical $3\sigma$ detection limits of $20-22$ mag in the $r^\prime$ band and $17.5-19.5$ mag in $K_s$. 

\subsection{LT observations}
\label{sec:LT}
The Liverpool Telescope \citep[LT;][]{2004SPIE.5489..679S} is a  2-metre fully robotic telescope on the Canary island of La Palma, Spain.  
A total of 19 galaxies were observed using the IO:O imaging camera. 
IO:O has a $10 \times 10$ arcmin$^{2}$ 
FoV and was operated with a $2 \times 2$ binning, providing
a pixel scale of $0\farcs3$\,pix$^{-1}$. The observations were made between
01:38 and 05:34 UT on 2019 Aug 15. For all fields, $2 \times 150$ s 
exposures in $r$ band were obtained, and for some of the
highest-probability candidates we also obtained $2 \times 150$ s
exposures in $i$ band. Reduced images were provided by the IO:O pipeline and stacked with SWarp.\footnote{\url{https://www.astromatic.net/software/swarp}} Image subtraction of these data were performed using our own subtraction tools rather than \texttt{hotpants}, and detection limits were measured 
performing PSF photometry at fixed positions over a grid around the centre of each image to determine the median and standard deviation of the sky background as measured in each aperture. Using these measurements, with the calibration tied to the PS1 photometric standards in each field, we derived 3$\sigma$ limiting magnitudes of $20.3$ in both bands.

\subsection{Galaxy catalogue incompleteness} 
\label{sect:incompleteness}

While searches for EM counterparts targeting known galaxies within the localisation region of a GW are eminently feasible for nearby events, such an approach becomes less effective as distance increases. There are two reasons for this: firstly, the density of galaxies per unit area on the sky increases such that tiling the GW map becomes more efficient; secondly, the completeness of galaxy catalogues drops off precipitously beyond $200-300$ Mpc. Nonetheless, for \gwtrig we used targeted deep optical and NIR observations of some of the most likely host galaxies in the probability map as reported by the \texttt{HOGWARTS} ranking tool. We thus needed to determine the completeness of the galaxy catalogues that \texttt{HOGWARTS} used for the position and distance of \gwtrig.

\begin{figure}
\centering
\includegraphics[width=\linewidth]{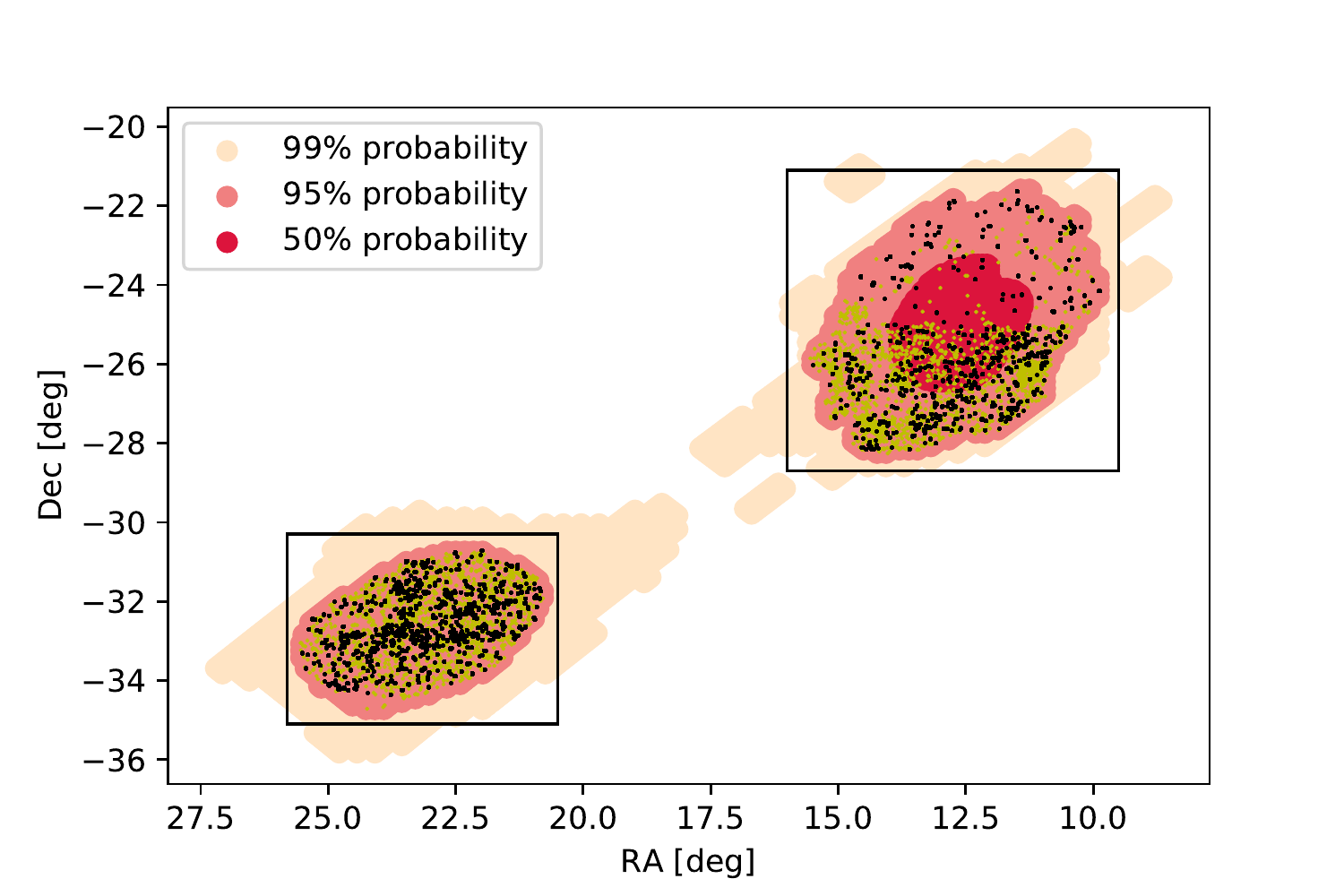}
\caption{The 50, 95 and 99\% probability regions for \gwtrig are shown, and boxes indicate the northern and southern regions of the map as discussed in the text. Galaxies which have a spectroscopic redshift in NED, and lie within the 95\% contour at a distance $\pm 3\sigma$ that of \gwtrig are marked in black; galaxies with an inconsistent spectroscopic redshift are plotted in yellow. The inhomogeneous coverage of NED in the northern contour is clearly visible.
}
\label{fig:ned_galaxies}
\end{figure}

To assess the completeness of GLADE, we queried NED for any galaxies within the 95\% probability region of the \texttt{LALInference} skymap, and with a listed spectroscopic redshift. This resulted in 5,209 galaxies, of which 1,376 have a spectroscopic redshift within $3\sigma$ of the \gwtrig\ distance luminosity marginalized over the whole sky ($267\pm52$\,Mpc). We plot the positions of these galaxies in Fig. \ref{fig:ned_galaxies}. What is apparent from Fig. \ref{fig:ned_galaxies} is that the completeness of the NED database varies across the map, with a sharp drop in the number of galaxies above Dec~$=-25^\circ$. This is almost certainly due to the lack of coverage of the 2dF
galaxy redshift survey above Dec~$=-25^\circ$ \citep{2001MNRAS.328.1039C}.

We attempt to quantify this varying incompleteness in NED in 
order to determine what fraction of stellar luminosity and 
mass we have covered in the galaxy targeted search. To this 
end, we selected galaxy candidates from the PS1 $3\pi$
catalogue, as this is the deepest, most homogeneous 
public imaging catalogue available over the whole skymap. 
PS1 reaches a limiting magnitude of 98\% completeness 
for point sources of 22.5--23\,mag in each of \gps, \rps, \ips, 
with extended source completeness being about 0.5 mag brighter 
\citep{chambers16}. For reference, an apparent magnitude limit 
of ${\sim}22$\,\,mag corresponds to an absolute magnitude
${\sim}{-}15.1$ mag at a 
distance of 267 Mpc. This absolute magnitude is comparable to 
that of the Small Magellanic Cloud, and so at the distance of 
\gwtrig PS1 is essentially complete to all galaxies of
relevance. While we may miss some very low surface brightness 
dwarf galaxies, these contain so little stellar mass that they 
can be ignored for our purposes (see Section\,\ref{sec:probability} 
for a discussion of this).

To create our galaxy candidate catalogue for \gwtrig, we queried the PS1 database 
\citep{Flewelling16}
for all sources within the northern 95\% localisation region. 
In order to select only extended objects, we require that $g_\mathrm{PSF}-g_\mathrm{Kron}>0.1$~mag, and $r_\mathrm{PSF}-r_\mathrm{Kron}>0.1$~mag, and in addition that the source has $n_\mathrm{detections}>10$ within the PS1 catalog. Finally, we limit ourselves to the brightest galaxies in the field, setting a threshold of $r_\mathrm{Kron}<20$ mag, which is equivalent to an absolute magnitude of ${\sim}{-}17$ mag at the distance of \gwtrig.
We also mask out regions in our catalogue around the Sculptor Galaxy NGC 253 and globular cluster NGC 288, which both contain a large number of spurious detections in the PS1 catalog.

\begin{figure}
\centering
\includegraphics[width=\linewidth]{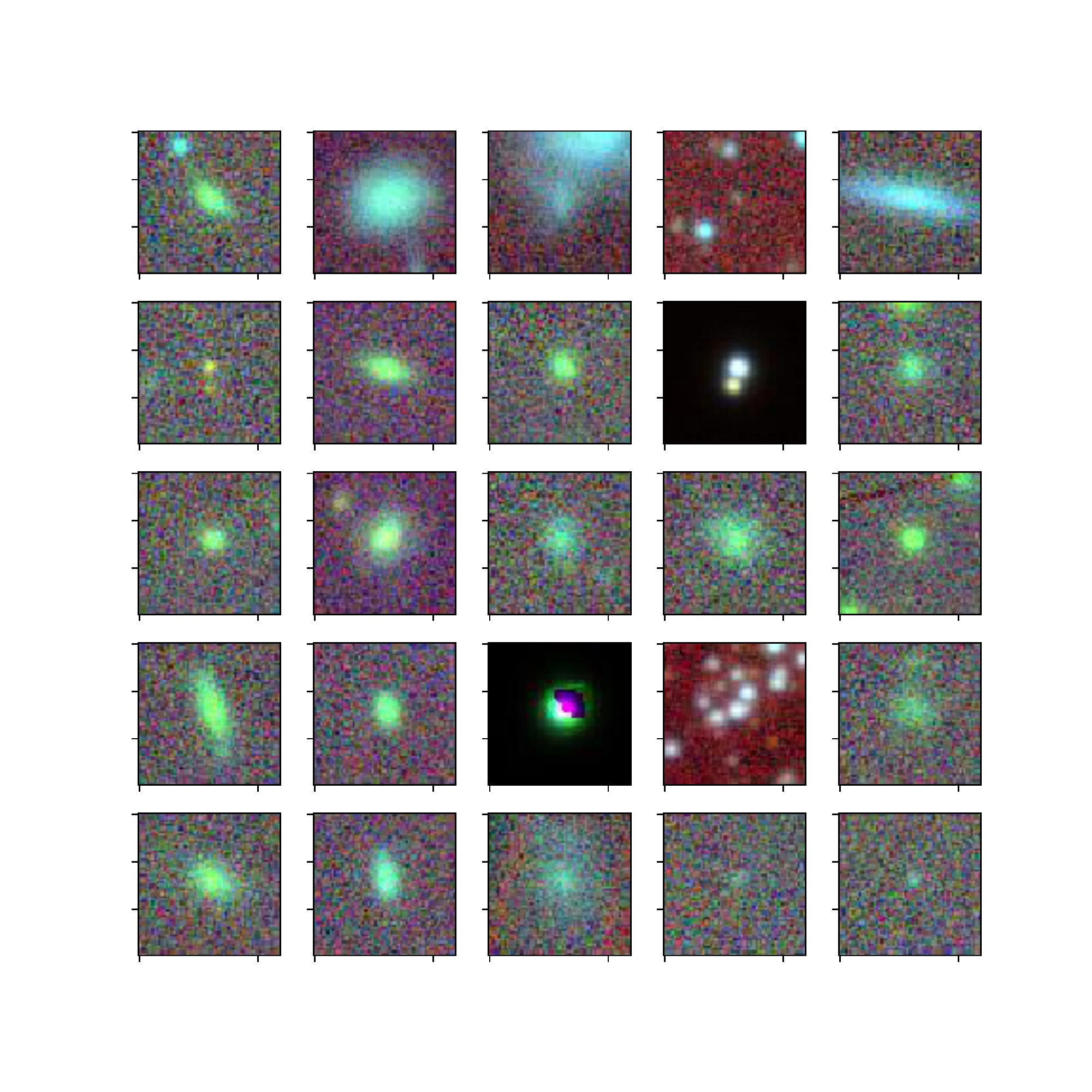}
\caption{Example 15\arcsec$\times$15\arcsec\ Pan-STARRS {\it gri} cutouts around extended sources identified by our cuts.}
\label{fig:ps1_galaxies}
\end{figure}

\begin{figure}
\centering
\includegraphics[width=\linewidth]{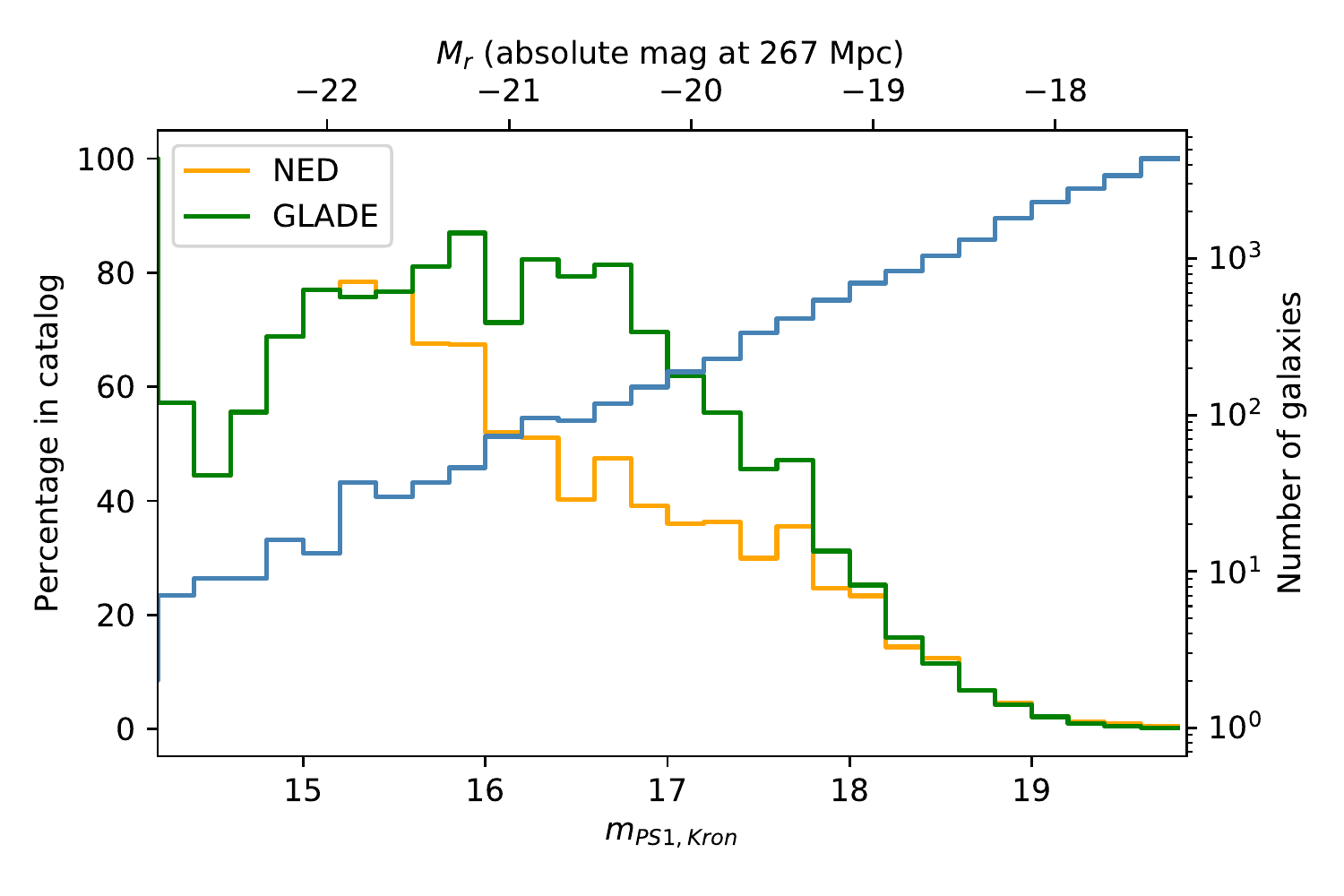}
\caption{Orange and green lines/left axis: The percentage of sources in the PS1 catalogue that have an associated NED and GLADE, respectively, cross-matched galaxy with spectroscopic redshift as a function of magnitude. Blue line/right axis: Histogram of galaxy counts as a function of magnitude, from our Pan-STARRS-derived extended source catalog.}
\label{fig:completeness}
\end{figure}

Visual inspection of a random sample of sources from our extended source catalogue confirms that the majority ($\gtrsim$90\%) are indeed galaxies (Fig. \ref{fig:ps1_galaxies}). The small number of sources brighter than $r=14$ mag in the catalogue all appear to be saturated, bright stars rather than galaxies, so we impose a brightness cut-off at $r=14$ mag. We are finally left with 23,466 candidate galaxies within the northern 95\% localisation probability region of \gwtrig. 
In  order to better assess the issue of completeness, we cross-matched our PS1 galaxy catalogue against the NED and GLADE galaxy lists, requiring a matching radius of $<1\farcs5$. We show the fraction of galaxies that have an associated NED or GLADE counterpart in Fig. \ref{fig:completeness}.
The GLADE completeness reaches a maximum of $\sim$80\% for galaxies brighter than $r\sim16.5$ mag, but drops rapidly at fainter magnitudes, with a completeness of only $\sim$50\% at $r=17.5$ mag, and $\lesssim$20\% for $r<18$ mag; the NED completeness is substantially lower between $r\sim 15.5$--17.5 mag.

The completeness of galaxy catalogs in the context of gravitational wave searches was also recently considered by \cite{kulkarni18b}. These authors employed a different approach to this work, by using nearby supernovae with known distances to generate a random sample of galaxies. From this, \citeauthor{kulkarni18b} then determine the fraction with extant spectroscopic redshifts. While  \citeauthor{kulkarni18b} look at a somewhat closer distance ($<200$ Mpc), it is nonetheless encouraging that their ``Relative Completeness Fraction'' of 75\% (which is implicitly weighted by host mass), is comparable to our completeness for the most luminous galaxies.

\subsection{Probability covered by our targeted search}
\label{sec:probability}

As demonstrated in the previous section, while the GLADE catalogue is somewhat incomplete in terms of galaxy number in the localisation volume, it contains the majority of the most luminous (and hence most massive) galaxies. In order to quantitatively estimate the efficiency of our galaxy-targeted search, in terms of covering the GW localisation probability, we proceed here to assign a definite probability of being the actual host of S190814bv to all catalogued galaxies in the volume, accounting for the mentioned incompleteness. The full list of targeted galaxies and the corresponding observations are reported in Table~\ref{tab:targeted-galaxies-LK-Pgal}. Let us consider galaxies as point-like objects, and let $i$ be an index that runs on all galaxies that are located within the volume $V_\alpha$ that contains a given fraction $\alpha$ of the GW 3D posterior localisation probability density $P_\mathrm{3D}(\mathrm{RA},\mathrm{Dec},d_\mathrm{L})$ (i.e.~the 3D skymap). We assume the probability $P_{\mathrm{gal},i}$ that the merger has taken place within galaxy $i$ to be proportional to the product between $P_\mathrm{3D}(\mathbf{x_i})$,
namely the GW localisation probability density per unit volume at the galaxy position $\mathbf{x}_i = (\mathrm{RA}_i,\mathrm{Dec}_i,d_{\mathrm{L},i})$, and $R_{\mathrm{NS-BH},i}$, that is the NS-BH merger rate in galaxy $i$ (which in principle depends on its present properties and on its history). \citet{Artale2019} have shown, combining state-of-the art compact binary population synthesis models and cosmological simulations, that the NS-BH rate in galaxies at low redshift correlates almost linearly with the galaxy total stellar mass ($R_\mathrm{NS-BH}\propto M^{0.8}$), with some scatter driven by differences in galaxy merger histories, specific star formation rate and metallicity evolution. Based on these results, for simplicity we assume $R_{\mathrm{NS-BH},i}\propto M_i$ and we use the galaxy $K_{\rm s}$-band luminosity $L_\mathrm{K}$ as a proxy for galaxy mass, so that $R_{\mathrm{NS-BH},i}\propto L_{\mathrm{K},i}$. 
This leads to
\begin{equation}
    P_{\mathrm{gal},i} = A\, P_\mathrm{3D}(\mathbf{x}_i)\, L_{\mathrm{K},i}
    \label{eq:single_gal_prob},
\end{equation}
which is similar to the galaxy ranking score used by \citet{arcavi17a}, but with $B$-band replaced by $K_{\rm s}$-band luminosity  (which is a better tracer of galaxy mass).
In order to compute the normalisation constant $A$, we impose the condition
\begin{equation}
    \alpha = \sum_{i=1}^{N} P_{\mathrm{gal},i} = A \sum_{i=1}^{N} P_\mathrm{3D}(\mathbf{x}_i)\, L_{\mathrm{K},i},
    \label{eq:Ploc_normalisation _1},
\end{equation}
where $N$ is the total number of galaxies in the volume $V_\alpha$ (equality \ref{eq:Ploc_normalisation _1} is strictly valid only when $\alpha=1$, but in practice it remains correct to an excellent approximation for $\alpha$ close to one). Since our catalogue only contains a fraction of the actual galaxies in the volume, we need to split the sum on the RHS of Eq.~\ref{eq:Ploc_normalisation _1} into two terms
\begin{equation}
    \sum_{i=1}^{N} P_\mathrm{3D}(\mathbf{x}_i)\, L_{\mathrm{K},i} = \sum_{i=1}^{N_\mathrm{cat}} P_\mathrm{3D}(\mathbf{x}_i)\, L_{\mathrm{K},i} + \sum_{i=N_\mathrm{cat}+1}^{N} P_\mathrm{3D}(\mathbf{x}_i)\, L_{\mathrm{K},i},
\end{equation}
where $N_\mathrm{cat}$ is the number of GLADE galaxies within $V_\alpha$. Assuming the remaining $K_{\rm s}$-band luminosity (present in the volume, but missing from the catalogue) to be uniformly distributed within the volume, we can approximate the last term as
\begin{equation}
    \sum_{i=N_\mathrm{cat}+1}^{N} P_\mathrm{3D}(\mathbf{x}_i)\, L_{\mathrm{K},i} \sim \left\langle P_\mathrm{3D}\right\rangle_{V_\alpha}(L_\mathrm{TOT}-L_\mathrm{cat}),
\end{equation}
where $L_\mathrm{cat}$ is the total $K_{\rm s}$-band luminosity in GLADE galaxies, $L_\mathrm{TOT}$ is the total $K_{\rm s}$-band luminosity in the localisation volume, and $ \left\langle P_\mathrm{3D}\right\rangle_{V_\alpha}$ is the GW 3D localisation  probability density averaged over the volume, i.e.~$ \left\langle P_\mathrm{3D}\right\rangle_{V_\alpha} = \alpha/\left | V_\alpha \right |$, where $|V_\alpha|$ is the extent of the localisation volume (e.g.~in comoving $\mathrm{Mpc^{3}})$.

This finally gives the normalisation constant as
\begin{equation}
    A = \frac{\alpha}{\sum_{i=1}^{N_\mathrm{cat}} P_\mathrm{3D}(\mathbf{x}_i)\, L_{\mathrm{K},i} + \alpha(L_\mathrm{TOT}-L_\mathrm{cat})/\left | V_\alpha \right |}.
    \label{eq:P_gal_normalisation}
\end{equation}
The total $K_{\rm s}$-band luminosity in the localisation  volume can be estimated as $L_\mathrm{TOT} \sim |V_\alpha| \times j$, where $j\sim (7\pm 1.5) \times 10^8 $L$_\odot\, h\,\mathrm{Mpc^{-3}}$ is the $K_{\rm s}$-band local luminosity density \citep{Hill2010}. 

In order to construct the volume that contains 95\% of the localisation probability from the latest public 3D GW localisation probability density based on GW parameter estimation \citep[the \texttt{LALInference} skymap;][] {2019GCN.25333....1T},
we employ a 3D greedy binning approach. The 3D bins are defined by dividing the sky into tiles using a healpix grid (with $N_\mathrm{side}=1024$), and further dividing the distance coordinate into 3000 linearly-spaced bins between $0$ and $700$ Mpc. The probability contained in each bin is assigned based on the 3D skymap (following \citealt{singer16b}). The bins are then summed in order of decreasing probability density until the enclosed probability equals $95\%$, which defines the desired localisation volume. The extent of the obtained volume is $|V_{95\%}|\approx 1.4\times 10^5\,\mathrm{Mpc^3}$, which gives $L_\mathrm{TOT}\sim (6.9 \pm 1.5)  \times 10^{13} $L$_\odot$. 1061 GLADE galaxies fall within this volume. Only 45\% of these have a $K_{\rm s}$-band measurement reported in the catalogue, due to the 2MASS magnitude limit. To circumvent this problem we utilize our own VINROUGE observations to obtain $K_{\rm s}$-band magnitudes for a large fraction of the galaxies. These data were processed through the VISTA Data Flow System \citep{GonzalezFernandez18}, which provides outputs in the same form as for other VISTA public surveys, including catalogue counts and photometric calibration (per tile) for each observation. We therefore 
determine and extract $K_{\rm s}$-band magnitudes for sources on each tile and cross match the resulting catalogues with our GLADE output. This results in 876 matches, providing an 82\% completeness. 
For the brightest galaxies we use the $K_{\rm s}-J$ colour from 2MASS and the redshifts reported in GLADE to k-correct the VISTA magnitudes, in order to compute the corresponding luminosities. For the remaining galaxies, we use the median k-correction (which amounts to $0.10$ mag). We finally compute the $K_{\rm s}$-band luminosity as
\begin{equation}
    \log(\mathrm{L_K/L_\odot}) = 0.4(3.27-\tilde K)-\log(1+z)+2\log(d_\mathrm{L}/d_0),
    \label{eq:K_band_lum}
\end{equation}
where 3.27 is the absolute $K_{\rm s}$-band magnitude of the Sun \citep{Willmer2018} (in the Vega system), $\tilde K$ is the k-corrected $K_{\rm s}$-band magnitude of the galaxy, and $d_0=10\,\mathrm{pc}$. Summing over all galaxies with a $K_{\rm s}$-band magnitude measurement, we obtain $\mathrm{L_{cat}}\sim 7.8 \times 10^{13}\,\mathrm{L_\odot}$, i.e.~the GLADE catalogue for S190814bv is essentially complete in terms of $K_{\rm s}$-band luminosity.
\begin{figure}
    \centering
    \includegraphics[width=\columnwidth]{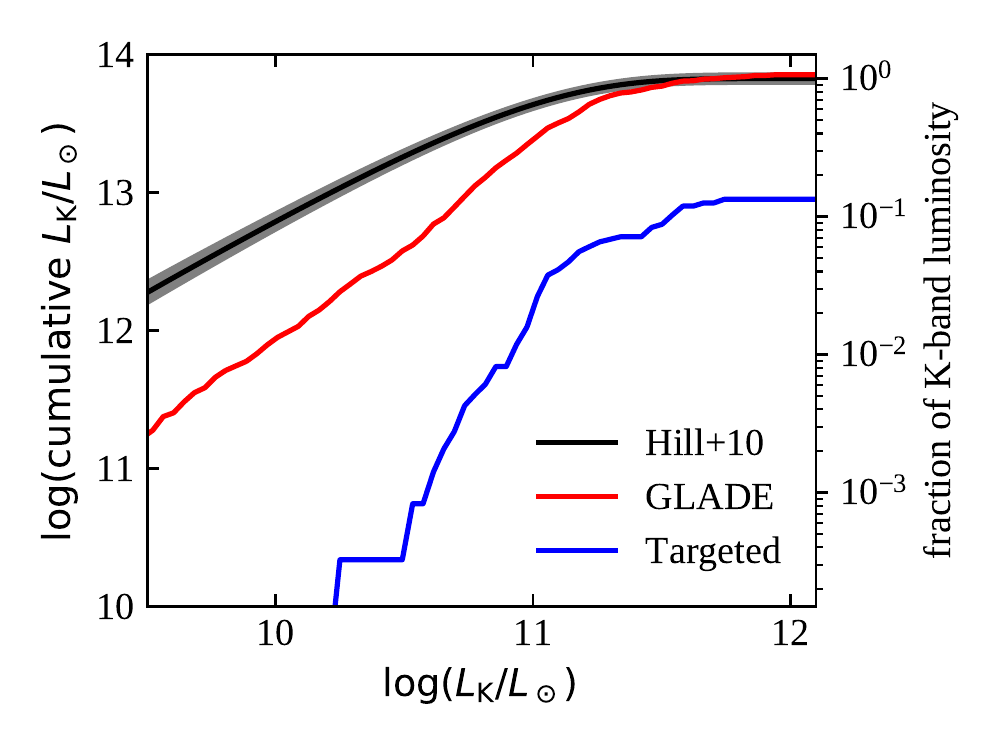}
    \caption{Cumulative $K_{\rm s}$-band luminosity distribution of GLADE galaxies in the S190814bv localisation region, and of our targeted galaxies, compared to the expected distribution in the 95\% localisation volume assuming the $K_{\rm s}$-band luminosity density and distribution from \citet{Hill2010}. The grey area in the latter distribution represents the $1\sigma$ uncertainty.}
    \label{fig:Kband_lum_distrib}
\end{figure}
This can be seen in Fig.~\ref{fig:Kband_lum_distrib}, which compares the cumulative $K_{\rm s}$-band luminosity distribution of   GLADE   galaxies in the S190814bv 95\% localisation volume (red line) and that of our targeted galaxies (blue line) with the expected distribution in the same volume, based on a Schechter fit to the local galaxy luminosity distribution \citep{Hill2010}. This comparison indicates that, despite the incompleteness of current catalogues, galaxy-targeting-based searches are still viable out to these distances, as already suggested, for example, by \citet{Hanna2014}, \citet{evans16} and \citet{gehrels16}.

\begin{figure}
    \centering
    \includegraphics[width=\columnwidth]{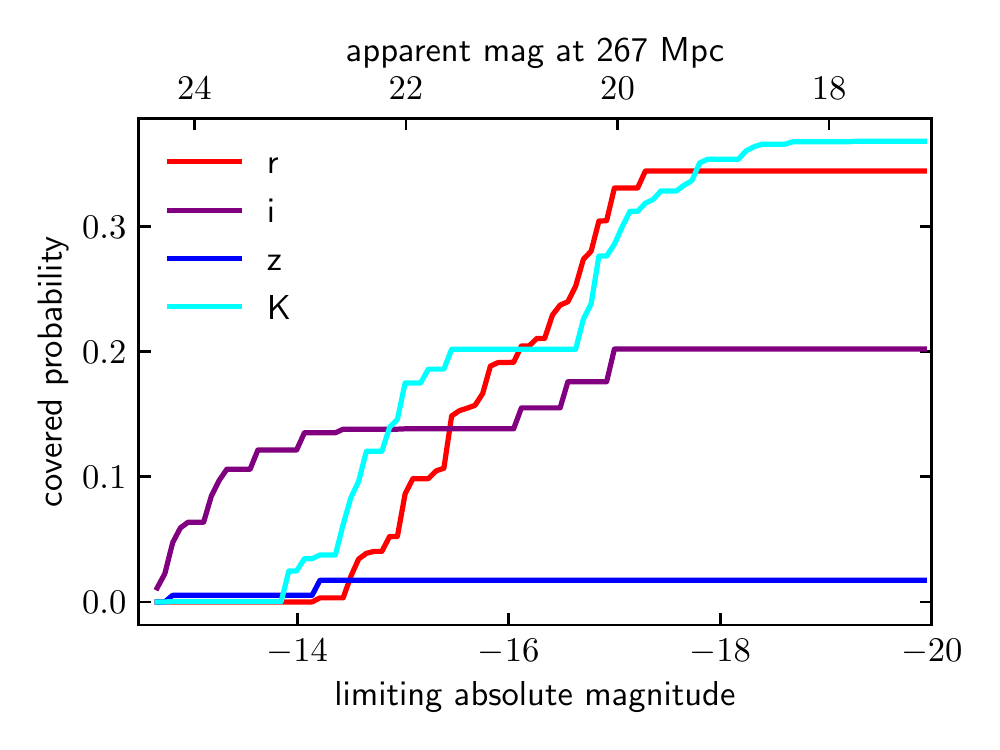}
    \caption{Probability (as defined in Eq.~\ref{eq:single_gal_prob}) covered by our galaxy-targeted search as a function of limiting absolute magnitude in different bands.}
    \label{fig:galaxy-targeted_coverage_vs_maglimit}
\end{figure}

All the galaxies in our targeted search (see Table~\ref{tab:targeted-galaxies-LK-Pgal}) apart from five have a measured $K_{\rm s}$-band magnitude reported in GLADE (from 2MASS). In other words, even though the \texttt{HOGWARTS} code selects the galaxies based on their $B$-band luminosity, the resulting sample is generally bright in $K_{\rm s}$-band as well. We compute $L_\mathrm{K}$ (following Eq.~\ref{eq:K_band_lum}) and therefore
$P_{\mathrm{gal},i}$ (Eq.~\ref{eq:single_gal_prob}) using our VISTA magnitudes, as explained above.
The resulting distribution of covered probability as a function of limiting magnitude in different bands is shown in Fig.~\ref{fig:galaxy-targeted_coverage_vs_maglimit}. The sum of the probabilities over the targeted galaxies in our search amounts to $\sim 50\%$. This does not enable us to place stringent limits on the properties of the putative EM counterpart of S190814bv using the galaxy-targeted search alone, but it nevertheless shows that targeted searches still have a reasonable chance of detecting a counterpart at $\sim$250 Mpc.

\section{Discussion}
\label{sect:discussion}

NS-BHs are hybrid merger events that offer insights into a range of behaviours that are not accessible through other mergers. They have both a larger total mass and a larger chirp mass than NS-NS systems. Thus they should produce a stronger GW signal that can be observed out to greater distances. No extant NS-BH systems are known and their range of masses, and astrophysical rates still therefore have very few observational constraints. Population synthesis models show NS-BH systems may be somewhat rarer than NS-NS \citep[e.g.,][]{ratesdoc,Dominik:2014,EldridgeStanway:2016,Kruckow:2018,GiacobboMapelli:2018,Neijssel:2019}, but with significant uncertainties \citep{Belcynskietal2016}. There is also tentative evidence that they may contribute to the known population of cosmological short GRBs \citep{Gompertz20}.

During the last phase of the NS-BH coalescence, the NS can be partially or totally disrupted by the BH tidal field or swallowed directly by the BH without any significant mass left outside the merger remnant \citep{Lattimer1976,Shibata2011}. EM emission is expected when the disruption occurs before the NS reaches the innermost stable circular orbit (ISCO) of the BH. 
Tidal disruption depends on the mass ratio of the two compact objects, on the BH spin, and on the NS Equation of State \citep[EoS, ][]{Shibata2011,Kyutoku11,Foucart12,Foucart18c}. 
Simulations in Newtonian gravity show that the NS can also be disrupted over several orbits \citep{rosswog05,davies05}. The properties of the progenitors dictate the mass ejected in tidal tails, and the potential formation of a disc wind. In turn, the properties of the ejecta (mass, electron fraction, entropy, and expansion velocity) determine the nucleosynthetic outcome, and hence the contribution that such binaries may make to the heavy element budget of the Universe \citep[see e.g.][]{Rosswogetal2017,Just15,Robertsetal2017}. The presence and properties of ejecta and disk determine also the possible formation of a relativistic jet and hence electromagnetic emission as a short GRB. The EM counterparts could also be much more varied than in the NS-NS case \citep{rosswog17}.

Finally, the combination of a GW-detected NS-BH binary with an EM counterpart would enable a standard siren measurement of the Hubble constant and other cosmological parameters \citep[e.g.,][]{schutz86, nissanke10} out to larger distances than attainable via NS-NS binaries.

In what follows, using our limits, we place constraints on the properties of the putative KN and GRB jet that might have been associated with S190814bv.

\subsection{Constraints on kilonova emission}
\label{sec:discussKN}

\subsubsection{Comparison to AT2017gfo-like kilonovae}

Currently, the only KN detected alongside a GW trigger is AT2017gfo, the KN that accompanied GW170817. While we have to be cautious since that source was classified as a NS-NS merger \citep{lvc_170817}, it is nonetheless prudent to compare it to our limits for S190814bv because it is the only high-confidence KN to date. Foreground\footnote{The typical value of $E(B-V)$ over the skymap is $<0.1$.} and host galaxy extinction is assumed to be negligible in this analysis.

\begin{figure}
    \centering
    \includegraphics[width=\columnwidth]{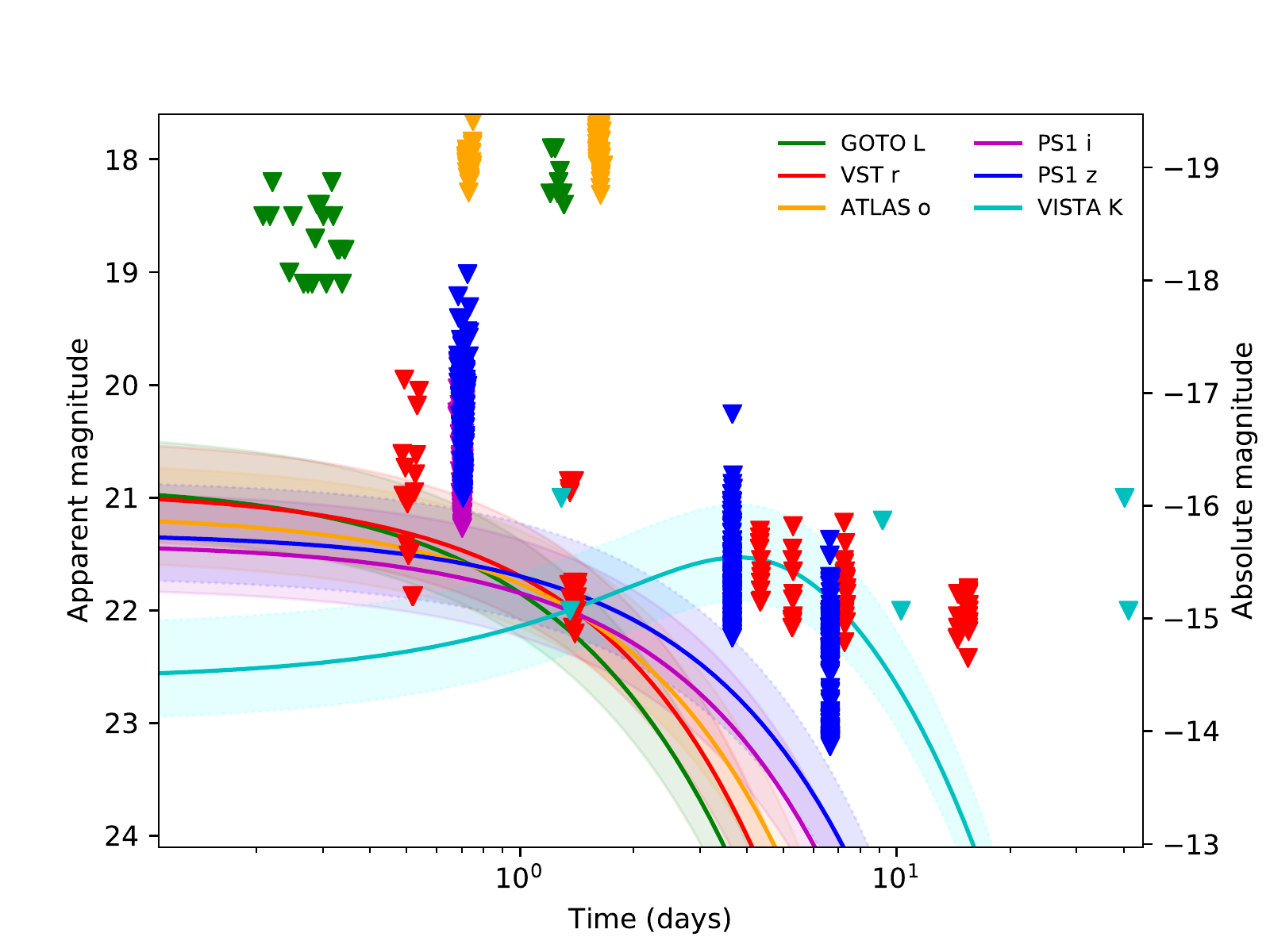}
    \caption{3$\sigma$ or 50\% completeness upper limits from the wide-field instrument follow-up campaign. The data used are referenced in Section\,\ref{sec:discussKN}. The kilonova models, representing an AT2017gfo-like evolution, are shifted to the luminosity distance measure from the LVC skymap (267 Mpc; \texttt{LALInference.v1.fits}), and the shaded regions represent the 1$\sigma$ confidence interval ($\pm$ 52~Mpc). Absolute magnitudes assume a distance of 267~Mpc. Foreground extinction is not included.}
    \label{fig:widefield_summary}
\end{figure}

Figure~\ref{fig:widefield_summary} presents our wide-field follow-up limits (ATLAS, GOTO, PS1, VISTA and VST), plotted against phenomenological fits\footnote{as described on the ENGRAVE webpage, \url{http://www.engrave-eso.org},
these \cite{bazin11} model fits are purely phenomenological \citep[cf.][]{gompertz18}, and describe the temporal evolution of AT2017gfo when shifted to the luminosity distance (267~Mpc) of \gwtrig{}.} to the AT2017gfo light curve based on data from \citet{andreoni17}, \citet{arcavi17b}, \citet{chornock17}, \citet{cowperthwaite17}, \citet{drout17}, \citet{evans17}, \citet{kasliwal17}, \citet{pian17}, \citet{smartt17}, \citet{tanvir17}, \citet{troja17}, \citet{utsumi17} and \citet{valenti17}.
We find that some of the early VST observations were deep enough to detect a KN of similar brightness to AT2017gfo if one occurred within the 1$\sigma$ distance confidence interval. The first VISTA-deep observation also constrains an AT2017gfo-like KN down to the S190814bv distance, and several PS1 frames constrain the near end of the distance distribution. However the large distance to 
this event precludes a strong statement on whether an AT2017gfo-like event would have been detected by PS1, VISTA or VST. 
Our deepest limits do exclude KNe (within the relevant frames) similar to those which have been claimed
to accompany GRB 130603B \citep{tanvir13,berger13}, GRB 050709 \citep{Jin16}, GRB 060614 \citep{yang15}, and GRB 150101B \citep{gompertz18,troja18b}. These were all brighter than AT2017gfo at similar epochs to our sampling \citep{gompertz18}.
Note, though, that some claimed KNe are fainter than AT2017gfo, such as those accompanying GRB\,160821B \citep{lamb19b, Troja2019} and GRB\,070809 \citep{jin20}.

\begin{figure}
\centering
\includegraphics[width=\linewidth]{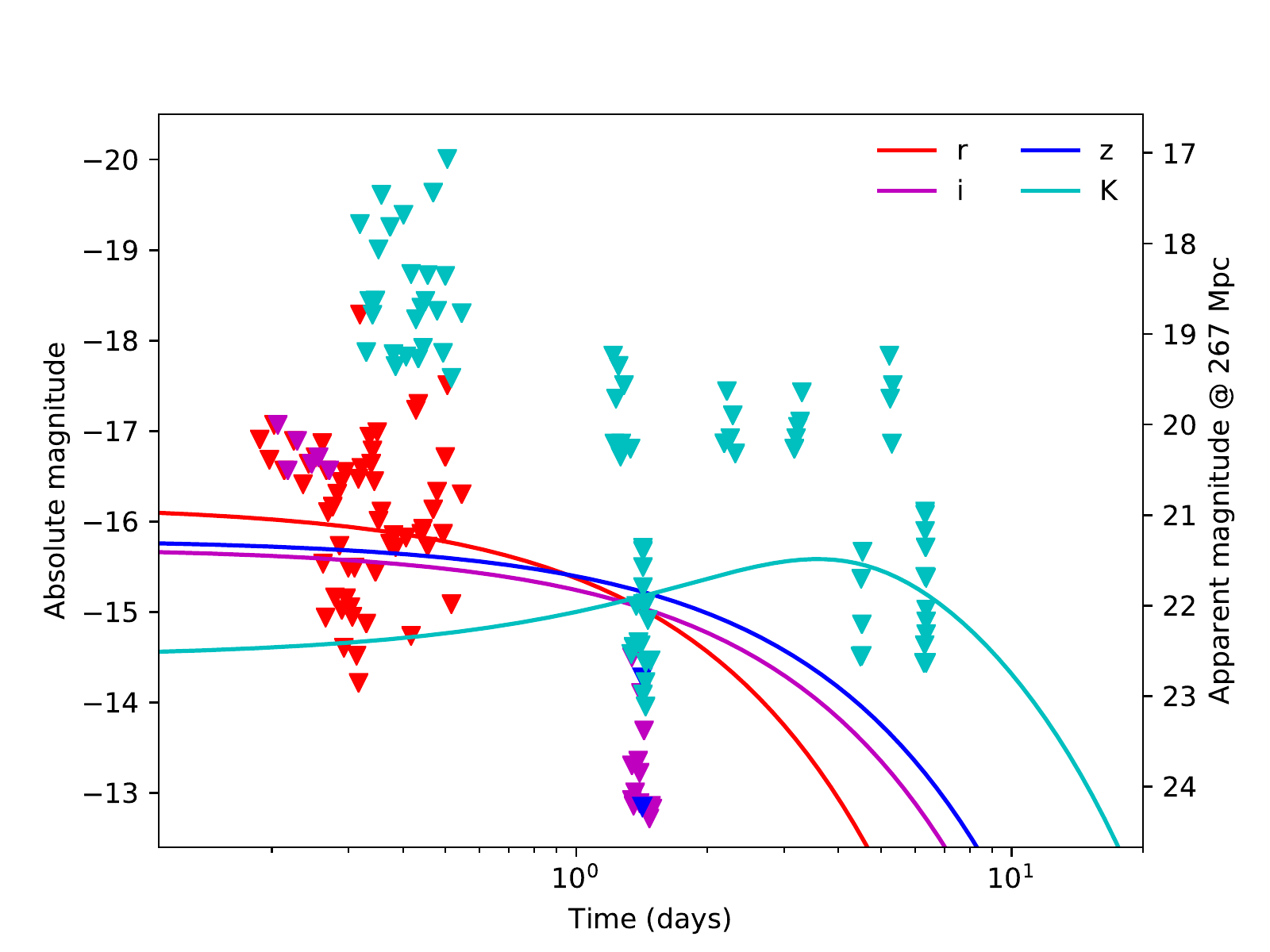}
\caption{Our galaxy targeted limits for S190814bv, alongside the equivalent AT2017gfo KN models. Due to the different distances of the observed galaxies, the data and models are presented in absolute magnitudes. For limits below the model lines, our observations would have uncovered a transient comparable to AT2017gfo, had it been present. We also show the apparent magnitude of the data and models when shifted to the luminosity distance of S190814bv (267~Mpc). Foreground extinction is not included.
}
\label{fig:gal_target_lims}
\end{figure}

Our galaxy-targeted observations are able to place significantly deeper limits over a fraction of the error box. In particular, early observations, which exceeded $r>22$ mag, are well below the expected brightness of an AT2017gfo-like KN. Data taken in the IR on timescales of $5-10$ days reaching $K>21$ mag are also 
competitive. This is illustrated in Fig.~\ref{fig:gal_target_lims}, which shows the comparison (in absolute magnitudes) between the galaxy-targeted limits and AT2017gfo. Solid lines in the figure show our phenomenological fits to the AT2017gfo light curve in the listed bands, converted to absolute magnitudes at the distances of the targeted galaxies. Downward pointing triangles show the absolute limiting magnitudes of our galaxy-targeted observations.
As can be seen, early observations from the WHT and TNG in the $r-$band
(\S\ref{sec:WHT} and \S\ref{sec:TNG}),
and later observations in the IR from HAWK-I are the most constraining (\S\ref{sec:VLT}),
and we are confident in these cases that our observations would have uncovered a KN similar to AT2017gfo if it had been present in the targeted galaxies.

\begin{figure}
\centering
\includegraphics[width=\linewidth]{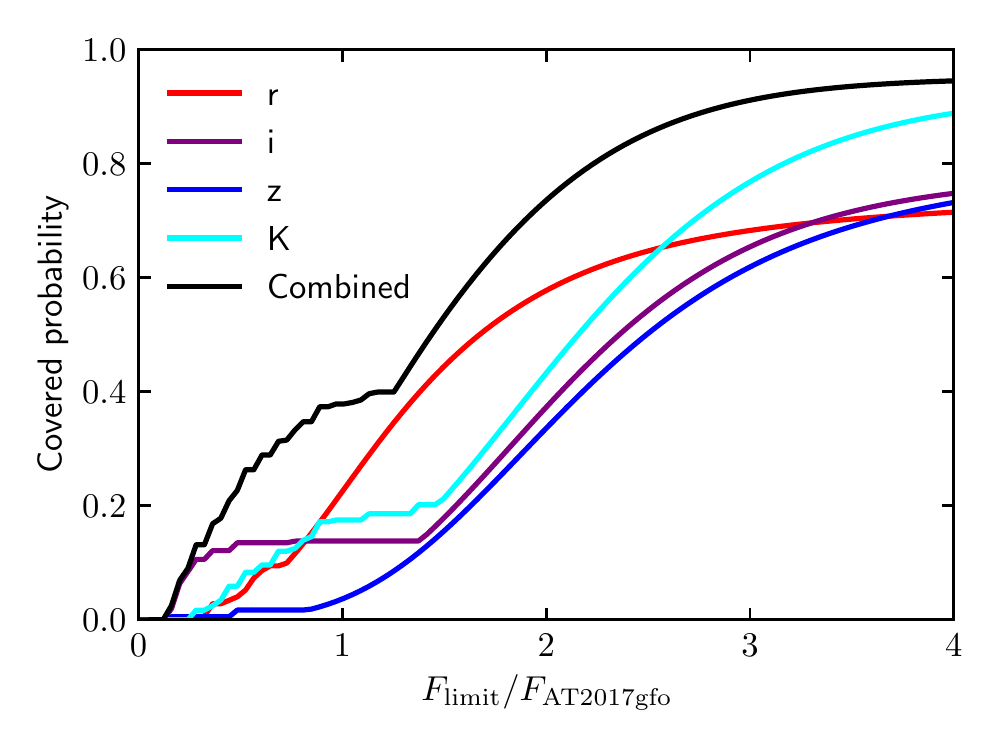}
\caption{The covered probability (defined as described in Appendix\,\ref{sec:from_maglims_to_param_lims}) at which we are sensitive to KN of a given brightness relative to AT2017gfo, based on both our galaxy targeted observations and wide-field limits. Coloured lines show the covered probabilities in four different bands, listed in the legend. The black line shows the combined covered probability.}
\label{fig:gal_target_coverage_frac}
\end{figure}

In order to assess more quantitatively the ability of our observations to uncover a putative AT2017gfo-like transient, we can combine our wide-field and galaxy-targeted observations as described in Appendix\,\ref{sec:from_maglims_to_param_lims}. Assuming as our EM counterpart model an AT2017gfo-like event whose flux is scaled by a constant factor, we can derive the covered probability as a function of the ratio between our limiting flux in the most constraining observations and that of AT2017gfo, which is shown in Figure~\ref{fig:gal_target_coverage_frac}, where we show the covered probability in the $r$, $i$, $z$ and $K$ bands (red, purple, blue and cyan lines, respectively), and the combined probability (black line) that corresponds to having a constraining observation in at least one band. For the few, highest probability galaxies the most constraining limit arises from our galaxy targeted programme, but for the majority of the localisation volume the most constraining limits are through wide field observations, in particular from VST ($r$ band), PS1 ($z-$band) and VISTA ($K_{\rm s}$-band). Our search is therefore sensitive to an AT2017gfo-like KN over $\sim 40\%$ of the localisation probability (as defined in \S\ref{sec:probability}), and over $\sim 80\%$ to a transient with the same temporal behaviour, but brighter by a factor of 2. The most constraining observations (due to both depth and coverage) are those in the $r$ and $K_{\rm s}$ bands.

\subsubsection{Constraints on the ejecta and on the binary properties}
There are fundamental differences between the merger of two NSs \citep[e.g.][]{Ciolfietal2017,radice18d,Wollaegeretal2018,Shibata&Hotokezaka2019} and that of a NS and a BH \citep[e.g.][]{Foucart13, Kyutoku13,Fernandezetal2017,Tanakaetal2014}. 

In the latter case less extreme values for the NS-BH mass ratio, larger BH spin and stiffer NS EoS favor the disruption of the NS before the ISCO, enabling the formation of the accretion disk, tidal tails, and unbound ejecta. This material forms
different components from which EM signals can originate~\citep{Rantsiou08,Pannarale2014,Foucart12, hinderer18, Foucart19,Barbieri19}. 
The KN emission for a given merger is a function of the mass deposited in the various components of the KN, including low-electron-fraction tidal tails \citep{Foucart14,Kiuchi15,Roberts17,Kyutoku18} and the neutrino- and viscosity-driven less neutron-rich winds \citep{Fernandez13,Just15}.
It is therefore relevant to compare the observational limits on any KN emission from \gwtrig{} with the expectations of NS-BH models. 

\begin{figure}
    \centering
    \includegraphics[width=\columnwidth]{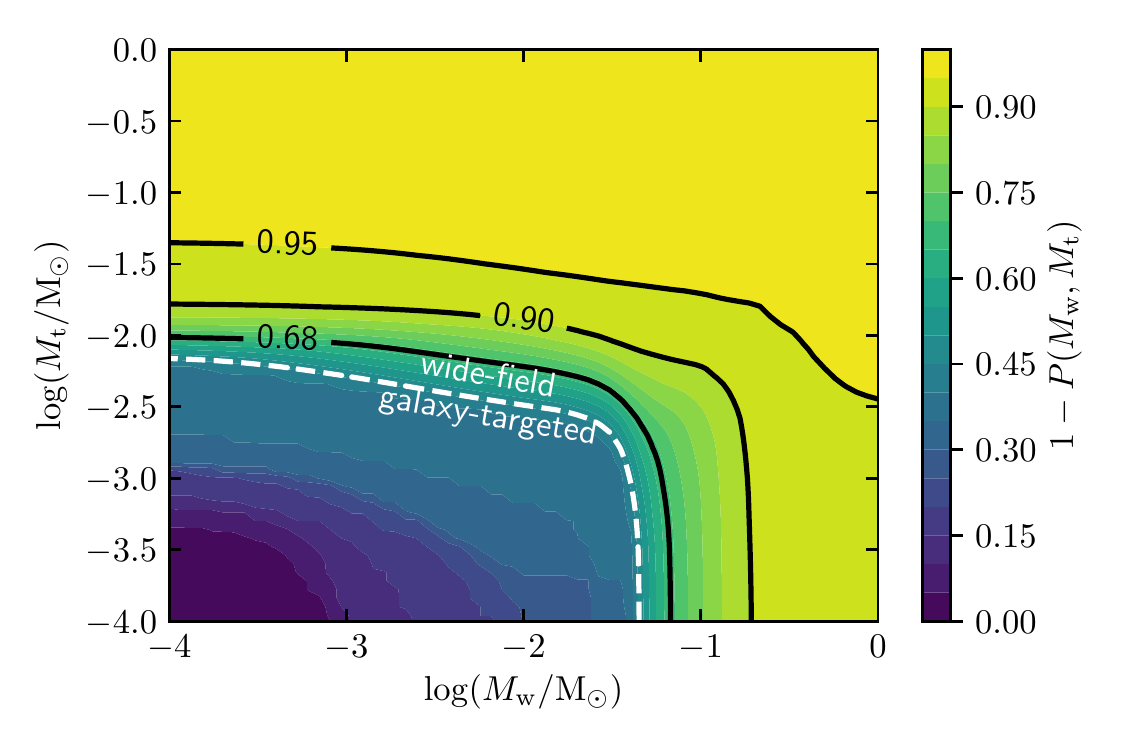}
    \caption{Limits on the tidal ejecta mass ($M_\mathrm{t}$) and secular disk wind mass ($M_\mathrm{w}$) that we obtain by comparing the NS-BH KN model from \citet{Barbieri19} to the limits derived from our search (both galaxy-targeted and wide-field). The colour map shows the confidence level 
    at which we can exclude each pair $(M_\mathrm{w},M_\mathrm{t})$.}
    \label{fig:Mt_Mw_limits}
\end{figure}

We use, for that purpose, the multi-component, anisotropic, NS-BH-specific KN model presented in \citet{Barbieri19}, which builds on the NS-NS KN model of \citet{perego17}. In this model, three outflow components produce KN emission: (1) the tidal ejecta (which are concentrated close to the orbital plane and have the shape of a crescent); two disk-related winds, namely (2) the neutrino-driven wind from the inner part of the accretion disk and (3) the viscosity-driven wind that results from small scale turbulence of magnetic origin inside the disk.
For simplicity, since the neutrino-driven wind (2) is expected to unbind only a small fraction of the disk mass in NS-BH remnants, we neglect that component. As a further simplification, we fix the average (root mean square) velocity $v$ and the (grey) opacity $\kappa$ of the remaining two components to plausible values, namely $v_\mathrm{t}=0.3 c$ and $\kappa_\mathrm{t}=15\,\mathrm{cm^2g^{-1}}$ for the tidal ejecta and $v_\mathrm{w}=0.1c $ and $\kappa_\mathrm{w}=5\,\mathrm{cm^{2}g^{-1}}$ for the viscous disk wind, based on their expected velocity and composition: the tidal ejecta are typically expected to retain a very low electron fraction $Y_\mathrm{e}<0.2$ (e.g.~\citealt{Fernandezetal2017}) leading to efficient r-process nucleosynthesis of Lanthanides and hence a high opacity $\kappa_\mathrm{t}>10\,\mathrm{cm^2\,g^{-1}}$ (\citealt{Tanaka2019arXiv}); disk wind outflows feature a wider range of $Y_\mathrm{e}$, due to viscous heating and neutrino irradiation from the inner part of the disk. Differently from NS-NS mergers, though, the absence of shocks and of intense neutrino production by a meta-stable neutron star remnant are likely to cause the disk wind to remain significantly neutron-rich (e.g.~\citealt{Fernandez13,Just15}). This justifies our choice of $\kappa_\mathrm{w}=5\,\mathrm{cm^{2}g^{-1}}$, which is appropriate for outflows with intermediate $Y_\mathrm{e}\sim 0.25$ -- 0.35 \citep{Tanaka2019arXiv}. We assume $\theta_\mathrm{v}=30^\circ$ as the viewing angle\footnote{This parameter has only a minor influence on the light curve, so this assumption does not affect our results significantly.} -- measured with respect to the total angular momentum axis -- which is the most likely value for a GW-detected inspiral  \citep[see][]{Schutz2011}. We are left with the total masses of the two components, $M_\mathrm{t}$ and $M_\mathrm{w}$, as free parameters. By requiring the resulting light curves to be compatible with our upper limits (both from the galaxy-targeted and from the wide-field searches), following the method outlined in Appendix\,\ref{sec:from_maglims_to_param_lims}, we obtain the constraints on $M_\mathrm{t}$ and $M_\mathrm{w}$ shown in Figure~\ref{fig:Mt_Mw_limits}. The colour map in the figure shows the confidence level at which we can exclude each pair $(M_\mathrm{w},M_\mathrm{t})$, denoted as $1-P(M_\mathrm{w},M_\mathrm{t})$. The region to the lower left of the white dashed line is constrained only by galaxy-targeted observations, while the outer region is constrained mostly by wide-field observations. Large tidal ejecta masses $M_\mathrm{t}>0.05\,\mathrm{M_\odot}$ are excluded with high confidence $>95\%$, and we can exclude the region $M_\mathrm{t}>0.01\,\mathrm{M_\odot}$ and $M_\mathrm{w}>0.1\,\mathrm{M_\odot}$ at approximately one sigma confidence.

\begin{figure}
    \centering
    \includegraphics[width=\columnwidth]{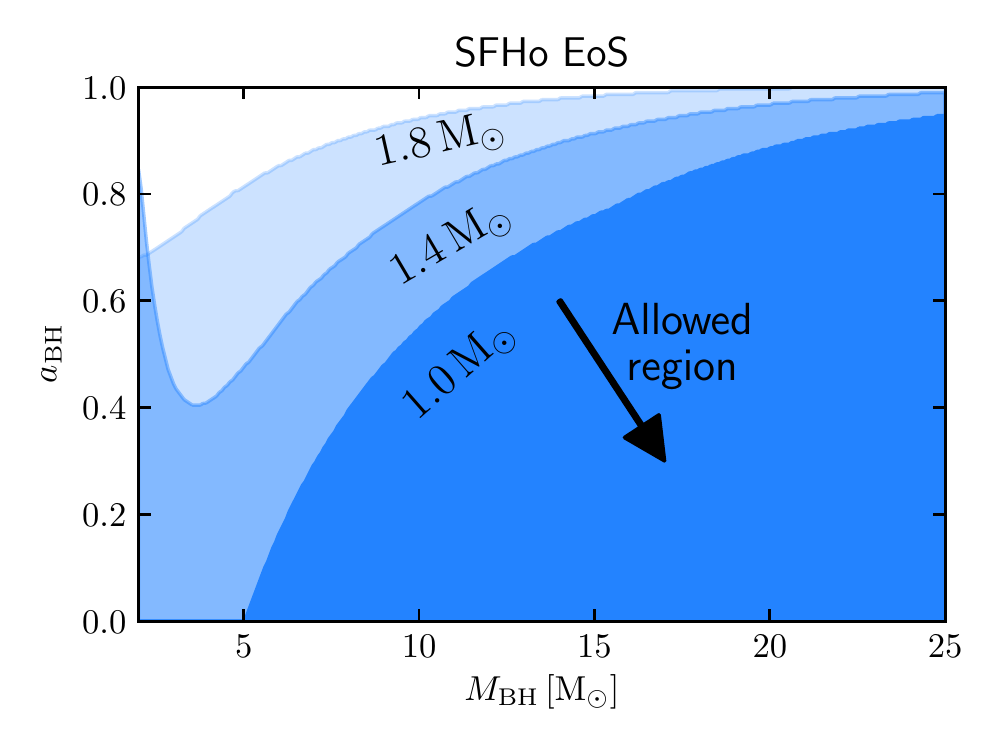}
    \caption{Constraints on the 
    BH spin ($a_\mathrm{BH}$) and 
    the BH mass ($M_\mathrm{BH}$) 
    of the NS-BH binary, assuming remnant disk and tidal ejecta mass limits of $M_\mathrm{disk}<0.3\,\mathrm{M_\odot}$ and $M_\mathrm{t}<10^{-2}\mathrm{M_\odot}$ 
    , which correspond to approximately 1 sigma exclusion confidence (see text). 
    The SFHo EoS has been adopted to compute the NS tidal deformability.}
    \label{fig:BHNS_allowed_params_SFHo}
\end{figure}

\begin{figure}
    \centering
    \includegraphics[width=\columnwidth]{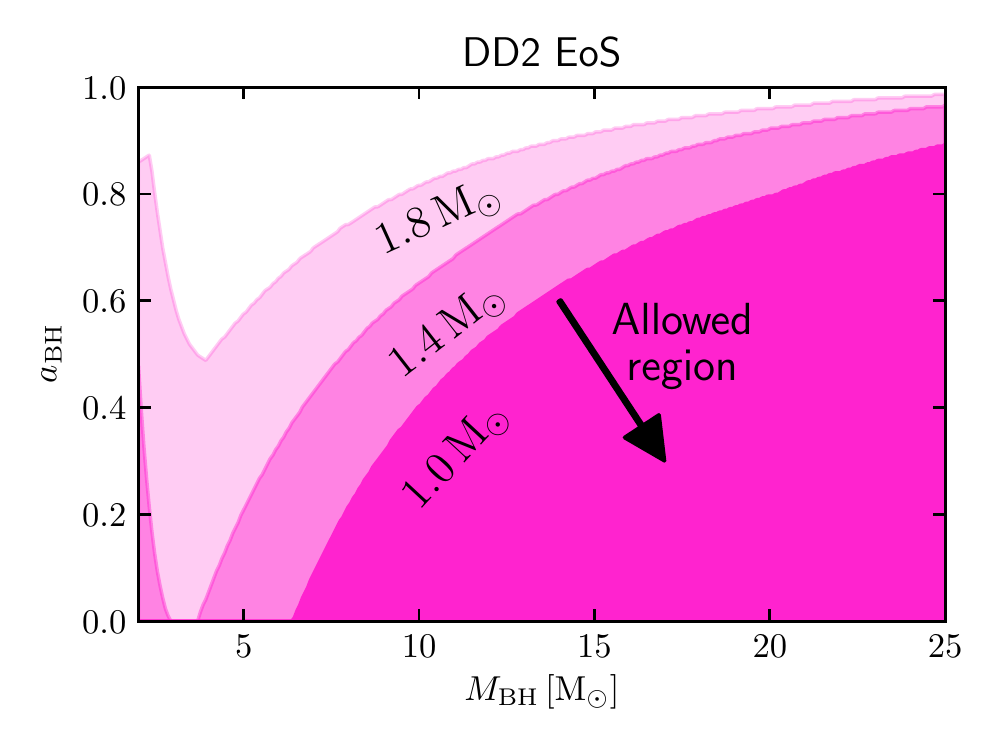}
    \caption{Same as Fig.~\ref{fig:BHNS_allowed_params_SFHo}, but for the DD2 equation of state.}
    \label{fig:BHNS_allowed_params_DD2}
\end{figure}

By employing numerical-relativity-based fitting formulae that link the properties of the outflows to those of the progenitor binary \citep{Foucart18c, Kawaguchi2016}, the limits can also be translated into constraints on the NS-BH binary intrinsic properties, again following \citet[][see also \citealt{Barbierietal2020EPJA}]{Barbieri19}. By assuming the disk wind mass to be $30\%$ of the total disk mass \citep[e.g.][]{Fernandez13,Just15,fernandez19}, we take our representative limits on the disk and tidal ejecta masses (corresponding to the region excluded at $1\sigma$ confidence in Fig.~\ref{fig:Mt_Mw_limits}) to be $M_\mathrm{disk}<0.3\,\mathrm{M_\odot}$ and $M_\mathrm{t}<10^{-2}\mathrm{M_\odot}$, respectively. Figure~\ref{fig:BHNS_allowed_params_SFHo} shows the NS-BH parameter space allowed by our limits, for three possible NS masses, assuming the SFHo NS EoS \citep{Steiner.etal:2013}. Different colour shades show the allowed parameter region  of the binary for a fixed NS mass (reported near the edge of the region). Figure~\ref{fig:BHNS_allowed_params_DD2} shows the corresponding limits assuming the DD2 EoS \citep{Typel.etal:2010,Hempel.etal:2012}, which is stiffer than SFHo.
These two EoSs are representative of the uncertainties in the NS EoS obtained from present nuclear and astrophysical constraints \citep[e.g.][]{Oertel.etal:2017}, as well as from constraints derived from GW170817 \citep{lvc_nseos}.

Based on these results, we can therefore exclude that the progenitor NS-BH
binary produced a large amount of ejecta. This is consistent with the negligible probability for remnant material left after the merger as reported by the LIGO/Virgo Collaboration in low latency \citep{2019GCN.25324....1L}. This indicates that most likely the NS was not disrupted by tidal forces during the final part of the inspiral towards its BH companion, which disfavors high (aligned) BH spins and small mass ratio (or both), as shown quantitatively in Figures~\ref{fig:Mt_Mw_limits}, \ref{fig:BHNS_allowed_params_SFHo} and \ref{fig:BHNS_allowed_params_DD2}. At very low mass ratios $q=M_\mathrm{BH}/M_\mathrm{NS}$, the maximum allowed BH spin actually increases with decreasing BH mass. Although the NS tidal disruption is more likely for lower BH masses, the fraction of unbound mass (on which our constraint is tighter) is much smaller \citep{Foucart19}. As a caveat, we note that the dynamical ejecta fitting formula by \citet{Kawaguchi2016} was calibrated only for mass ratios $3\leq M_\mathrm{BH}/M_\mathrm{NS}\leq 7$ (but see \citealt{Foucart19}).

We note that a small or absent amount of mass left outside the remnant BH was found to be the most likely outcome of BH-NS mergers in the population synthesis simulations described in \citet{Zappaetal2019}.

\subsection{Constraints of GRB afterglow-like emission}

\begin{figure}
    \centering
    \includegraphics[width=\columnwidth]{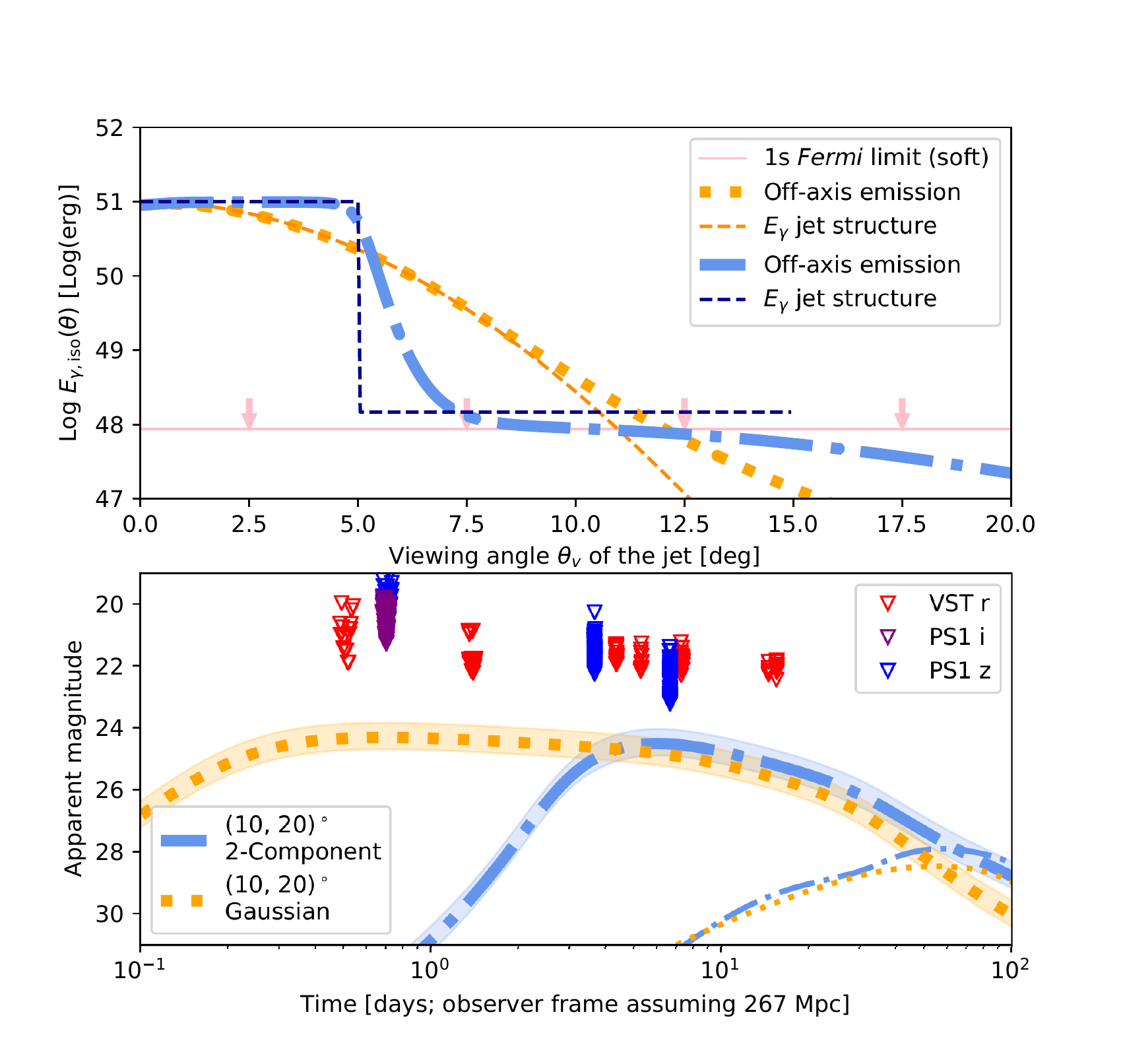}
    \caption{The upper limits from \textit{Fermi} GBM \citep{2019GCN.25326....1K} provide an inclination limit for a GRB with a GRB\,170817A-like jet structure (pink line with downward arrows, top panel). We assume two possible structures following the phenomenology of those in \cite{lamb19}, namely a Gaussian (orange) and a two-component structure (blue). The parameters are reported in the text. 
    The bottom panel shows the $r$ band afterglow for each structure (Gaussian as an orange dotted line, two-component as a blue dash-dotted line) at an inclination of 10$^\circ$ and 20$^\circ$ (thinner lines) assuming an ambient number density $n=10^{-3}$ cm$^{-3}$, microphysical parameters $\varepsilon_B = {\varepsilon_e}^2 = 0.01$, and $p=2.15$. Upper limits in the $r$ band from VST, and $i$ and $z$ band limits from PS1 are shown as triangles. Models are calculated for a luminosity distance 267\,Mpc, the shaded region on the $10^\circ$ afterglow indicates a $\pm 52$\,Mpc uncertainty in the luminosity distance.}
    \label{fig:GRB_prompt_afterglow_limits}
\end{figure}

By assuming that S190814bv launched a short GRB jet, and that all short GRBs have a similar jet structure to that seen in GW170817, we can use the upper limits on any prompt $\gamma$-ray emission to constrain the inclination of the system \citep[e.g.][]{salafia19,saleem2019,song19}. We employ the two jet structures of \cite{lamb19} \citep[see also][]{resmi18, salafia19}, both of which are compatible with the afterglow of GRB\,170817A, namely a Gaussian and a two-component structure. For both structures, the central-core isotropic-equivalent kinetic energy is $E_{\rm{K, iso}}=10^{52}$ erg and the Lorentz factor is $\Gamma = 100$. The two-component structure has `wings' with 10\% of the core kinetic isotropic equivalent energy and $\Gamma=5$. The core half-opening angles are $\theta_\mathrm{c}=0.09$ rad for the the Gaussian structure, and $\theta_\mathrm{c}=0.07$ rad for the two-component structure.
Figure~\ref{fig:GRB_prompt_afterglow_limits} shows the reported $Fermi$/GBM upper-limit
(\citealt{2019GCN.25326....1K} -- pink line), assuming a soft, $\sim$1 s duration burst \citep[see][]{goldstein2016} at 267\,Mpc.
Using the two jet structure models described above, 
the observed isotropic equivalent $\gamma$-ray energy for an off-axis observer can be found using the method in \citealt{ioka19} \citep[which is equivalent to that described in][]{salafia15}.
We assume a 10\% efficiency for energy dissipated as $\gamma$-rays by the jet and include opacity due to pair-production where the Lorentz factor is $\Gamma \lesssim 20-30$ following the method in \citet[][see also \citealt{matsumoto19}]{lamb16, lamb17}. The top panel of Fig.~\ref{fig:GRB_prompt_afterglow_limits} shows the resulting isotropic-equivalent emitted gamma-ray energy, as a function of the viewing angle, for the Gaussian model (orange dotted line) and the two-component model (blue dash-dotted line) respectively.
If there had been a successful GRB\,170817A-like jet, 
the figure shows that the system should be inclined at $>10^\circ$, or with a $(\theta_\mathrm{v}-\theta_\mathrm{c})\gtrsim 5^\circ$ where $\theta_\mathrm{c}$ is the jet's core opening angle and $\theta_\mathrm{v}$ is the viewing angle from the central axis.
The cosmological population of short GRBs typically have an isotropic $\gamma$-ray energy in the range $10^{49}\lesssim E_{\gamma,{\rm iso}} \lesssim 10^{52}$\,erg \citep{fong2015}, thus a successful-GRB producing jet (if any) may have had a lower efficiency or core energy than those assumed here and the lower-limit on the off-axis angle could be smaller.

The afterglow for each structure is shown in the bottom panel of Fig.~\ref{fig:GRB_prompt_afterglow_limits} at an inclination of $10^\circ$ and $20^\circ$ (thick and thin lines, respectively).
The $r$ band VST upper limits are shown as red triangles and the $i$- and $z$-band PS1 upper limits as purple and blue triangles, respectively.
For the afterglow light curves we assume an ambient density $n=10^{-3}$\,cm$^{-3}$, microphysical parameters $\varepsilon_\mathrm{B} = {\varepsilon_\mathrm{e}}^2 = 0.01$,
and an electron distribution index of $p=2.15$.
In Figure \ref{fig:GRB_prompt_afterglow_limits} we only show the $r$ band light-curve, noting that the difference in magnitude for $r - z$ is $\delta m_{r-z} \sim 0.2$ for our model parameters.
Our model afterglow light curve is too faint to be constrained by the upper limits. However, for these parameters we can rule out an environment with an ambient density $n\gtrsim 1$\,cm$^{-3}$ for a system inclined at $(\theta_\mathrm{v}-\theta_\mathrm{c})\sim5^\circ$, where we have assumed our energy and microphysical parameters are typical \citep[e.g.][]{fong2015,Gompertz15}.

\subsection{Comparison to other studies}

S190814bv has also been the target of further follow-up reported by other groups.

\cite{Gomezetal2019} present a study of 96 GLADE galaxies in the 50\% error region, which represent 70\% of the integrated luminosity in the covered region. They estimate that they cover 25\% of all galaxies in the overall localisation area, and are complete down to 0.75 $L^*$ \citep{Schechter1976} in the region they cover. The typical limiting magnitude they obtain after image subtraction is $i=22.2$ mag, which is equivalent to $M_i=-14.9$ mag and therefore fainter than AT2017gfo at the distance of S190814bv, at an observing time of $\approx36$ hrs after the GRB. These limits are comparable to the ones we reach with VST in $r$ at a similar time. They rule out KNe with $M_{ej}>0.01$ M$_\odot$ in their observations, but the incomplete coverage prevents their observations from being constraining at a high confidence.

\cite{Andreonietal2019} present the results of the EM counterpart search by the GROWTH collaboration, using both DECam wide-field tiling observations and targeted spectroscopic and photometric observations of detected candidate transients. All candidates are found to be unrelated SNe. The wide-field coverage is very complete ($>98\%$), with the most constraining limit reached at 3.4 days, $z>22.3$ mag. This limit is comparable to our PS1 $z$ band limits at a similar time. Comparing their limits to different models, they constrain, depending on model and distance, the ejecta mass to $M_{ej}<0.03\cdots0.1$ M$_\odot$, which is consistent with our results.

\cite{Watson2020arXiv} present shallow wide-field observations with the DDOTI imager, covering the entire main probability region (90\% of the total probability) down to an unfiltered limit of $w>17.9$ mag half a day after trigger. They find no candidate transients. They are able to rule out typical on-axis sGRBs but would not have detected a KN similar to AT2017gfo at the distance of S190814bv.

\cite{Dobieetal2019} present their search for radio transients using ASKAP. They find a single significant transient, AT2019osy, which they suggest is likely associated with a low luminosity AGN. ENGRAVE observations of AT2019osy will be presented in a companion paper in preparation.

\cite{Antieretal2020} report on rapid early follow-up by the Global Rapid Advanced Network Devoted to the Multi-messenger Addicts (GRANDMA), reaching limits of $17-18.5$ mag within the first two hours, earlier but shallower than the GOTO limits presented in this work. These place no constraints on the potential KN emission associated with S190814bv.

Finally, during the revision of this manuscript a preprint was circulated by \citet{Vieiraetal2020}, describing the wide-field optical search by the Canada-France-Hawaii Telescope (CFHT). The search reaches limits comparable to ours, despite covering a lower total localisation probability. The corresponding constraints on the putative KN ejecta masses are similar to ours, even though we caution that they are obtained using non BHNS-specific KN models.

\subsection{Ruling out identified transients as counterparts}
\label{sec:ruling_out_candidates}

The worldwide intensive efforts to identify a counterpart to S190814bv had led to the identification of multiple transients within the error localisation, even though this remains one of the smallest regions available. In part this is due to the deep observations that were capable of identifying transients sources fainter than 22 mag. In addition to the transients identified here through our searches, additional counterparts have been found by other groups \citep{Andreonietal2019,Gomezetal2019,Dobieetal2019,Vieiraetal2020}. In total approximately 75 unique optical transients were identified. In principle, each of these should be considered a potential counterpart unless it can be ruled out through follow-up observations. There are various routes that such an approach can take. Firm reasons for rejection include:

\begin{enumerate}
\item The identification of the transient source in imaging taken prior to the detection of the GW event (Pre.Det)\footnote{The definition in parenthesis is that used in Table C3.}. 
\item A spectrum of the source or host galaxy that places the source outside of the plausible 3D-GW volume (i.e. too distant, or too close, (Spec.Host.z)). 
\item A spectrum which identifies the source as a different kind of transient event, for which the progenitors are known (SN). 
\item The source is actually moving, normally because it is an asteroid, but in one case a high proper motion star (Ast, HPM). 
\end{enumerate}

In addition there are further indications which can be used to disfavour sources, but offer a less secure rejection of their association with S190814bv such as

\begin{enumerate}
\setcounter{enumi}{4}
\item A photometric redshift which is inconsistent with the 3D-GW volume (Phot.Host.z).
\item A lightcurve which does not match the expectations for the counterparts of NS-NS or NS-BH mergers, but is in keeping with a supernova (SN?)
\item No obvious underlying host galaxy (No.Host)
\item A source which is nuclear in its host galaxy and therefore likely to be related to AGN activity (AGN?).
\end{enumerate}

These latter scenarios (5-8) are not as robust as 1-4 since each has potential pitfalls. Photometric redshifts often have significant associated uncertainty and are prone to catastrophic failure. It is possible that a photometric redshift which formally places the event outside the GW horizon could be in error. The use of lightcurves requires some assumptions as to the nature of the electromagnetic emission from the GW event. Since we have only a single well sampled kilonova, and a handful of events identified superposed to short-GRB afterglows, this provides a limited observational comparison. Furthermore, we have no kilonova clearly associated to black-hole neutron star mergers. Nonetheless there are strong reasons to expect low ejecta masses and hence strong limits on the associated luminosities and timescales, hence the photometric evolution can provide a constraint. The lack of an obvious host galaxy would at first sight suggest a distant object, or a large kick to the progenitor. In most cases we would expect to be able to identify a host within $\sim 100$ kpc of the transient location. The absence of such a host would disfavour an association with the GW event. However, it should also be noted that some short GRBs arise from an apparently ``hostless” population \citep[e.g.][]{berger10,tunnicliffe14}, and such an event could be missed. Finally, while most nuclear activity is due to either AGN activity or nuclear starbursts, there are suggestions that mergers could be driven at much higher rates within accretion discs around supermassive black holes \citep{Bartosetal2017,Stoneetal2017}. Hence, while nuclear events would apparently be disfavoured as counterparts this should not rule them out. Indeed, the transient AT2019osy, identified as a nuclear radio transient, was an object of interest in the error region of S190814bv. For this reason we separate events which are ruled out for one of the reasons 5-8 from those firmly ruled out via 1-4. 

In Appendix\,\ref{sect:candidates} we present a summary of all transient sources identified in the error region of S190814bv by our searches and those of others. We also indicate the reasons that each of these can be rejected (or not). Of the 73 sources presented there 36 are ruled out robustly, 29 are unlikely based on weaker constraints, the remaining 8 have little information to make such distinctions. There is some value in ascertaining if any of these eight events could be plausible counterparts through future observations, for example to obtain host redshifts. However, we also note that these events are not ruled out due to a paucity of observational constraints, rather than any particular diagnostics which would indicate they are likely related to S190814bv. Indeed, given the small error localisation of S190814bv and its relatively high distance, this list of transients provides some indication of the challenge that will remain in identifying robust EM counterparts to GW sources even in the 4-detector era. 

\section{Conclusions}
\label{sect:conclusions}

S190814bv was unique amongst the GW detections to date in having an exceptionally small error box. 
This in turn made it plausible to search for EM emission via targeting of known galaxies. 
However, such searches were hampered by the large distance to the event (\GWDist\ compared to $\approx 40$~Mpc for GW170817).
Although unprecedented in previous observations, systems like S190814bv may well be more common in the future thanks to both the increasing sensitivity (hence range) of the detectors, and the addition of further 
GW observatories such as KAGRA in the Kamiokande underground site, Japan, and LIGO-India \citep{LVC2018}. 
Hence, it is relevant to consider what may be the most effective route to the identification of counterparts in this era (corresponding to O4 and beyond). 
It is striking that at these distances relatively sensitive wide-field searches such as PS1, VST and VISTA do not, in general, reach sufficiently deep limits to constrain a KN comparable in luminosity to AT2017gfo.
This suggests that the current generation of wide-field facilities may not be especially well suited to the majority of candidates in the future, where observations may need to reach $r > 23$ mag  to probe a reasonable fraction of KN parameter space (see \citealt{SaguesCarracedo2020} and \citealt{Coughlin2020} who reach similar conclusions). 
While some wide-field facilities may be able to attain sufficient depth over a significant fraction of future events (e.g. DECam, BlackGEM, Rubin Observatory LSST) it may well be the case that events at $\sim 300$ Mpc may only be detectable by 8 m class observatories. 
The requirement to observe such events will depend sensitively on where the true event rate of NS-NS and NS-BH lies. 
At the higher end, optical/IR observers can focus on more nearby events. 
However, should the event rate lie at the lower end it is necessary to consider if 8 m telescope resources may be needed to identify counterparts, and ELT-like resources required for their follow-up. 
It therefore continues to be case that the effort should be expended 
on extending the GW detector network such that the 3D probability volumes for the GW events can become tractable for such observations to be plausible.

\begin{acknowledgements}
Based on observations collected at the European Southern Observatory under ESO programmes 1102.D-0353(E), 1102.D-0353(F), 1102.D-0353(Q), 1102.D-0353(G), 0103.D-0070(A), 0103.D-0070(B), 0103.D-0703(A), 0103.D-0722(A), 0103.A-9099(A), 198.D-2010(D) and 60.A-9285(A). 

 ATLAS is primarily funded through NEO NASA grants NN12AR55G, 80NSSC18K0284, and 80NSSC18K1575. 
 The ATLAS science products have been
 made possible through the contributions of the University of Hawaii
 IfA, the Queen's University Belfast, the Space
 Telescope Science Institute, and the South African Astronomical Observatory. 
PanSTARRS is primarily funded through NEO NASA grants NASA Grants NNX08AR22G,  NNX14AM74G. The Pan-STARRS science products for LIGO-Virgo follow-up are 
made possible through the contributions of the University of Hawaii
IfA and the Queen's University Belfast.

The Gravitational-wave Optical Transient Observer (GOTO) project acknowledges the support of the Monash-Warwick Alliance; Warwick University; Monash University; Sheffield University; Leicester University; Armagh Observatory \& Planetarium; the National Astronomical Research Institute of Thailand (NARIT); University of Portsmouth; Turku University and the Instituto de Astrof\'isica de Canarias (IAC).

Part of the funding for GROND was generously granted from the Leibniz-Prize to Prof.\ G. Hasinger (DFG grant HA 1850/28-1). 

The Liverpool Telescope is operated on the island of La Palma by Liverpool John Moores University in the Spanish Observatorio del Roque de los Muchachos of the Instituto de Astrof\'isica de Canarias with financial support from the UK Science and Technology Facilities Council.

The WHT and its override programme are operated on the island of La Palma by the Isaac Newton Group of Telescopes in the Spanish Observatorio del Roque de los Muchachos of the Instituto de Astrof\'isica de Canarias; part of these data were taken under program (19A)N3. 

FEB thanks CONICYT Basal AFB-170002 and Chile's Ministry of Economy fund IC120009. 

MGB, PDA and AM acknowledge support from ASI grant I/004/11/3. 
MBr, EC, AP and SPi acknowledge support from MIUR (PRIN 2017 grant 20179ZF5KS). 
EB, EM and MT acknowledge funding from GRAWITA. 

SHB is indebted to the Danish National Research Foundation (DNRF132) for support. SCa acknowledges support from grant MAE0065741. 

EC acknowledges the support of the H2020 OPTICON programme 730890.

TWC acknowledges the Humboldt Foundation and Marie Sklodowska-Curie grant 842471. MDP thanks Istanbul University for support. PAE acknowledges UKSA support. RAJEF is supported by an STFC studentship. MF is supported by a Royal Society - SFI University Research Fellowship. LG was funded by the EU H2020 programme under MSCA grant no.\ 839090. 

CG, JH and LI were supported by a research grant from VILLUM FONDEN (project 16599). 
CG and LI were supported by a research grant from VILLUM FONDEN (25501).

GGh acknowledges the PRIN MIUR "Figaro" for financial support. 
AGo acknowledges financial support from the Slovenian Research Agency (grants P1-0031, I0-0033, and J1-8136). 

BPG, AJL and JDL acknowledge support from ERC grant 725246 (TEDE, PI Levan). 

SGG acknowledges support by FCT Funda\c{c}\~ao para a Ci\^encia e Tecnologia and by Project PTDC/FIS-AST-31546. GGr acknowledges the ESCAPE H2020 project no.\ 824064. MG is supported by the Polish NCN MAESTRO grant 2014/14/A/ST9/00121. PJG acknowledges support from NOVA and from the South African NRF SARChI grant 111692. 

CPG and MS acknowledge support from EU/FP7-ERC grant no.\ 615929.

KEH acknowledges support by a Project Grant from The Icelandic Research Fund. YDH acknowledges support from the China Scholarships Council. JJ acknowledges support from NOVA and NWO-FAPESP grant for instrumentation. AJ acknowledges funding from the European Research Council (ERC). ZPJ was supported by the Foundation for Distinguished Young Scholars of Jiangsu Province (no.\ BK20180050). PGJ acknowledges funding from the ERC under Consolidator Grant agreement no.\ 647208. DAK acknowledges Spanish research project RTI2018-098104-J-I00 (GRBPhot). 
SKl acknowledges support by DFG grant Kl 766/16-3. ECK acknowledges support from the GREAT research environment. GPL acknowledges support from STFC via grant ST/N000757/1. GL was supported by a research grant (19054) from VILLUM FONDEN. KM acknowledges support from the ERC (grant no.\ 758638). IM is partially supported by OzGrav (ARC project CE17010000). MM acknowledges support from ERC through ERC-2017-CoG no.\ 770017. MJM acknowledges the National Science Centre, Poland, grant 2018/30/E/ST9/00208.

BMJ and DW are supported by Independent Research Fund Denmark grant DFF-7014-00017.

MN is supported by a Royal Astronomical Society Research Fellowship. ANG acknowledges support by grant DFG Kl 766/16-3. PTOB acknowledges funding from STFC. SRO gratefully acknowledges the support of the Leverhulme Trust. FO acknowledges the support of the H2020 Hemera program, grant no.\ 730970. MAPT was supported by grants RYC-2015-17854 and AYA2017-83216-P. EP aknowledges financial support from INAF. GP is supported by the Millennium Science Initiative through grant IC120009. MLP is partially supported by a "Linea 2" project of the Catania University. JQV acknowledges support from CONICYT folio 21180886. TMR acknowledges the support of the Vilho, Yrjo and Kalle Vaisala Foundation. ARo acknowledges support from Premiale LBT 2013. SR is supported by VR grants 2016-03657\_3 and the research environment grant GREAT, Dnr.\ 2016-06012, and the Swedish National Space board, Dnr.\ 107/16. OSS acknowledges the Italian Ministry of Research (MIUR) grant 1.05.06.13. LSa acknowledges the Irish Research Council Scholarship no.\ GOIPG/2017/1525. SJS acknowledges support from STFC Grant ST/P000312/1. 

ERS and DS acknowledge funding from UK STFC CG ST/P000495/1. 

RLCS acknowledges funding from STFC. 

DS acknowledges support from STFC via grant ST/T007184/1.
SDV acknowledges the support of the CNES. LW supported by Polish NCN DAINA 2017/27/L/ST9/03221.

The Cosmic DAWN center is funded by the Danish National Research Foundation.

\end{acknowledgements}

{\it Author contributions}. KA contributed to GOTO candidate vetting. LA contributed to data interpretation and discussion. CB produced the kilonova light curve tables used to put limits on the progenitor binary properties. FEB helped with the interpretation and contributed to the manuscript. SB served on the on-call operations team triggering VLT observations and is the EFOSC2 Instrument Scientist. MGB coordinated the working group that interfaces with external facilities. KB contributed to HST follow-up observations. MTB obtained difference images and magnitude limits for the targeted galaxy search. MBr contributed to governance as a member of the ENGRAVE Governing Council (GC), provided comments on the manuscript and contributed to editing the manuscript. EB contributed to governance as a GC member. SHB served on the on-call operations team triggering VLT observations. MBu contributed to modelling discussions and provided comments on the manuscript. SCh contributed with comments on the manuscript. EC contributed to the VST transient search and galaxy targeted search. AJCT used complementary data to study many galaxies inside the 90\% probability contour. KCC co-leads the Pan-STARRS science surveys. SCa provided comments on the manuscript. TWC is the PI of the GROND GW project, provided GROND images and served on the on-call operations team. RCi contributed to theoretical modelling and interpretation. AC served on the on-call operations team triggering VLT observations, is the contact person for INTEGRAL and neutrino telescopes and provided comments on the manuscript. CMC obtained and reduced the data obtained with the Liverpool Telescope. SCo contributed to governance as an Executive Committee (EC) member and supervised the development of the manuscript. RCu collected, mined, and analysed the GOTO data. FDA provided comments on the manuscript. PDA contributed to governance as a GC member, served on the on-call operations team and obtained and reduced TNG images. GDC is a member of the infrastructure and outreach working groups. MDV provided comments on the manuscript. LD co-leads the ATLAS science surveys. MDP provided comments on and edited the manuscript. VD served on the on-call operations and writing teams and provided comments on the manuscript. VSD is a core member of the GOTO collaboration. MJD is a member of the GOTO operations team and assisted with the GOTO contribution. NER searched for transients with the VST. PAE contributed to infrastructure and provided an interface to the Neil Gehrels Swift Observatory team. RAJEF served on the on-call operations team triggering VLT observations. AF served on the on-call operations and writing teams. MF was the event advocate for S190814bv and contributed to obtaining observations, analysis, and paper writing. ASF led HST follow-up of AT2019osy.  JPUF served on the on-call operations team. LG served on the on-call operations team triggering VLT observations and provided comments on the manuscript. CG served on the on-call operations team. DKG is a PI of the GOTO project and contributed to the observations and data analysis. FG has contributed to reduction and calibration of VST images. GGh contributed to Sec. 4. JHG served on the on-call operations team. AGo commented on the manuscript. BPG is a member of the GOTO team, and provided the AT2017gfo model fits, analysis comparing them to S190814bv (Sec 4.1), and Figs.\ 9 and 10. CGF managed the rapid reduction pipeline for the VISTA imaging. SGG served on the on-call operations team triggering VLT observations and EPO team. AGr was the GRAWITA contact person and PI and image reducer of VST data. GGr provided Fig.~1 and the VST coverage. MG contributed to HAWK-I observations. PJG coordinated observations with MeerLICHT and contributed to the manuscript. CPG served on the on-call operations team. TH contributed to GOTO candidate vetting. KEH contributed to the observations, data analysis and manuscript editing. JH contributed to governance as a GC member, served on the on-call operations team and contributed to editing the manuscript. YDH used complementary data to study many galaxies inside the 90\% probability contour. MEH manages and leads the Pan-STARRS data reductions. CI is FORS2 instrument scientist and provided comments on the manuscript. LI served on the on-call operations team triggering VLT observations and contributed to the observation strategy. JJ served on the on-call operations team triggering VLT observations and contributed to data reduction. AJ contributed to discussion and analysis. ZPJ served on the on-call operations team triggering VLT observations and contributed to discussions and analysis. PGJ contributed to governance as a GC member and is the PI and a coordinator and observer of the WHT. EK contributed to the reduction and analysis of the HAWK-I data. DAK created the GCN BibTeX entries, performed the study comparison in Section 4.3, and wrote part of the introduction. SKi contributed with data analysis of a candidate counterpart. SKl was involved in the GROND data reduction and analysis. ECK carried out image subtraction of HAWK-I and TNG data and served on the on-call operations team triggering VLT observations. MK contributed to GOTO candidate vetting. RK provided comments on the manuscript and served on the on-call operations and writing teams. HK reduced FORS2 imaging data. GPL contributed the GRB prompt and afterglow emission theory, modelling, text and a figure. GL scanned FORS2 images and coordinated the polarimetry working group. AJL contributed to management as the chair of the EC, coordinated observations with VLT/WHT and contributed to analysis and writing. FL is a member of the EPO working group. TBL carried out observations with Pan-STARRS. JDL served on the on-call operations team and enacted GOTO observations. EMag manages and leads the Pan-STARRS data reductions. KM contributed to management as an EC member, served on the on-call operations team, coordinated part of the VLT observations,  contributed to data reduction and analysis, and provided comments on the manuscript. EMai contributed observations within the GRAWITA context. IM contributed to astrophysical modelling and interpretation. MM contributed to paper writing and host galaxy interpretation. SMa contributed to the analysis of HAWK-I data. ORMB provided limiting magnitudes of ATLAS and PS1 observations and the table of PS1 counterparts. AM is a member of the imaging working group and contributed to observations and data reduction. MJM performed SED fitting and contributed to interpretation of ALMA results. BMJ worked on infrastructure, data (notably VISTA) and analysis. SMo was an observer at the WHT and served on the on-call operations team. LNi contributed to VST and TNG data management and analysis. MN served on the on-call operations team. ANG performed GROND data reduction and analysis. LNu is a core member of the GOTO collaboration. PTOB contributed to GOTO discussion and interpretation. SRO served on the on-call operations team. FO served on the on-call operations team. EPa contributed follow-up observations with VST and TNG. BP contributed to the activities of the working group that interfaces with external facilities (LVC). PA contributed to the development of the NS-BH kilonova model and provided text for the manuscript. DAP contributed to gathering and analysis of Liverpool Telescope observations. EPi contributed to governance as a GC member provided comments on the manuscript. GP served on the on-call operations team. SPi is a member of the spectroscopy team and coordinator of the EPO working group. SPo is a core member of the GOTO collaboration. AP provided comments on the manuscript. MAPT contributed WHT observations. MLP provided comments on the manuscript. JQV provided comments on the manuscript. FR contributed obtaining limiting magnitudes of difference images. GR contributed to the GOTO observations. ARa is the PI of the GROND ToO time project. ARe provided Pan-STARRS and ATLAS science analysis. TMR contributed to the reduction and analysis of the HAWK-I data. SSR created and ran the subtraction pipeline for the VISTA imaging. ARo contributed to follow-up with TNG and VST. SR contributed to editing the introduction and Section 4.1. NBS contributed to FORS and HAWK-I observations and data reduction and served on the on-call operations team triggering VLT observations. ASC served on the on-call operations team triggering VLT observations. OSS developed and implemented the framework for the comparison of observational limits and theoretical models, led the writing of secs. 3.7, 4, and Appendix A, produced figures 7, 8, 11, 12, 13, 14 and B.1, and compiled Tables B.1 and B.2. LSa contributed the ranked galaxy lists from the HOGWARTs algorithm and web application. RS discussed the results and commented on the manuscript. SS provided commens on the manuscript. LSb contributed to the design of X-shooter observations. PScha was a member of the writing team on-call for this event and contributed to editing the manuscript. PSchi is the VST PI and provided guaranteed time VST observations. ASBS carried out observations with Pan-STARRS. TS is a GROND team member and reduced GROND images. SJS contributed to governance as the chair of the GC, co-leads the Pan-STARRS and ATLAS surveys, and contributed manuscript text. MS served on the on-call operations team. KWS developed and ran the Pan-STARRS and ATLAS transient science servers at QUB. JS contributed to governance as a GC member and provided comments on the manuscript. SS helped compile the table of PS1 candidates. ERS contributed as a member of the GOTO team. RLCS was a member of the writing team. DS contributed to governance as a GC member, is the GOTO PI \& liason, was the WHT UK PI, contributed data analysis and provided content. GS served on the on-call writing team. CWS provided Pan-STARRS and ATLAS science analysis. NRT is PI of VINROUGE and contributed to governance as a GC member, contributed to VISTA analysis and paper writing. VT contributed to data analysis. JLT co-leads the ATLAS science surveys. MT served on the on-call operations triggering VLT observations and writing teams. KU processed GOTO data -- image stacking and photometric calibration. AJvdH provided comments on the manuscript. SDV contributed to management as an EC member, served on the on-call operations team, and coordinated part of the VLT observations. NAW provided comments on the manuscript. DW served on the writing team on-call for this event, on the on-call operations team and contributed to the interpretation of the data. KWie served on the on-call operations team. KWii developed and maintained the computing environment and software for the imaging working group. LW served on the on-call operations team. SY contributed to transient detection in VST data. SXY provided comments on the manuscript. DRY built and helps manage much of the ENGRAVE communication infrastructure. 

\bibliographystyle{aa}
\bibliography{newref.bib}

\begin{thebibliography}{258}
\expandafter\ifx\csname natexlab\endcsname\relax\def\natexlab#1{#1}\fi

\bibitem[{Abadie {et~al.}(2010)}]{ratesdoc}
Abadie, J. {et~al.} 2010, Classical and Quantum Gravity, 27, 173001

\bibitem[{{Abazajian} {et~al.}(2009){Abazajian}, {Adelman-McCarthy},
  {Ag{\"u}eros}, {Allam}, {Allende Prieto}, {An}, {Anderson}, {Anderson},
  {Annis}, {Bahcall}, \& et~al.}]{abazajian09}
{Abazajian}, K.~N., {Adelman-McCarthy}, J.~K., {Ag{\"u}eros}, M.~A., {et~al.}
  2009, \apjs, 182, 543

\bibitem[{{Abbott} {et~al.}(2018{\natexlab{a}}){Abbott}, {Abbott}, {Abbott},
  {Abernathy}, {Acernese}, {Ackley}, {Adams}, {Adams}, {Addesso}, {Adhikari},
  {Adya}, {Affeldt}, {Agathos}, {Agatsuma}, {Aggarwal}, {Aguiar}, {Aiello},
  {Ain}, {Ajith}, {Akutsu}, {Allen}, {Allocca}, {Altin}, {Ananyeva},
  {Anderson}, {Anderson}, {Ando}, {Appert}, {Arai}, {Araya}, {Araya}, {Areeda},
  {Arnaud}, {Arun}, {Asada}, {Ascenzi}, {Ashton}, {Aso}, {Ast}, {Aston},
  {Astone}, {Atsuta}, {Aufmuth}, {Aulbert}, {Avila-Alvarez}, {Awai}, {Babak},
  {Bacon}, {Bader}, {Baiotti}, {Baker}, {Baldaccini}, {Ballardin}, {Ballmer},
  {Barayoga}, {Barclay}, {Barish}, {Barker}, {Barone}, {Barr}, {Barsotti},
  {Barsuglia}, {Barta}, {Bartlett}, {Barton}, {Bartos}, {Bassiri}, {Basti},
  {Batch}, {Baune}, {Bavigadda}, {Bazzan}, {B{\'e}csy}, {Beer}, {Bejger},
  {Belahcene}, {Belgin}, {Bell}, {Berger}, {Bergmann}, {Berry}, {Bersanetti},
  {Bertolini}, {Betzwieser}, {Bhagwat}, {Bhandare}, {Bilenko}, {Billingsley},
  {Billman}, {Birch}, {Birney}, {Birnholtz}, {Biscans}, {Bisht}, {Bitossi},
  {Biwer}, {Bizouard}, {Blackburn}, {Blackman}, {Blair}, {Blair}, {Blair},
  {Bloemen}, {Bock}, {Boer}, {Bogaert}, {Bohe}, {Bondu}, {Bonnand }, {Boom},
  {Bork}, {Boschi}, {Bose}, {Bouffanais}, {Bozzi}, {Bradaschia}, {Brady},
  {Braginsky}, {Branchesi}, {Brau}, {Briant}, {Brillet}, {Brinkmann},
  {Brisson}, {Brockill}, {Broida}, {Brooks}, {Brown}, {Brown}, {Brown},
  {Brunett}, {Buchanan}, {Buikema}, {Bulik}, {Bulten}, {Buonanno}, {Buskulic},
  {Buy}, {Byer}, {Cabero}, {Cadonati}, {Cagnoli}, {Cahillane}, {Calder{\'o}n
  Bustillo}, {Callister}, {Calloni}, {Camp}, {Cannon}, {Cao}, {Cao}, {Capano},
  {Capocasa}, {Carbognani}, {Caride}, {Casanueva Diaz}, {Casentini}, {Caudill},
  {Cavagli{\`a}}, {Cavalier}, {Cavalieri}, {Cella}, {Cepeda}, {Cerboni
  Baiardi}, {Cerretani}, {Cesarini}, {Chamberlin}, {Chan}, {Chao}, {Charlton},
  {Chassande-Mottin}, {Cheeseboro}, {Chen}, {Chen}, {Cheng}, {Chincarini},
  {Chiummo}, {Chmiel}, {Cho}, {Cho}, {Chow}, {Christensen}, {Chu}, {Chua},
  {Chua}, {Chung}, {Ciani}, {Clara}, {Clark}, {Cleva}, {Cocchieri}, {Coccia},
  {Cohadon}, {Colla}, {Collette}, {Cominsky}, {Constancio}, {Conti}, {Cooper},
  {Corbitt}, {Cornish}, {Corsi}, {Cortese}, {Costa}, {Coughlin}, {Coughlin},
  {Coulon}, {Countryman}, {Couvares}, {Covas}, {Cowan}, {Coward}, {Cowart},
  {Coyne}, {Coyne}, {Creighton}, {Creighton}, {Cripe}, {Crowder}, {Cullen},
  {Cumming}, {Cunningham}, {Cuoco}, {Dal Canton}, {Danilishin}, {D'Antonio},
  {Danzmann}, {Dasgupta}, {Da Silva Costa}, {Dattilo}, {Dave}, {Davier},
  {Davies}, {Davis}, {Daw}, {Day}, {Day}, {De}, {DeBra}, {Debreczeni},
  {Degallaix}, {De Laurentis}, {Del{\'e}glise}, {Del Pozzo}, {Denker}, {Dent},
  {Dergachev}, {De Rosa}, {DeRosa}, {DeSalvo}, {Devine}, {Dhurandhar},
  {D{\'\i}az}, {Fiore}, {Giovanni}, {Girolamo}, {Lieto}, {Pace}, {Palma},
  {Virgilio}, {Doctor}, {Doi}, {Dolique}, {Donovan}, {Dooley}, {Doravari},
  {Dorrington}, {Douglas}, {Dovale {\'A}lvarez}, {Downes}, {Drago}, {Drever},
  {Driggers}, {Du}, {Ducrot}, {Dwyer}, {Eda}, {Edo}, {Edwards}, {Effler},
  {Eggenstein}, {Ehrens}, {Eichholz}, {Eikenberry}, {Eisenstein}, {Essick},
  {Etienne}, {Etzel}, {Evans}, {Evans}, {Everett}, {Factourovich}, {Fafone},
  {Fair}, {Fairhurst}, {Fan}, {Farinon}, {Farr}, {Farr}, {Fauchon-Jones},
  {Favata}, {Fays}, {Fehrmann}, {Fejer}, {Fern{\'a}ndez Galiana}, {Ferrante},
  {Ferreira}, {Ferrini}, {Fidecaro}, {Fiori}, {Fiorucci}, {Fisher}, {Flaminio},
  {Fletcher}, {Fong}, {Forsyth}, {Fournier}, {Frasca}, {Frasconi}, {Frei},
  {Freise}, {Frey}, {Frey}, {Fries}, {Fritschel}, {Frolov}, {Fujii},
  {Fujimoto}, {Fulda}, {Fyffe}, {Gabbard}, {Gadre}, {Gaebel}, {Gair},
  {Gammaitoni}, {Gaonkar}, {Garufi}, {Gaur}, {Gayathri}, {Gehrels}, {Gemme},
  {Genin}, {Gennai}, {George}, {Gergely}, {Germain}, {Ghonge}, {Ghosh},
  {Ghosh}, {Ghosh}, {Giaime}, {Giardina}, {Giazotto}, {Gill}, {Glaefke},
  {Goetz}, {Goetz}, {Gondan}, {Gonz{\'a}lez}, {Gonzalez Castro}, {Gopakumar},
  {Gorodetsky}, {Gossan}, {Gosselin}, {Gouaty}, {Grado}, {Graef}, {Granata},
  {Grant}, {Gras}, {Gray}, {Greco}, {Green}, {Groot}, {Grote}, {Grunewald},
  {Guidi}, {Guo}, {Gupta}, {Gupta}, {Gushwa}, {Gustafson}, {Gustafson},
  {Hacker}, {Hagiwara}, {Hall}, {Hall}, {Hammond}, {Haney}, {Hanke}, {Hanks},
  {Hanna}, {Hannam}, {Hanson}, {Hardwick}, {Harms}, {Harry}, {Harry}, {Hart},
  {Hartman}, {Haster}, {Haughian}, {Hayama}, {Healy}, {Heidmann}, {Heintze},
  {Heitmann}, {Hello}, {Hemming}, {Hendry}, {Heng}, {Hennig}, {Henry},
  {Heptonstall}, {Heurs}, {Hild}, {Hirose}, {Hoak}, {Hofman}, {Holt}, {Holz},
  {Hopkins}, {Hough}, {Houston}, {Howell}, {Hu}, {Huerta}, {Huet}, {Hughey},
  {Husa}, {Huttner}, {Huynh-Dinh}, {Indik}, {Ingram}, {Inta}, {Ioka}, {Isa},
  {Isac}, {Isi}, {Isogai}, {Itoh}, {Iyer}, {Izumi}, {Jacqmin}, {Jani},
  {Jaranowski}, {Jawahar}, {Jim{\'e}nez-Forteza}, {Johnson}, {Jones}, {Jones},
  {Jonker}, {Ju}, {Junker}, {Kagawa}, {Kajita}, {Kakizaki}, {Kalaghatgi},
  {Kalogera}, {Kamiizumi}, {Kanda}, {Kand hasamy}, {Kanemura}, {Kaneyama},
  {Kang}, {Kanner}, {Karki}, {Karvinen}, {Kasprzack}, {Kataoka},
  {Katsavounidis}, {Katzman}, {Kaufer}, {Kaur}, {Kawabe}, {Kawai}, {Kawamura},
  {K{\'e}f{\'e}lian}, {Keitel}, {Kelley}, {Kennedy}, {Key}, {Khalili}, {Khan},
  {Khan}, {Khan}, {Khazanov}, {Kijbunchoo}, {Kim}, {Kim}, {Kim}, {Kim}, {Kim},
  {Kim}, {Kimbrell}, {Kimura}, {King}, {King}, {Kirchhoff}, {Kissel}, {Klein},
  {Kleybolte}, {Klimenko}, {Koch}, {Koehlenbeck}, {Kojima}, {Kokeyama},
  {Koley}, {Komori}, {Kondrashov}, {Kontos}, {Korobko}, {Korth}, {Kotake},
  {Kowalska}, {Kozak}, {Kr{\"a}mer}, {Kringel}, {Krishnan}, {Kr{\'o}lak},
  {Kuehn}, {Kumar}, {Kumar}, {Kumar}, {Kuo}, {Kuroda}, {Kutynia}, {Kuwahara},
  {Lackey}, {Landry}, {Lang}, {Lange}, {Lantz}, {Lanza}, {Lartaux-Vollard},
  {Lasky}, {Laxen}, {Lazzarini}, {Lazzaro}, {Leaci}, {Leavey}, {Lebigot},
  {Lee}, {Lee}, {Lee}, {Lee}, {Lee}, {Lehmann}, {Lenon}, {Leonardi}, {Leong},
  {Leroy}, {Letendre}, {Levin}, {Li}, {Libson}, {Littenberg}, {Liu},
  {Lockerbie}, {Lombardi}, {London}, {Lord}, {Lorenzini}, {Loriette},
  {Lormand}, {Losurdo}, {Lough}, {Lousto}, {Lovelace}, {L{\"u}ck}, {Lundgren},
  {Lynch}, {Ma}, {Macfoy}, {Machenschalk}, {MacInnis}, {Macleod},
  {Maga{\~n}a-Sandoval}, {Majorana}, {Maksimovic}, {Malvezzi}, {Man}, {Mandic},
  {Mangano}, {Mano}, {Mansell}, {Manske}, {Mantovani}, {Marchesoni}, {Marchio},
  {Marion}, {M{\'a}rka}, {M{\'a}rka}, {Markosyan}, {Maros}, {Martelli},
  {Martellini}, {Martin}, {Martynov}, {Mason}, {Masserot}, {Massinger},
  {Masso-Reid}, {Mastrogiovanni}, {Matichard}, {Matone}, {Matsumoto},
  {Matsushima}, {Mavalvala}, {Mazumder}, {McCarthy}, {McClelland },
  {McCormick}, {McGrath}, {McGuire}, {McIntyre}, {McIver}, {McManus}, {McRae},
  {McWilliams}, {Meacher}, {Meadors}, {Meidam}, {Melatos}, {Mendell},
  {Mendoza-Gand ara}, {Mercer}, {Merilh}, {Merzougui}, {Meshkov}, {Messenger},
  {Messick}, {Metzdorff}, {Meyers}, {Mezzani}, {Miao}, {Michel}, {Michimura},
  {Middleton}, {Mikhailov}, {Milano}, {Miller}, {Miller}, {Miller}, {Miller},
  {Millhouse}, {Minenkov}, {Ming}, {Mirshekari}, {Mishra}, {Mitrofanov},
  {Mitselmakher}, {Mittleman}, {Miyakawa}, {Miyamoto}, {Miyamoto}, {Miyoki},
  {Moggi}, {Mohan}, {Mohapatra}, {Montani}, {Moore}, {Moore}, {Moraru},
  {Moreno}, {Morii}, {Morisaki}, {Moriwaki}, {Morriss}, {Mours}, {Mow-Lowry},
  {Mueller}, {Muir}, {Mukherjee}, {Mukherjee}, {Mukherjee}, {Mukund},
  {Mullavey}, {Munch}, {Muniz}, {Murray}, {Mytidis}, {Nagano}, {Nakamura},
  {Nakamura}, {Nakano}, {Nakano}, {Nakano}, {Nakao}, {Napier}, {Nardecchia},
  {Narikawa}, {Naticchioni}, {Nelemans}, {Nelson}, {Neri}, {Nery}, {Neunzert},
  {Newport}, {Newton}, {Nguyen}, {Ni}, {Nielsen}, {Nissanke}, {Nitz}, {Noack},
  {Nocera}, {Nolting}, {Normandin}, {Nuttall}, {Oberling}, {Ochsner}, {Oelker},
  {Ogin}, {Oh}, {Oh}, {Ohashi}, {Ohishi}, {Ohkawa}, {Ohme}, {Okutomi},
  {Oliver}, {Ono}, {Ono}, {Oohara}, {Oppermann}, {Oram}, {O'Reilly},
  {O'Shaughnessy}, {Ottaway}, {Overmier}, {Owen}, {Pace}, {Page}, {Pai}, {Pai},
  {Palamos}, {Palashov}, {Palomba}, {Pal-Singh}, {Pan}, {Pankow}, {Pannarale},
  {Pant}, {Paoletti}, {Paoli}, {Papa}, {Paris}, {Parker}, {Pascucci},
  {Pasqualetti}, {Passaquieti}, {Passuello}, {Patricelli}, {Pearlstone},
  {Pedraza}, {Pedurand }, {Pekowsky}, {Pele}, {Pe{\~n}a Arellano}, {Penn},
  {Perez}, {Perreca}, {Perri}, {Pfeiffer}, {Phelps}, {Piccinni}, {Pichot},
  {Piergiovanni}, {Pierro}, {Pillant}, {Pinard}, {Pinto}, {Pitkin}, {Poe},
  {Poggiani}, {Popolizio}, {Post}, {Powell}, {Prasad}, {Pratt}, {Predoi},
  {Prestegard}, {Prijatelj}, {Principe}, {Privitera}, {Prodi}, {Prokhorov},
  {Puncken}, {Punturo}, {Puppo}, {P{\"u}rrer}, {Qi}, {Qin}, {Qiu}, {Quetschke},
  {Quintero}, {Quitzow-James}, {Raab}, {Rabeling}, {Radkins}, {Raffai}, {Raja},
  {Rajan}, {Rakhmanov}, {Rapagnani}, {Raymond}, {Razzano}, {Re}, {Read},
  {Regimbau}, {Rei}, {Reid}, {Reitze}, {Rew}, {Reyes}, {Rhoades}, {Ricci},
  {Riles}, {Rizzo}, {Robertson}, {Robie}, {Robinet}, {Rocchi}, {Rolland},
  {Rollins}, {Roma}, {Romano}, {Romie}, {Rosi{\'n}ska}, {Rowan}, {R{\"u}diger},
  {Ruggi}, \& {Ryan}}]{LVC2018}
{Abbott}, B.~P., {Abbott}, R., {Abbott}, T.~D., {et~al.} 2018{\natexlab{a}},
  Living Reviews in Relativity, 21, 3

\bibitem[{{Abbott} {et~al.}(2016){Abbott}, {Abbott}, {Abbott}, {Abernathy},
  {Acernese}, {Ackley}, {Adams}, {Adams}, {Addesso}, {Adhikari}, \&
  et~al.}]{lvc_150914}
{Abbott}, B.~P., {Abbott}, R., {Abbott}, T.~D., {et~al.} 2016, Physical Review
  Letters, 116, 061102

\bibitem[{{Abbott} {et~al.}(2017{\natexlab{a}}){Abbott}, {Abbott}, {Abbott},
  {Acernese}, {Ackley}, {Adams}, {Adams}, {Addesso}, {Adhikari}, {Adya}, \&
  et~al.}]{lvc_h0}
{Abbott}, B.~P., {Abbott}, R., {Abbott}, T.~D., {et~al.} 2017{\natexlab{a}},
  \nat, 551, 85

\bibitem[{{Abbott} {et~al.}(2017{\natexlab{b}}){Abbott}, {Abbott}, {Abbott},
  {Acernese}, {Ackley}, {Adams}, {Adams}, {Addesso}, {Adhikari}, {Adya}, \&
  et~al.}]{lvc17grb}
{Abbott}, B.~P., {Abbott}, R., {Abbott}, T.~D., {et~al.} 2017{\natexlab{b}},
  \apjl, 848, L13

\bibitem[{{Abbott} {et~al.}(2017{\natexlab{c}}){Abbott}, {Abbott}, {Abbott},
  {Acernese}, {Ackley}, {Adams}, {Adams}, {Addesso}, {Adhikari}, {Adya}, \&
  et~al.}]{lvc_170817}
{Abbott}, B.~P., {Abbott}, R., {Abbott}, T.~D., {et~al.} 2017{\natexlab{c}},
  Physical Review Letters, 119, 161101

\bibitem[{{Abbott} {et~al.}(2017{\natexlab{d}}){Abbott}, {Abbott}, {Abbott},
  {Acernese}, {Ackley}, {Adams}, {Adams}, {Addesso}, {Adhikari}, {Adya}, \&
  et~al.}]{lvc_mma}
{Abbott}, B.~P., {Abbott}, R., {Abbott}, T.~D., {et~al.} 2017{\natexlab{d}},
  \apjl, 848, L12

\bibitem[{{Abbott} {et~al.}(2018{\natexlab{b}}){Abbott}, {Abbott}, {Abbott},
  {Acernese}, {Ackley}, {Adams}, {Adams}, {Addesso}, {Adhikari}, {Adya}, \&
  et~al.}]{lvc_nseos}
{Abbott}, B.~P., {Abbott}, R., {Abbott}, T.~D., {et~al.} 2018{\natexlab{b}},
  Physical Review Letters, 121, 161101

\bibitem[{{Abbott} {et~al.}(2018{\natexlab{c}}){Abbott}, {Abdalla}, {Allam},
  {Amara}, {Annis}, {Asorey}, {Avila}, {Ballester}, {Banerji}, {Barkhouse},
  {Baruah}, {Baumer}, {Bechtol}, {Becker}, {Benoit-L{\'e}vy}, {Bernstein},
  {Bertin}, {Blazek}, {Bocquet}, {Brooks}, {Brout}, {Buckley-Geer}, {Burke},
  {Busti}, {Campisano}, {Cardiel-Sas}, {Carnero Rosell}, {Carrasco Kind},
  {Carretero}, {Castander}, {Cawthon}, {Chang}, {Chen}, {Conselice}, {Costa},
  {Crocce}, {Cunha}, {D'Andrea}, {da Costa}, {Das}, {Daues}, {Davis}, {Davis},
  {De Vicente}, {DePoy}, {DeRose}, {Desai}, {Diehl}, {Dietrich}, {Dodelson},
  {Doel}, {Drlica-Wagner}, {Eifler}, {Elliott}, {Evrard}, {Farahi}, {Fausti
  Neto}, {Fernandez}, {Finley}, {Flaugher}, {Foley}, {Fosalba}, {Friedel},
  {Frieman}, {Garc{\'\i}a-Bellido}, {Gaztanaga}, {Gerdes}, {Giannantonio},
  {Gill}, {Glazebrook}, {Goldstein}, {Gower}, {Gruen}, {Gruendl}, {Gschwend},
  {Gupta}, {Gutierrez}, {Hamilton}, {Hartley}, {Hinton}, {Hislop}, {Hollowood},
  {Honscheid}, {Hoyle}, {Huterer}, {Jain}, {James}, {Jeltema}, {Johnson},
  {Johnson}, {Kacprzak}, {Kent}, {Khullar}, {Klein}, {Kovacs}, {Koziol},
  {Krause}, {Kremin}, {Kron}, {Kuehn}, {Kuhlmann}, {Kuropatkin}, {Lahav},
  {Lasker}, {Li}, {Li}, {Liddle}, {Lima}, {Lin}, {L{\'o}pez-Reyes}, {MacCrann},
  {Maia}, {Maloney}, {Manera}, {March}, {Marriner}, {Marshall}, {Martini},
  {McClintock}, {McKay}, {McMahon}, {Melchior}, {Menanteau}, {Miller},
  {Miquel}, {Mohr}, {Morganson}, {Mould}, {Neilsen}, {Nichol}, {Nogueira},
  {Nord}, {Nugent}, {Nunes}, {Ogand o}, {Old}, {Pace}, {Palmese},
  {Paz-Chinch{\'o}n}, {Peiris}, {Percival}, {Petravick}, {Plazas}, {Poh},
  {Pond}, {Porredon}, {Pujol}, {Refregier}, {Reil}, {Ricker}, {Rollins},
  {Romer}, {Roodman}, {Rooney}, {Ross}, {Rykoff}, {Sako}, {Sanchez}, {Sanchez},
  {Santiago}, {Saro}, {Scarpine}, {Scolnic}, {Serrano}, {Sevilla-Noarbe},
  {Sheldon}, {Shipp}, {Silveira}, {Smith}, {Smith}, {Smith}, {Soares-Santos},
  {Sobreira}, {Song}, {Stebbins}, {Suchyta}, {Sullivan}, {Swanson}, {Tarle},
  {Thaler}, {Thomas}, {Thomas}, {Troxel}, {Tucker}, {Vikram}, {Vivas},
  {Walker}, {Wechsler}, {Weller}, {Wester}, {Wolf}, {Wu}, {Yanny}, {Zenteno},
  {Zhang}, {Zuntz}, {DES Collaboration}, {Juneau}, {Fitzpatrick}, {Nikutta},
  {Nidever}, {Olsen}, {Scott}, \& {NOAO Data Lab}}]{Abbottetal2018ApJS}
{Abbott}, T.~M.~C., {Abdalla}, F.~B., {Allam}, S., {et~al.} 2018{\natexlab{c}},
  \apjs, 239, 18

\bibitem[{{Ackley} {et~al.}(2019){Ackley}, {Dyer}, {Eyles}, {Gallowey},
  {Gompertz}, {Kennedy}, {Levan}, {Lyman}, {Mong}, \&
  {Obradovic}}]{2019GCN.25337....1A}
{Ackley}, K., {Dyer}, M., {Eyles}, R., {et~al.} 2019, GRB Coordinates Network,
  25337, 1

\bibitem[{{Acosta-Pulido} {et~al.}(2002){Acosta-Pulido}, {Ballesteros},
  {Barreto}, {Correa}, {Delgado}, {Dominguez-Tagle}, {Hernand ez}, {Lopez},
  {Manchado}, {Manescau}, {Moreno}, {Prada}, {Redondo}, {Sanchez}, \&
  {Tenegi}}]{Acosta-Pulido2002}
{Acosta-Pulido}, J., {Ballesteros}, E., {Barreto}, M., {et~al.} 2002, The
  Newsletter of the Isaac Newton Group of Telescopes, 6, 22

\bibitem[{{Ageron} {et~al.}(2019){Ageron}, {Baret}, {Coleiro}, {Colomer},
  {Dornic}, {Kouchner}, \& {Pradier}}]{2019GCN.25330....1A}
{Ageron}, M., {Baret}, B., {Coleiro}, A., {et~al.} 2019, GRB Coordinates
  Network, 25330, 1

\bibitem[{{Alam} {et~al.}(2015){Alam}, {Albareti}, {Allende Prieto}, {Anders},
  {Anderson}, {Anderton}, {Andrews}, {Armengaud}, {Aubourg}, {Bailey}, {Basu},
  {Bautista}, {Beaton}, {Beers}, {Bender}, {Berlind}, {Beutler}, {Bhardwaj},
  {Bird}, {Bizyaev}, {Blake}, {Blanton}, {Blomqvist}, {Bochanski}, {Bolton},
  {Bovy}, {Shelden Bradley}, {Brandt}, {Brauer}, {Brinkmann}, {Brown},
  {Brownstein}, {Burden}, {Burtin}, {Busca}, {Cai}, {Capozzi}, {Carnero
  Rosell}, {Carr}, {Carrera}, {Chambers}, {Chaplin}, {Chen}, {Chiappini},
  {Chojnowski}, {Chuang}, {Clerc}, {Comparat}, {Covey}, {Croft}, {Cuesta},
  {Cunha}, {da Costa}, {Da Rio}, {Davenport}, {Dawson}, {De Lee}, {Delubac},
  {Deshpande}, {Dhital}, {Dutra-Ferreira}, {Dwelly}, {Ealet}, {Ebelke},
  {Edmondson}, {Eisenstein}, {Ellsworth}, {Elsworth}, {Epstein}, {Eracleous},
  {Escoffier}, {Esposito}, {Evans}, {Fan}, {Fern{\'a}ndez-Alvar}, {Feuillet},
  {Filiz Ak}, {Finley}, {Finoguenov}, {Flaherty}, {Fleming}, {Font-Ribera},
  {Foster}, {Frinchaboy}, {Galbraith-Frew}, {Garc{\'\i}a},
  {Garc{\'\i}a-Hern{\'a}ndez}, {Garc{\'\i}a P{\'e}rez}, {Gaulme}, {Ge},
  {G{\'e}nova-Santos}, {Georgakakis}, {Ghezzi}, {Gillespie}, {Girardi},
  {Goddard}, {Gontcho}, {Gonz{\'a}lez Hern{\'a}ndez}, {Grebel}, {Green},
  {Grieb}, {Grieves}, {Gunn}, {Guo}, {Harding}, {Hasselquist}, {Hawley},
  {Hayden}, {Hearty}, {Hekker}, {Ho}, {Hogg}, {Holley-Bockelmann}, {Holtzman},
  {Honscheid}, {Huber}, {Huehnerhoff}, {Ivans}, {Jiang}, {Johnson},
  {Kinemuchi}, {Kirkby}, {Kitaura}, {Klaene}, {Knapp}, {Kneib}, {Koenig},
  {Lam}, {Lan}, {Lang}, {Laurent}, {Le Goff}, {Leauthaud}, {Lee}, {Lee},
  {Licquia}, {Liu}, {Long}, {L{\'o}pez-Corredoira}, {Lorenzo-Oliveira},
  {Lucatello}, {Lundgren}, {Lupton}, {Mack}, {Mahadevan}, {Maia}, {Majewski},
  {Malanushenko}, {Malanushenko}, {Manchado}, {Manera}, {Mao}, {Maraston},
  {Marchwinski}, {Margala}, {Martell}, {Martig}, {Masters}, {Mathur},
  {McBride}, {McGehee}, {McGreer}, {McMahon}, {M{\'e}nard}, {Menzel},
  {Merloni}, {M{\'e}sz{\'a}ros}, {Miller}, {Miralda-Escud{\'e}}, {Miyatake},
  {Montero-Dorta}, {More}, {Morganson}, {Morice-Atkinson}, {Morrison},
  {Mosser}, {Muna}, {Myers}, {Nand ra}, {Newman}, {Neyrinck}, {Nguyen},
  {Nichol}, {Nidever}, {Noterdaeme}, {Nuza}, {O'Connell}, {O'Connell},
  {O'Connell}, {Ogando}, {Olmstead}, {Oravetz}, {Oravetz}, {Osumi}, {Owen},
  {Padgett}, {Padmanabhan}, {Paegert}, {Palanque-Delabrouille}, {Pan},
  {Parejko}, {P{\^a}ris}, {Park}, {Pattarakijwanich}, {Pellejero-Ibanez},
  {Pepper}, {Percival}, {P{\'e}rez-Fournon}, {Ṕrez-Ra`fols}, {Petitjean},
  {Pieri}, {Pinsonneault}, {Porto de Mello}, {Prada}, {Prakash},
  {Price-Whelan}, {Protopapas}, {Raddick}, {Rahman}, {Reid}, {Rich}, {Rix},
  {Robin}, {Rockosi}, {Rodrigues}, {Rodr{\'\i}guez-Torres}, {Roe}, {Ross},
  {Ross}, {Rossi}, {Ruan}, {Rubi{\~n}o-Mart{\'\i}n}, {Rykoff},
  {Salazar-Albornoz}, {Salvato}, {Samushia}, {S{\'a}nchez}, {Santiago},
  {Sayres}, {Schiavon}, {Schlegel}, {Schmidt}, {Schneider}, {Schultheis},
  {Schwope}, {Sc{\'o}ccola}, {Scott}, {Sellgren}, {Seo}, {Serenelli}, {Shane},
  {Shen}, {Shetrone}, {Shu}, {Silva Aguirre}, {Sivarani}, {Skrutskie},
  {Slosar}, {Smith}, {Sobreira}, {Souto}, {Stassun}, {Steinmetz}, {Stello},
  {Strauss}, {Streblyanska}, {Suzuki}, {Swanson}, {Tan}, {Tayar}, {Terrien},
  {Thakar}, {Thomas}, {Thomas}, {Thompson}, {Tinker}, {Tojeiro}, {Troup},
  {Vargas-Maga{\~n}a}, {Vazquez}, {Verde}, {Viel}, {Vogt}, {Wake}, {Wang},
  {Weaver}, {Weinberg}, {Weiner}, {White}, {Wilson}, {Wisniewski},
  {Wood-Vasey}, {Ye`che}, {York}, {Zakamska}, {Zamora}, {Zasowski}, {Zehavi},
  {Zhao}, {Zheng}, {Zhou}, {Zhou}, {Zou}, \& {Zhu}}]{2015ApJS..219...12A}
{Alam}, S., {Albareti}, F.~D., {Allende Prieto}, C., {et~al.} 2015, \apjs, 219,
  12

\bibitem[{{Alvarez-Muniz} {et~al.}(2019){Alvarez-Muniz}, {Pedreira}, {Zas},
  {Hampert}, \& {Schimp}}]{2019GCN.25409....1A}
{Alvarez-Muniz}, J., {Pedreira}, F., {Zas}, E., {Hampert}, K.~H., \& {Schimp},
  M. 2019, GRB Coordinates Network, 25409, 1

\bibitem[{{Andreoni} {et~al.}(2017){Andreoni}, {Ackley}, {Cooke}, {Acharyya},
  {Allison}, {Anderson}, {Ashley}, {Baade}, {Bailes}, {Bannister}, {Beardsley},
  {Bessell}, {Bian}, {Bland}, {Boer}, {Booler}, {Brandeker}, {Brown},
  {Buckley}, {Chang}, {Coward}, {Crawford}, {Crisp}, {Crosse}, {Cucchiara},
  {Cup{\'a}k}, {de Gois}, {Deller}, {Devillepoix}, {Dobie}, {Elmer}, {Emrich},
  {Farah}, {Farrell}, {Franzen}, {Gaensler}, {Galloway}, {Gendre}, {Giblin},
  {Goobar}, {Green}, {Hancock}, {Hartig}, {Howell}, {Horsley}, {Hotan},
  {Howie}, {Hu}, {Hu}, {James}, {Johnston}, {Johnston-Hollitt}, {Kaplan},
  {Kasliwal}, {Keane}, {Kenney}, {Klotz}, {Lau}, {Laugier}, {Lenc}, {Li},
  {Liang}, {Lidman}, {Luvaul}, {Lynch}, {Ma}, {Macpherson}, {Mao},
  {McClelland}, {McCully}, {M{\"o}ller}, {Morales}, {Morris}, {Murphy},
  {Noysena}, {Onken}, {Orange}, {Os{\l}owski}, {Pallot}, {Paxman}, {Potter},
  {Pritchard}, {Raja}, {Ridden-Harper}, {Romero-Colmenero}, {Sadler}, {Sansom},
  {Scalzo}, {Schmidt}, {Scott}, {Seghouani}, {Shang}, {Shannon}, {Shao},
  {Shara}, {Sharp}, {Sokolowski}, {Sollerman}, {Staff}, {Steele}, {Sun},
  {Suntzeff}, {Tao}, {Tingay}, {Towner}, {Thierry}, {Trott}, {Tucker},
  {V{\"a}is{\"a}nen}, {Krishnan}, {Walker}, {Wang}, {Wang}, {Wayth}, {Whiting},
  {Williams}, {Williams}, {Wolf}, {Wu}, {Wu}, {Yang}, {Yuan}, {Zhang}, {Zhou},
  \& {Zovaro}}]{andreoni17}
{Andreoni}, I., {Ackley}, K., {Cooke}, J., {et~al.} 2017, \pasa, 34, e069

\bibitem[{{Andreoni} {et~al.}(2019{\natexlab{a}}){Andreoni}, {Goldstein},
  {Ahumada}, {Anand}, {Bulla}, {Dahiwale}, {De}, {Dhawan}, {Kasliwal}, \&
  {Kong}}]{2019GCN.25362....1A}
{Andreoni}, I., {Goldstein}, D.~A., {Ahumada}, T., {et~al.} 2019{\natexlab{a}},
  GRB Coordinates Network, 25362, 1

\bibitem[{{Andreoni} {et~al.}(2019{\natexlab{b}}){Andreoni}, {Goldstein},
  {Dobie}, \& {Kasliwal}}]{2019GCN.25488....1A}
{Andreoni}, I., {Goldstein}, D.~A., {Dobie}, D., \& {Kasliwal}, M.~M.
  2019{\natexlab{b}}, GRB Coordinates Network, 25488, 1

\bibitem[{{Andreoni} {et~al.}(2019{\natexlab{c}}){Andreoni}, {Goldstein},
  {Kasliwal}, {Nugent}, {Zhou}, {Newman}, {Bulla}, {Foucart}, {Hotokezaka},
  {Nakar}, {Nissanke}, {Raaijmakers}, {Bloom}, {De}, {Jencson}, {Ward},
  {Ahumada}, {Anand}, {Buckley}, {Caballero-Garc{\'\i}a}, {Castro-Tirado},
  {Copperwheat}, {Coughlin}, {Cenko}, {Gromadzki}, {Hu}, {Karambelkar},
  {Perley}, {Sharma}, {Valeev}, {Cook}, {Fremling}, {Kumar}, {Taggart},
  {Bagdasaryan}, {Cooke}, {Dahiwale}, {Dhawan}, {Dobie}, {Gatkine}, {Golkhou},
  {Goobar}, {Guerra Chaves}, {Hankins}, {Kaplan}, {Kong}, {Kool}, {Mohite},
  {Sollerman}, {Tzanidakis}, {Webb}, \& {Zhang}}]{Andreonietal2019}
{Andreoni}, I., {Goldstein}, D.~A., {Kasliwal}, M.~M., {et~al.}
  2019{\natexlab{c}}, \apj, in press (arXiv:1910.13409v2), arXiv:1910.13409v2

\bibitem[{{Antier} {et~al.}(2020){Antier}, {Agayeva}, {Aivazyan}, {Alishov},
  {Arbouch}, {Baransky}, {Barynova}, {Bai}, {Basa}, {Beradze}, {Bertin},
  {Berthier}, {Bla{\v{z}}ek}, {Bo{\"e}r}, {Burkhonov}, {Burrell}, {Cailleau},
  {Chabert}, {Chen}, {Christensen}, {Coleiro}, {Cordier}, {Corre}, {Coughlin},
  {Coward}, {Crisp}, {Delattre}, {Dietrich}, {Ducoin}, {Duverne},
  {Marchal-Duval}, {Gendre}, {Eymar}, {Fock-Hang}, {Han}, {Hello}, {Howell},
  {Inasaridze}, {Ismailov}, {Kann}, {Kapanadze}, {Klotz}, {Kochiashvili},
  {Lachaud}, {Leroy}, {Le Van Su}, {Lin}, {Li}, {Lognone}, {Marron}, {Mo},
  {Moore}, {Natsvlishvili}, {Noysena}, {Perrigault}, {Peyrot}, {Samadov},
  {Sadibekova}, {Simon}, {Stachie}, {Teng}, {Thierry}, {Th{\"o}ne}, {Tillayev},
  {Turpin}, {de Ugarte Postigo}, {Vachier}, {Vardosanidze}, {Vasylenko},
  {Vidadi}, {Wang}, {Wang}, {Wei}, {Yan}, {Zhang}, {Zhang}, \&
  {Zhang}}]{Antieretal2020}
{Antier}, S., {Agayeva}, S., {Aivazyan}, V., {et~al.} 2020, \mnras, 492, 3904

\bibitem[{{Appenzeller} {et~al.}(1998){Appenzeller}, {Fricke}, {F{\"u}rtig},
  {G{\"a}ssler}, {H{\"a}fner}, {Harke}, {Hess}, {Hummel}, {J{\"u}rgens},
  {Kudritzki}, {Mantel}, {Meisl}, {Muschielok}, {Nicklas}, {Rupprecht},
  {Seifert}, {Stahl}, {Szeifert}, \& {Tarantik}}]{appenzeller98}
{Appenzeller}, I., {Fricke}, K., {F{\"u}rtig}, W., {et~al.} 1998, The
  Messenger, 94, 1

\bibitem[{{Arcavi} {et~al.}(2017{\natexlab{a}}){Arcavi}, {Hosseinzadeh},
  {Howell}, {McCully}, {Poznanski}, {Kasen}, {Barnes}, {Zaltzman}, {Vasylyev},
  {Maoz}, \& {Valenti}}]{arcavi17b}
{Arcavi}, I., {Hosseinzadeh}, G., {Howell}, D.~A., {et~al.} 2017{\natexlab{a}},
  \nat, 551, 64

\bibitem[{{Arcavi} {et~al.}(2017{\natexlab{b}}){Arcavi}, {McCully},
  {Hosseinzadeh}, {Howell}, {Vasylyev}, {Poznanski}, {Zaltzman}, {Maoz},
  {Singer}, {Valenti}, {Kasen}, {Barnes}, {Piran}, \& {Fong}}]{arcavi17a}
{Arcavi}, I., {McCully}, C., {Hosseinzadeh}, G., {et~al.} 2017{\natexlab{b}},
  \apjl, 848, L33

\bibitem[{{Arnaboldi} {et~al.}(2010){Arnaboldi}, {Petr-Gotzens}, {Rejkuba},
  {Neeser}, {Szeifert}, {Ivanov}, {Hummel}, {Hilker}, {Neumayer}, {M{\o}ller},
  {Nilsson}, {Venemans}, {Hatziminaoglou}, {Hussain}, {Stanke}, {Teixeira},
  {Ramsay}, {Retzlaff}, {Slijkhuis}, {Comer{\'o}n}, {Melnick}, {Romaniello},
  {Emerson}, {Sutherland}, {Irwin}, {Lewis}, {Hodgkin}, \&
  {Gonzales-Solares}}]{2010Msngr.139....6A}
{Arnaboldi}, M., {Petr-Gotzens}, M., {Rejkuba}, M., {et~al.} 2010, The
  Messenger, 139, 6

\bibitem[{{Arsenault} {et~al.}(2006){Arsenault}, {Hubin}, {Stroebele},
  {Fedrigo}, {Oberti}, {Kissler-Patig}, {Bacon}, {McDermid}, {Bonaccini-Calia},
  {Biasi}, {Gallieni}, {Riccardi}, {Donaldson}, {Lelouarn}, {Hackenberg},
  {Conzelman}, {Delabre}, {Stuik}, {Paufique}, {Kasper}, {Vernet}, {Downing},
  {Esposito}, {Duchateau}, {Franx}, {Myers}, \& {Goodsell}}]{Arsenault2006}
{Arsenault}, R., {Hubin}, N., {Stroebele}, S., {et~al.} 2006, The Messenger,
  123, 6

\bibitem[{{Artale} {et~al.}(2019){Artale}, {Mapelli}, {Giacobbo}, {Sabha},
  {Spera}, {Santoliquido}, \& {Bressan}}]{Artale2019}
{Artale}, M.~C., {Mapelli}, M., {Giacobbo}, N., {et~al.} 2019, \mnras, 487,
  1675

\bibitem[{{Barbieri} {et~al.}(2019){Barbieri}, {Salafia}, {Perego}, {Colpi}, \&
  {Ghirlanda}}]{Barbieri19}
{Barbieri}, C., {Salafia}, O.~S., {Perego}, A., {Colpi}, M., \& {Ghirlanda}, G.
  2019, \aap, 625, A152

\bibitem[{{Barbieri} {et~al.}(2020){Barbieri}, {Salafia}, {Perego}, {Colpi}, \&
  {Ghirlanda}}]{Barbierietal2020EPJA}
{Barbieri}, C., {Salafia}, O.~S., {Perego}, A., {Colpi}, M., \& {Ghirlanda}, G.
  2020, European Physical Journal A, 56, 8

\bibitem[{{Bartos} {et~al.}(2017){Bartos}, {Kocsis}, {Haiman}, \&
  {M{\'a}rka}}]{Bartosetal2017}
{Bartos}, I., {Kocsis}, B., {Haiman}, Z., \& {M{\'a}rka}, S. 2017, \apj, 835,
  165

\bibitem[{{Bauer} {et~al.}(2019){Bauer}, {Fruchter}, {Gonzalez Lopez},
  {Hjorth}, {Kangas}, {Kim}, {Levan}, {Malesani}, {Michalowski}, \&
  {Milvang-Jensen}}]{2019GCN.25801....1B}
{Bauer}, F.~E., {Fruchter}, A.~S., {Gonzalez Lopez}, J., {et~al.} 2019, GRB
  Coordinates Network, 25801, 1

\bibitem[{{Bazin} {et~al.}(2011){Bazin}, {Ruhlmann-Kleider},
  {Palanque-Delabrouille}, {Rich}, {Aubourg}, {Astier}, {Balland}, {Basa},
  {Carlberg}, {Conley}, {Fouchez}, {Guy}, {Hardin}, {Hook}, {Howell}, {Pain},
  {Perrett}, {Pritchet}, {Regnault}, {Sullivan}, {Fourmanoit},
  {Gonz{\'a}lez-Gait{\'a}n}, {Lidman}, {Perlmutter}, {Ripoche}, \&
  {Walker}}]{bazin11}
{Bazin}, G., {Ruhlmann-Kleider}, V., {Palanque-Delabrouille}, N., {et~al.}
  2011, \aap, 534, A43

\bibitem[{{Becker}(2015)}]{becker2015}
{Becker}, A. 2015, {HOTPANTS: High Order Transform of PSF ANd Template
  Subtraction}

\bibitem[{{Belczynski} {et~al.}(2016){Belczynski}, {Holz}, {Bulik}, \&
  {O'Shaughnessy}}]{Belcynskietal2016}
{Belczynski}, K., {Holz}, D.~E., {Bulik}, T., \& {O'Shaughnessy}, R. 2016,
  \nat, 534, 512

\bibitem[{{Benn} {et~al.}(2008){Benn}, {Dee}, \& {Ag{\'o}cs}}]{Benn2008}
{Benn}, C., {Dee}, K., \& {Ag{\'o}cs}, T. 2008, Society of Photo-Optical
  Instrumentation Engineers (SPIE) Conference Series, Vol. 7014, {ACAM: a new
  imager/spectrograph for the William Herschel Telescope}, 70146X

\bibitem[{{Berger}(2010)}]{berger10}
{Berger}, E. 2010, \apj, 722, 1946

\bibitem[{{Berger} {et~al.}(2013){Berger}, {Fong}, \& {Chornock}}]{berger13}
{Berger}, E., {Fong}, W., \& {Chornock}, R. 2013, \apjl, 774, L23

\bibitem[{{Bertin} \& {Arnouts}(1996)}]{bertin1996}
{Bertin}, E. \& {Arnouts}, S. 1996, \aaps, 117, 393

\bibitem[{{Bloom} {et~al.}(2012){Bloom}, {Richards}, {Nugent}, {Quimby},
  {Kasliwal}, {Starr}, {Poznanski}, {Ofek}, {Cenko}, {Butler}, {Kulkarni},
  {Gal-Yam}, \& {Law}}]{2012PASP..124.1175B}
{Bloom}, J.~S., {Richards}, J.~W., {Nugent}, P.~E., {et~al.} 2012, \pasp, 124,
  1175

\bibitem[{{Boksenberg}(1985)}]{Boksenberg1985}
{Boksenberg}, A. 1985, Vistas in Astronomy, 28, 531

\bibitem[{{Brocato} {et~al.}(2018){Brocato}, {Branchesi}, {Cappellaro},
  {Covino}, {Grado}, {Greco}, {Limatola}, {Stratta}, {Yang}, {Campana},
  {D'Avanzo}, {Getman}, {Melandri}, {Nicastro}, {Palazzi}, {Pian},
  {Piranomonte}, {Pulone}, {Rossi}, {Tomasella}, {Amati}, {Antonelli},
  {Ascenzi}, {Benetti}, {Bulgarelli}, {Capaccioli}, {Cella}, {Dadina}, {De
  Cesare}, {D'Elia}, {Ghirlanda}, {Ghisellini}, {Giuffrida}, {Iannicola},
  {Israel}, {Lisi}, {Longo}, {Mapelli}, {Marinoni}, {Marrese}, {Masetti},
  {Patricelli}, {Possenti}, {Radovich}, {Razzano}, {Salvaterra}, {Schipani},
  {Spera}, {Stamerra}, {Stella}, {Tagliaferri}, {Testa}, \&
  {Grawita-Gravitational Wave Inaf Team}}]{Brocato18}
{Brocato}, E., {Branchesi}, M., {Cappellaro}, E., {et~al.} 2018, \mnras, 474,
  411

\bibitem[{{Brunn} {et~al.}(2019){Brunn}, {Sagues Carracedo}, {Chen},
  {Copperwheat}, {D'Avanzo}, {Fraser}, {Heintz}, {Hjorth}, {Japelj}, \&
  {Jonker}}]{2019GCN.25384....1B}
{Brunn}, S.~H., {Sagues Carracedo}, A., {Chen}, T.-W., {et~al.} 2019, GRB
  Coordinates Network, 25384, 1

\bibitem[{{Cai} {et~al.}(2019){Cai}, {Yi}, {Xiao}, {Luo}, {Li}, {Li}, {Li},
  {Liao}, {Xiong}, \& {Liu}}]{2019GCN.25365....1C}
{Cai}, C., {Yi}, Q.~B., {Xiao}, S., {et~al.} 2019, GRB Coordinates Network,
  25365, 1

\bibitem[{{Capaccioli} \& {Schipani}(2011)}]{Capaccioli11}
{Capaccioli}, M. \& {Schipani}, P. 2011, The Messenger, 146, 2

\bibitem[{{Cartier} {et~al.}(2019){Cartier}, {Briceno}, {Olivares}, {Tucker},
  {Antilon}, {Rodriguez}, {Meza-Retamal}, {Quirola}, {Points}, \&
  {Allam}}]{2019GCN.25784....1C}
{Cartier}, R., {Briceno}, C., {Olivares}, F., {et~al.} 2019, GRB Coordinates
  Network, 25784, 1

\bibitem[{{Casali} {et~al.}(2006){Casali}, {Pirard}, {Kissler-Patig},
  {Moorwood}, {Bedin}, {Biereichel}, {Delabre}, {Dorn}, {Finger}, {Gojak},
  {Huster}, {Jung}, {Koch}, {Lizon}, {Mehrgan}, {Pozna}, {Silber}, {Sokar}, \&
  {Stegmeier}}]{casali06}
{Casali}, M., {Pirard}, J.-F., {Kissler-Patig}, M., {et~al.} 2006, in
  \procspie, Vol. 6269, Society of Photo-Optical Instrumentation Engineers
  (SPIE) Conference Series, 62690W

\bibitem[{{Castro-Tirado} {et~al.}(2019){Castro-Tirado}, {Valeev}, {Hu},
  {Fernandez-Garcia}, {Sokolov}, {Carrasco}, {Castellon}, {Tucker},
  {Soares-Santos}, \& {Geier}}]{2019GCN.25543....1C}
{Castro-Tirado}, A.~J., {Valeev}, A.~F., {Hu}, Y.-D., {et~al.} 2019, GRB
  Coordinates Network, 25543, 1

\bibitem[{{Chambers} {et~al.}(2016){Chambers}, {Magnier}, {Metcalfe},
  {Flewelling}, {Huber}, {Waters}, {Denneau}, {Draper}, {Farrow}, {Finkbeiner},
  {Holmberg}, {Koppenhoefer}, {Price}, {Saglia}, {Schlafly}, {Smartt},
  {Sweeney}, {Wainscoat}, {Burgett}, {Grav}, {Heasley}, {Hodapp}, {Jedicke},
  {Kaiser}, {Kudritzki}, {Luppino}, {Lupton}, {Monet}, {Morgan}, {Onaka},
  {Stubbs}, {Tonry}, {Banados}, {Bell}, {Bender}, {Bernard}, {Botticella},
  {Casertano}, {Chastel}, {Chen}, {Chen}, {Cole}, {Deacon}, {Frenk},
  {Fitzsimmons}, {Gezari}, {Goessl}, {Goggia}, {Goldman}, {Grebel}, {Hambly},
  {Hasinger}, {Heavens}, {Heckman}, {Henderson}, {Henning}, {Holman}, {Hopp},
  {Ip}, {Isani}, {Keyes}, {Koekemoer}, {Kotak}, {Long}, {Lucey}, {Liu},
  {Martin}, {McLean}, {Morganson}, {Murphy}, {Nieto-Santisteban}, {Norberg},
  {Peacock}, {Pier}, {Postman}, {Primak}, {Rae}, {Rest}, {Riess}, {Riffeser},
  {Rix}, {Roser}, {Schilbach}, {Schultz}, {Scolnic}, {Szalay}, {Seitz},
  {Shiao}, {Small}, {Smith}, {Soderblom}, {Taylor}, {Thakar}, {Thiel},
  {Thilker}, {Urata}, {Valenti}, {Walter}, {Watters}, {Werner}, {White},
  {Wood-Vasey}, \& {Wyse}}]{chambers16}
{Chambers}, K.~C., {Magnier}, E.~A., {Metcalfe}, N., {et~al.} 2016, arXiv
  e-prints [\eprint[arXiv]{1612.05560}]

\bibitem[{{Chen} {et~al.}(2019){Chen}, {Nicuesa Guelbenzu}, {Fraser}, {Bolmer},
  {Schweyer}, {Rau}, {Klose}, {Salmon}, {Smartt}, \&
  {Levan}}]{2019GCN.25372....1C}
{Chen}, T.-W., {Nicuesa Guelbenzu}, A., {Fraser}, M., {et~al.} 2019, GRB
  Coordinates Network, 25372, 1

\bibitem[{{Chornock} {et~al.}(2017){Chornock}, {Berger}, {Kasen},
  {Cowperthwaite}, {Nicholl}, {Villar}, {Alexander}, {Blanchard}, {Eftekhari},
  {Fong}, {Margutti}, {Williams}, {Annis}, {Brout}, {Brown}, {Chen}, {Drout},
  {Farr}, {Foley}, {Frieman}, {Fryer}, {Herner}, {Holz}, {Kessler}, {Matheson},
  {Metzger}, {Quataert}, {Rest}, {Sako}, {Scolnic}, {Smith}, \&
  {Soares-Santos}}]{chornock17}
{Chornock}, R., {Berger}, E., {Kasen}, D., {et~al.} 2017, \apjl, 848, L19

\bibitem[{{Ciolfi} {et~al.}(2017){Ciolfi}, {Kastaun}, {Giacomazzo}, {Endrizzi},
  {Siegel}, \& {Perna}}]{Ciolfietal2017}
{Ciolfi}, R., {Kastaun}, W., {Giacomazzo}, B., {et~al.} 2017, \prd, 95, 063016

\bibitem[{{Colless} {et~al.}(2001){Colless}, {Dalton}, {Maddox}, {Sutherland },
  {Norberg}, {Cole}, {Bland -Hawthorn}, {Bridges}, {Cannon}, {Collins},
  {Couch}, {Cross}, {Deeley}, {De Propris}, {Driver}, {Efstathiou}, {Ellis},
  {Frenk}, {Glazebrook}, {Jackson}, {Lahav}, {Lewis}, {Lumsden}, {Madgwick},
  {Peacock}, {Peterson}, {Price}, {Seaborne}, \&
  {Taylor}}]{2001MNRAS.328.1039C}
{Colless}, M., {Dalton}, G., {Maddox}, S., {et~al.} 2001, \mnras, 328, 1039

\bibitem[{{Coughlin} {et~al.}(2020){Coughlin}, {Dietrich}, {Antier}, {Bulla},
  {Foucart}, {Hotokezaka}, {Raaijmakers}, {Hinderer}, \&
  {Nissanke}}]{Coughlin2020}
{Coughlin}, M.~W., {Dietrich}, T., {Antier}, S., {et~al.} 2020, \mnras, 492,
  863

\bibitem[{{Coulter} {et~al.}(2017{\natexlab{a}}){Coulter}, {Kilpatrick},
  {Siebert}, {Foley}, {Shappee}, {Drout}, {Simon}, A., \& {Rest}}]{coulterGCN}
{Coulter}, D., {Kilpatrick}, C., {Siebert}, M., {et~al.} 2017{\natexlab{a}},
  GRB Coordinates Network, Circular Service, No.~21529, \#1 (2017), 21529

\bibitem[{{Coulter} {et~al.}(2017{\natexlab{b}}){Coulter}, {Foley},
  {Kilpatrick}, {Drout}, {Piro}, {Shappee}, {Siebert}, {Simon}, {Ulloa},
  {Kasen}, {Madore}, {Murguia-Berthier}, {Pan}, {Prochaska}, {Ramirez-Ruiz},
  {Rest}, \& {Rojas-Bravo}}]{coulter17}
{Coulter}, D.~A., {Foley}, R.~J., {Kilpatrick}, C.~D., {et~al.}
  2017{\natexlab{b}}, Science, 358, 1556

\bibitem[{{Covino} {et~al.}(2017){Covino}, {Wiersema}, {Fan}, {Toma},
  {Higgins}, {Melandri}, {D'Avanzo}, {Mundell}, {Palazzi}, {Tanvir},
  {Bernardini}, {Branchesi}, {Brocato}, {Campana}, {Alighieri}, {G{\"o}tz},
  {Fynbo}, {Gao}, {Gomboc}, {Gompertz}, {Greiner}, {Hjorth}, {Jin}, {Kaper},
  {Klose}, {Kobayashi}, {Kopac}, {Kouveliotou}, {Levan}, {Mao}, {Malesani},
  {Pian}, {Rossi}, {Salvaterra}, {Starling}, {Steele}, {Tagliaferri}, {Troja},
  {Horst}, \& {Wijers}}]{covino17}
{Covino}, S., {Wiersema}, K., {Fan}, Y.~Z., {et~al.} 2017, Nature Astronomy, 1,
  791

\bibitem[{{Cowperthwaite} {et~al.}(2017){Cowperthwaite}, {Berger}, {Villar},
  {Metzger}, {Nicholl}, {Chornock}, {Blanchard}, {Fong}, {Margutti},
  {Soares-Santos}, {Alexander}, {Allam}, {Annis}, {Brout}, {Brown}, {Butler},
  {Chen}, {Diehl}, {Doctor}, {Drout}, {Eftekhari}, {Farr}, {Finley}, {Foley},
  {Frieman}, {Fryer}, {Garc{\'{\i}}a-Bellido}, {Gill}, {Guillochon}, {Herner},
  {Holz}, {Kasen}, {Kessler}, {Marriner}, {Matheson}, {Neilsen}, {Quataert},
  {Palmese}, {Rest}, {Sako}, {Scolnic}, {Smith}, {Tucker}, {Williams},
  {Balbinot}, {Carlin}, {Cook}, {Durret}, {Li}, {Lopes}, {Louren{\c c}o},
  {Marshall}, {Medina}, {Muir}, {Mu{\~n}oz}, {Sauseda}, {Schlegel}, {Secco},
  {Vivas}, {Wester}, {Zenteno}, {Zhang}, {Abbott}, {Banerji}, {Bechtol},
  {Benoit-L{\'e}vy}, {Bertin}, {Buckley-Geer}, {Burke}, {Capozzi}, {Carnero
  Rosell}, {Carrasco Kind}, {Castander}, {Crocce}, {Cunha}, {D'Andrea}, {da
  Costa}, {Davis}, {DePoy}, {Desai}, {Dietrich}, {Drlica-Wagner}, {Eifler},
  {Evrard}, {Fernandez}, {Flaugher}, {Fosalba}, {Gaztanaga}, {Gerdes},
  {Giannantonio}, {Goldstein}, {Gruen}, {Gruendl}, {Gutierrez}, {Honscheid},
  {Jain}, {James}, {Jeltema}, {Johnson}, {Johnson}, {Kent}, {Krause}, {Kron},
  {Kuehn}, {Nuropatkin}, {Lahav}, {Lima}, {Lin}, {Maia}, {March}, {Martini},
  {McMahon}, {Menanteau}, {Miller}, {Miquel}, {Mohr}, {Neilsen}, {Nichol},
  {Ogando}, {Plazas}, {Roe}, {Romer}, {Roodman}, {Rykoff}, {Sanchez},
  {Scarpine}, {Schindler}, {Schubnell}, {Sevilla-Noarbe}, {Smith}, {Smith},
  {Sobreira}, {Suchyta}, {Swanson}, {Tarle}, {Thomas}, {Thomas}, {Troxel},
  {Vikram}, {Walker}, {Wechsler}, {Weller}, {Yanny}, \&
  {Zuntz}}]{cowperthwaite17}
{Cowperthwaite}, P.~S., {Berger}, E., {Villar}, V.~A., {et~al.} 2017, \apjl,
  848, L17

\bibitem[{{D{\'a}lya} {et~al.}(2018){D{\'a}lya}, {Galg{\'o}czi}, {Dobos},
  {Frei}, {Heng}, {Macas}, {Messenger}, {Raffai}, \& {de Souza}}]{dalya18}
{D{\'a}lya}, G., {Galg{\'o}czi}, G., {Dobos}, L., {et~al.} 2018, \mnras, 479,
  2374

\bibitem[{{D'Avanzo} {et~al.}(2018){D'Avanzo}, {Campana}, {Salafia}, {Ghirland
  a}, {Ghisellini}, {Melandri}, {Bernardini}, {Branchesi}, {Chassande-Mottin},
  {Covino}, {D'Elia}, {Nava}, {Salvaterra}, {Tagliaferri}, \&
  {Vergani}}]{Davanzoetal2018}
{D'Avanzo}, P., {Campana}, S., {Salafia}, O.~S., {et~al.} 2018, \aap, 613, L1

\bibitem[{{D'Avanzo} {et~al.}(2019{\natexlab{a}}){D'Avanzo}, {Melandri},
  {Izzo}, {Branchesi}, {D'Elia}, {Greco}, {Piranomonte}, {Rossi}, {Testa}, \&
  {Boschin}}]{2019GCN.25331....1D}
{D'Avanzo}, P., {Melandri}, A., {Izzo}, L., {et~al.} 2019{\natexlab{a}}, GRB
  Coordinates Network, 25331, 1

\bibitem[{{D'Avanzo} {et~al.}(2019{\natexlab{b}}){D'Avanzo}, {Rossi}, {Greco},
  {Piranomonte}, {Difabrizio}, {Bragaglia}, \&
  {Carosati}}]{2019GCN.25361....1D}
{D'Avanzo}, P., {Rossi}, A., {Greco}, G., {et~al.} 2019{\natexlab{b}}, GRB
  Coordinates Network, 25361, 1

\bibitem[{{Davies} {et~al.}(2005){Davies}, {Levan}, \& {King}}]{davies05}
{Davies}, M.~B., {Levan}, A.~J., \& {King}, A.~R. 2005, \mnras, 356, 54

\bibitem[{{De} {et~al.}(2019{\natexlab{a}}){De}, {Goldstein}, {Andreoni},
  {Cenko}, {Bloom}, {Perley}, \& {Kasliwal}}]{2019GCN.25348....1D}
{De}, K., {Goldstein}, D., {Andreoni}, I., {et~al.} 2019{\natexlab{a}}, GRB
  Coordinates Network, 25348, 1

\bibitem[{{De} {et~al.}(2019{\natexlab{b}}){De}, {Jencson}, {Kasliwal},
  {Goldstein}, \& {Andreoni}}]{2019GCN.25478....1D}
{De}, K., {Jencson}, J., {Kasliwal}, M.~M., {Goldstein}, D.~A., \& {Andreoni},
  I. 2019{\natexlab{b}}, GRB Coordinates Network, 25478, 1

\bibitem[{{Dichiari} {et~al.}(2019){Dichiari}, {Troja}, {Cenko}, {Gatkine},
  {Kutyrev}, {Durbak}, {Veilleux}, {Minervini}, \&
  {Ricci}}]{2019GCN.25374....1D}
{Dichiari}, S., {Troja}, E., {Cenko}, S.~B., {et~al.} 2019, GRB Coordinates
  Network, 25374, 1

\bibitem[{{Dimitriadis} {et~al.}(2019){Dimitriadis}, {Brown}, {Seibert},
  {Foley}, {Coulter}, {Kilpatrick}, {Siellez}, \&
  {Rojas-Bravo}}]{2019GCN.25395....1D}
{Dimitriadis}, G., {Brown}, J.~S., {Seibert}, M.~R., {et~al.} 2019, GRB
  Coordinates Network, 25395, 1

\bibitem[{{Dobie} {et~al.}(2018){Dobie}, {Kaplan}, {Murphy}, {Lenc}, {Mooley},
  {Lynch}, {Corsi}, {Frail}, {Kasliwal}, \& {Hallinan}}]{Dobieetal2018}
{Dobie}, D., {Kaplan}, D.~L., {Murphy}, T., {et~al.} 2018, \apjl, 858, L15

\bibitem[{{Dobie} {et~al.}(2019){Dobie}, {Stewart}, {Murphy}, {Lenc}, {Wang},
  {Kaplan}, {Andreoni}, {Banfield}, {Brown}, {Corsi}, {De}, {Goldstein},
  {Hallinan}, {Hotan}, {Hotokezaka}, {Jaodand}, {Karambelkar}, {Kasliwal},
  {McConnell}, {Mooley}, {Moss}, {Newman}, {Perley}, {Prakash}, {Pritchard},
  {Sadler}, {Sharma}, {Ward}, {Whiting}, \& {Zhou}}]{Dobieetal2019}
{Dobie}, D., {Stewart}, A., {Murphy}, T., {et~al.} 2019, \apjl, 887, L13

\bibitem[{{Dominik} {et~al.}(2015){Dominik}, {Berti}, {O'Shaughnessy},
  {Mandel}, {Belczynski}, {Fryer}, {Holz}, {Bulik}, \&
  {Pannarale}}]{Dominik:2014}
{Dominik}, M., {Berti}, E., {O'Shaughnessy}, R., {et~al.} 2015, \apj, 806, 263

\bibitem[{{Drout} {et~al.}(2017){Drout}, {Piro}, {Shappee}, {Kilpatrick},
  {Simon}, {Contreras}, {Coulter}, {Foley}, {Siebert}, {Morrell}, {Boutsia},
  {Di Mille}, {Holoien}, {Kasen}, {Kollmeier}, {Madore}, {Monson},
  {Murguia-Berthier}, {Pan}, {Prochaska}, {Ramirez-Ruiz}, {Rest}, {Adams},
  {Alatalo}, {Ba{\~n}ados}, {Baughman}, {Beers}, {Bernstein}, {Bitsakis},
  {Campillay}, {Hansen}, {Higgs}, {Ji}, {Maravelias}, {Marshall}, {Bidin},
  {Prieto}, {Rasmussen}, {Rojas-Bravo}, {Strom}, {Ulloa},
  {Vargas-Gonz{\'a}lez}, {Wan}, \& {Whitten}}]{drout17}
{Drout}, M.~R., {Piro}, A.~L., {Shappee}, B.~J., {et~al.} 2017, Science, 358,
  1570

\bibitem[{{Dyer} {et~al.}(2018){Dyer}, {Dhillon}, {Littlefair}, {Steeghs},
  {Ulaczyk}, {Chote}, {Galloway}, \& {Rol}}]{dyer18}
{Dyer}, M.~J., {Dhillon}, V.~S., {Littlefair}, S., {et~al.} 2018, in Society of
  Photo-Optical Instrumentation Engineers (SPIE) Conference Series, Vol. 10704,
  Observatory Operations: Strategies, Processes, and Systems VII, 107040C

\bibitem[{{Eichler} {et~al.}(1989){Eichler}, {Livio}, {Piran}, \&
  {Schramm}}]{eichler89}
{Eichler}, D., {Livio}, M., {Piran}, T., \& {Schramm}, D.~N. 1989, \nat, 340,
  126

\bibitem[{{Eldridge} \& {Stanway}(2016)}]{EldridgeStanway:2016}
{Eldridge}, J.~J. \& {Stanway}, E.~R. 2016, \mnras, 462, 3302

\bibitem[{{Evans} {et~al.}(2017){Evans}, {Cenko}, {Kennea}, {Emery}, {Kuin},
  {Korobkin}, {Wollaeger}, {Fryer}, {Madsen}, {Harrison}, {Xu}, {Nakar},
  {Hotokezaka}, {Lien}, {Campana}, {Oates}, {Troja}, {Breeveld}, {Marshall},
  {Barthelmy}, {Beardmore}, {Burrows}, {Cusumano}, {D'A{\`\i}}, {D'Avanzo},
  {D'Elia}, {de Pasquale}, {Even}, {Fontes}, {Forster}, {Garcia}, {Giommi},
  {Grefenstette}, {Gronwall}, {Hartmann}, {Heida}, {Hungerford}, {Kasliwal},
  {Krimm}, {Levan}, {Malesani}, {Melandri}, {Miyasaka}, {Nousek}, {O'Brien},
  {Osborne}, {Pagani}, {Page}, {Palmer}, {Perri}, {Pike}, {Racusin}, {Rosswog},
  {Siegel}, {Sakamoto}, {Sbarufatti}, {Tagliaferri}, {Tanvir}, \&
  {Tohuvavohu}}]{evans17}
{Evans}, P.~A., {Cenko}, S.~B., {Kennea}, J.~A., {et~al.} 2017, Science, 358,
  1565

\bibitem[{{Evans} {et~al.}(2016{\natexlab{a}}){Evans}, {Kennea}, {Barthelmy},
  {Beardmore}, {Burrows}, {Campana}, {Cenko}, {Gehrels}, {Giommi}, {Gronwall},
  {Marshall}, {Malesani}, {Markwardt}, {Mingo}, {Nousek}, {O'Brien}, {Osborne},
  {Pagani}, {Page}, {Palmer}, {Perri}, {Racusin}, {Siegel}, {Sbarufatti}, \&
  {Tagliaferri}}]{evans16}
{Evans}, P.~A., {Kennea}, J.~A., {Barthelmy}, S.~D., {et~al.}
  2016{\natexlab{a}}, \mnras, 460, L40

\bibitem[{{Evans} {et~al.}(2019){Evans}, {Kennea}, {Tohuvavohu}, {Barthelmy},
  {Beardmore}, {Bernardini}, {Breeveld}, {Brown}, {Burrows}, \&
  {Campana}}]{2019GCN.25400....1E}
{Evans}, P.~A., {Kennea}, J.~A., {Tohuvavohu}, A., {et~al.} 2019, GRB
  Coordinates Network, 25400, 1

\bibitem[{{Evans} {et~al.}(2016{\natexlab{b}}){Evans}, {Osborne}, {Kennea},
  {Campana}, {O'Brien}, {Tanvir}, {Racusin}, {Burrows}, {Cenko}, \&
  {Gehrels}}]{evans16a}
{Evans}, P.~A., {Osborne}, J.~P., {Kennea}, J.~A., {et~al.} 2016{\natexlab{b}},
  \mnras, 455, 1522

\bibitem[{{Fern{\'a}ndez} {et~al.}(2017){Fern{\'a}ndez}, {Foucart}, {Kasen},
  {Lippuner}, {Desai}, \& {Roberts}}]{Fernandezetal2017}
{Fern{\'a}ndez}, R., {Foucart}, F., {Kasen}, D., {et~al.} 2017, Classical and
  Quantum Gravity, 34, 154001

\bibitem[{{Fern{\'a}ndez} \& {Metzger}(2013)}]{Fernandez13}
{Fern{\'a}ndez}, R. \& {Metzger}, B.~D. 2013, \mnras, 435, 502

\bibitem[{{Fern{\'a}ndez} {et~al.}(2019){Fern{\'a}ndez}, {Tchekhovskoy},
  {Quataert}, {Foucart}, \& {Kasen}}]{fernandez19}
{Fern{\'a}ndez}, R., {Tchekhovskoy}, A., {Quataert}, E., {Foucart}, F., \&
  {Kasen}, D. 2019, \mnras, 482, 3373

\bibitem[{{Flesch}(2015)}]{MILLIQUAS}
{Flesch}, E.~W. 2015, \pasa, 32, e010

\bibitem[{{Flewelling} {et~al.}(2016){Flewelling}, {Magnier}, {Chambers},
  {Heasley}, {Holmberg}, {Huber}, {Sweeney}, {Waters}, {Calamida}, {Casertano},
  {Chen}, {Farrow}, {Hasinger}, {Henderson}, {Long}, {Metcalfe}, {Narayan},
  {Nieto-Santisteban}, {Norberg}, {Rest}, {Saglia}, {Szalay}, {Thakar},
  {Tonry}, {Valenti}, {Werner}, {White}, {Denneau}, {Draper}, {Hodapp},
  {Jedicke}, {Kaiser}, {Kudritzki}, {Price}, {Wainscoat}, {Builders},
  {Chastel}, {McLean}, {Postman}, \& {Shiao}}]{Flewelling16}
{Flewelling}, H.~A., {Magnier}, E.~A., {Chambers}, K.~C., {et~al.} 2016, arXiv
  e-prints, arXiv:1612.05243

\bibitem[{{Fong} \& {Berger}(2013)}]{fong13b}
{Fong}, W. \& {Berger}, E. 2013, \apj, 776, 18

\bibitem[{{Fong} {et~al.}(2015){Fong}, {Berger}, {Margutti}, \&
  {Zauderer}}]{fong2015}
{Fong}, W., {Berger}, E., {Margutti}, R., \& {Zauderer}, B.~A. 2015, \apj, 815,
  102

\bibitem[{{Foucart}(2012)}]{Foucart12}
{Foucart}, F. 2012, \prd, 86, 124007

\bibitem[{{Foucart} {et~al.}(2013){Foucart}, {Deaton}, {Duez}, {Kidder},
  {MacDonald}, {Ott}, {Pfeiffer}, {Scheel}, {Szilagyi}, \&
  {Teukolsky}}]{Foucart13}
{Foucart}, F., {Deaton}, M.~B., {Duez}, M.~D., {et~al.} 2013, \prd, 87, 084006

\bibitem[{{Foucart} {et~al.}(2014){Foucart}, {Deaton}, {Duez}, {O'Connor},
  {Ott}, {Haas}, {Kidder}, {Pfeiffer}, {Scheel}, \& {Szilagyi}}]{Foucart14}
{Foucart}, F., {Deaton}, M.~B., {Duez}, M.~D., {et~al.} 2014, \prd, 90, 024026

\bibitem[{{Foucart} {et~al.}(2019){Foucart}, {Duez}, {Kidder}, {Nissanke},
  {Pfeiffer}, \& {Scheel}}]{Foucart19}
{Foucart}, F., {Duez}, M.~D., {Kidder}, L.~E., {et~al.} 2019, arXiv e-prints
  [\eprint[arXiv]{1903.09166}]

\bibitem[{{Foucart} {et~al.}(2018){Foucart}, {Hinderer}, \&
  {Nissanke}}]{Foucart18c}
{Foucart}, F., {Hinderer}, T., \& {Nissanke}, S. 2018, \prd, 98, 081501

\bibitem[{{Freiburghaus} {et~al.}(1999){Freiburghaus}, {Rosswog}, \&
  {Thielemann}}]{freiburghaus99}
{Freiburghaus}, C., {Rosswog}, S., \& {Thielemann}, F.-K. 1999, \apjl, 525,
  L121

\bibitem[{{Freudling} {et~al.}(2013){Freudling}, {Romaniello}, {Bramich},
  {Ballester}, {Forchi}, {Garc{\'\i}a-Dabl{\'o}}, {Moehler}, \&
  {Neeser}}]{Freudling13}
{Freudling}, W., {Romaniello}, M., {Bramich}, D.~M., {et~al.} 2013, \aap, 559,
  A96

\bibitem[{{Gaia Collaboration} {et~al.}(2016){Gaia Collaboration}, {Brown},
  {Vallenari}, {Prusti}, {de Bruijne}, {Mignard}, {Drimmel}, {Babusiaux},
  {Bailer-Jones}, {Bastian}, {Biermann}, {Evans}, {Eyer}, {Jansen}, {Jordi},
  {Katz}, {Klioner}, {Lammers}, {Lindegren}, {Luri}, {O'Mullane}, {Panem},
  {Pourbaix}, {Randich}, {Sartoretti}, {Siddiqui}, {Soubiran}, {Valette}, {van
  Leeuwen}, {Walton}, {Aerts}, {Arenou}, {Cropper}, {H{\o}g}, {Lattanzi},
  {Grebel}, {Holland}, {Huc}, {Passot}, {Perryman}, {Bramante}, {Cacciari},
  {Casta{\~n}eda}, {Chaoul}, {Cheek}, {De Angeli}, {Fabricius}, {Guerra},
  {Hern{\'a}ndez}, {Jean-Antoine-Piccolo}, {Masana}, {Messineo}, {Mowlavi},
  {Nienartowicz}, {Ord{\'o}{\~n}ez-Blanco}, {Panuzzo}, {Portell}, {Richards},
  {Riello}, {Seabroke}, {Tanga}, {Th{\'e}venin}, {Torra}, {Els},
  {Gracia-Abril}, {Comoretto}, {Garcia-Reinaldos}, {Lock}, {Mercier},
  {Altmann}, {Andrae}, {Astraatmadja}, {Bellas-Velidis}, {Benson}, {Berthier},
  {Blomme}, {Busso}, {Carry}, {Cellino}, {Clementini}, {Cowell}, {Creevey},
  {Cuypers}, {Davidson}, {De Ridder}, {de Torres}, {Delchambre}, {Dell'Oro},
  {Ducourant}, {Fr{\'e}mat}, {Garc{\'\i}a-Torres}, {Gosset}, {Halbwachs},
  {Hambly}, {Harrison}, {Hauser}, {Hestroffer}, {Hodgkin}, {Huckle}, {Hutton},
  {Jasniewicz}, {Jordan}, {Kontizas}, {Korn}, {Lanzafame}, {Manteiga},
  {Moitinho}, {Muinonen}, {Osinde}, {Pancino}, {Pauwels}, {Petit},
  {Recio-Blanco}, {Robin}, {Sarro}, {Siopis}, {Smith}, {Smith}, {Sozzetti},
  {Thuillot}, {van Reeven}, {Viala}, {Abbas}, {Abreu Aramburu}, {Accart},
  {Aguado}, {Allan}, {Allasia}, {Altavilla}, {{\'A}lvarez}, {Alves},
  {Anderson}, {Andrei}, {Anglada Varela}, {Antiche}, {Antoja}, {Ant{\'o}n},
  {Arcay}, {Bach}, {Baker}, {Balaguer-N{\'u}{\~n}ez}, {Barache}, {Barata},
  {Barbier}, {Barblan}, {Barrado y Navascu{\'e}s}, {Barros}, {Barstow},
  {Becciani}, {Bellazzini}, {Bello Garc{\'\i}a}, {Belokurov}, {Bendjoya},
  {Berihuete}, {Bianchi}, {Bienaym{\'e}}, {Billebaud}, {Blagorodnova},
  {Blanco-Cuaresma}, {Boch}, {Bombrun}, {Borrachero}, {Bouquillon}, {Bourda},
  {Bouy}, {Bragaglia}, {Breddels}, {Brouillet}, {Br{\"u}semeister},
  {Bucciarelli}, {Burgess}, {Burgon}, {Burlacu}, {Busonero}, {Buzzi}, {Caffau},
  {Cambras}, {Campbell}, {Cancelliere}, {Cantat-Gaudin}, {Carlucci},
  {Carrasco}, {Castellani}, {Charlot}, {Charnas}, {Chiavassa}, {Clotet},
  {Cocozza}, {Collins}, {Costigan}, {Crifo}, {Cross}, {Crosta}, {Crowley},
  {Dafonte}, {Damerdji}, {Dapergolas}, {David}, {David}, {De Cat}, {de Felice},
  {de Laverny}, {De Luise}, {De March}, {de Martino}, {de Souza}, {Debosscher},
  {del Pozo}, {Delbo}, {Delgado}, {Delgado}, {Di Matteo}, {Diakite},
  {Distefano}, {Dolding}, {Dos Anjos}, {Drazinos}, {Duran}, {Dzigan},
  {Edvardsson}, {Enke}, {Evans}, {Eynard Bontemps}, {Fabre}, {Fabrizio},
  {Faigler}, {Falc{\~a}o}, {Farr{\`a}s Casas}, {Federici}, {Fedorets},
  {Fern{\'a}ndez-Hern{\'a}ndez}, {Fernique}, {Fienga}, {Figueras}, {Filippi},
  {Findeisen}, {Fonti}, {Fouesneau}, {Fraile}, {Fraser}, {Fuchs}, {Gai},
  {Galleti}, {Galluccio}, {Garabato}, {Garc{\'\i}a-Sedano}, {Garofalo},
  {Garralda}, {Gavras}, {Gerssen}, {Geyer}, {Gilmore}, {Girona}, {Giuffrida},
  {Gomes}, {Gonz{\'a}lez-Marcos}, {Gonz{\'a}lez-N{\'u}{\~n}ez},
  {Gonz{\'a}lez-Vidal}, {Granvik}, {Guerrier}, {Guillout}, {Guiraud},
  {G{\'u}rpide}, {Guti{\'e}rrez-S{\'a}nchez}, {Guy}, {Haigron},
  {Hatzidimitriou}, {Haywood}, {Heiter}, {Helmi}, {Hobbs}, {Hofmann}, {Holl},
  {Holland }, {Hunt}, {Hypki}, {Icardi}, {Irwin}, {Jevardat de Fombelle},
  {Jofr{\'e}}, {Jonker}, {Jorissen}, {Julbe}, {Karampelas}, {Kochoska},
  {Kohley}, {Kolenberg}, {Kontizas}, {Koposov}, {Kordopatis}, {Koubsky},
  {Krone-Martins}, {Kudryashova}, {Kull}, {Bachchan}, {Lacoste-Seris}, {Lanza},
  {Lavigne}, {Le Poncin-Lafitte}, {Lebreton}, {Lebzelter}, {Leccia}, {Leclerc},
  {Lecoeur-Taibi}, {Lemaitre}, {Lenhardt}, {Leroux}, {Liao}, {Licata},
  {Lindstr{\o}m}, {Lister}, {Livanou}, {Lobel}, {L{\"o}ffler}, {L{\'o}pez},
  {Lorenz}, {MacDonald}, {Magalh{\~a}es Fernandes}, {Managau}, {Mann},
  {Mantelet}, {Marchal}, {Marchant}, {Marconi}, {Marinoni}, {Marrese},
  {Marschalk{\'o}}, {Marshall}, {Mart{\'\i}n-Fleitas}, {Martino}, {Mary},
  {Matijevi{\v{c}}}, {Mazeh}, {McMillan}, {Messina}, {Michalik}, {Millar},
  {Mirand a}, {Molina}, {Molinaro}, {Molinaro}, {Moln{\'a}r}, {Moniez},
  {Montegriffo}, {Mor}, {Mora}, {Morbidelli}, {Morel}, {Morgenthaler},
  {Morris}, {Mulone}, {Muraveva}, {Musella}, {Narbonne}, {Nelemans},
  {Nicastro}, {Noval}, {Ord{\'e}novic}, {Ordieres-Mer{\'e}}, {Osborne},
  {Pagani}, {Pagano}, {Pailler}, {Palacin}, {Palaversa}, {Parsons}, {Pecoraro},
  {Pedrosa}, {Pentik{\"a}inen}, {Pichon}, {Piersimoni}, {Pineau}, {Plachy},
  {Plum}, {Poujoulet}, {Pr{\v{s}}a}, {Pulone}, {Ragaini}, {Rago}, {Rambaux},
  {Ramos-Lerate}, {Ranalli}, {Rauw}, {Read}, {Regibo}, {Reyl{\'e}}, {Ribeiro},
  {Rimoldini}, {Ripepi}, {Riva}, {Rixon}, {Roelens}, {Romero-G{\'o}mez},
  {Rowell}, {Royer}, {Ruiz-Dern}, {Sadowski}, {Sagrist{\`a} Sell{\'e}s},
  {Sahlmann}, {Salgado}, {Salguero}, {Sarasso}, {Savietto}, {Schultheis},
  {Sciacca}, {Segol}, {Segovia}, {Segransan}, {Shih}, {Smareglia}, {Smart},
  {Solano}, {Solitro}, {Sordo}, {Soria Nieto}, {Souchay}, {Spagna}, {Spoto},
  {Stampa}, {Steele}, {Steidelm{\"u}ller}, {Stephenson}, {Stoev}, {Suess},
  {S{\"u}veges}, {Surdej}, {Szabados}, {Szegedi-Elek}, {Tapiador}, {Taris},
  {Tauran}, {Taylor}, {Teixeira}, {Terrett}, {Tingley}, {Trager}, {Turon},
  {Ulla}, {Utrilla}, {Valentini}, {van Elteren}, {Van Hemelryck}, {van
  Leeuwen}, {Varadi}, {Vecchiato}, {Veljanoski}, {Via}, {Vicente}, {Vogt},
  {Voss}, {Votruba}, {Voutsinas}, {Walmsley}, {Weiler}, {Weingrill}, {Wevers},
  {Wyrzykowski}, {Yoldas}, {{\v{Z}}erjal}, {Zucker}, {Zurbach}, {Zwitter},
  {Alecu}, {Allen}, {Allende Prieto}, {Amorim}, {Anglada-Escud{\'e}},
  {Arsenijevic}, {Azaz}, {Balm}, {Beck}, {Bernstein}, {Bigot}, {Bijaoui},
  {Blasco}, {Bonfigli}, {Bono}, {Boudreault}, {Bressan}, {Brown}, {Brunet},
  {Bunclark}, {Buonanno}, {Butkevich}, {Carret}, {Carrion}, {Chemin},
  {Ch{\'e}reau}, {Corcione}, {Darmigny}, {de Boer}, {de Teodoro}, {de Zeeuw},
  {Delle Luche}, {Domingues}, {Dubath}, {Fodor}, {Fr{\'e}zouls}, {Fries},
  {Fustes}, {Fyfe}, {Gallardo}, {Gallegos}, {Gardiol}, {Gebran}, {Gomboc},
  {G{\'o}mez}, {Grux}, {Gueguen}, {Heyrovsky}, {Hoar}, {Iannicola}, {Isasi
  Parache}, {Janotto}, {Joliet}, {Jonckheere}, {Keil}, {Kim}, {Klagyivik},
  {Klar}, {Knude}, {Kochukhov}, {Kolka}, {Kos}, {Kutka}, {Lainey}, {LeBouquin},
  {Liu}, {Loreggia}, {Makarov}, {Marseille}, {Martayan}, {Martinez-Rubi},
  {Massart}, {Meynadier}, {Mignot}, {Munari}, {Nguyen}, {Nordlander}, {Ocvirk},
  {O'Flaherty}, {Olias Sanz}, {Ortiz}, {Osorio}, {Oszkiewicz}, {Ouzounis},
  {Palmer}, {Park}, {Pasquato}, {Peltzer}, {Peralta}, {P{\'e}turaud},
  {Pieniluoma}, {Pigozzi}, {Poels}, {Prat}, {Prod'homme}, {Raison}, {Rebordao},
  {Risquez}, {Rocca-Volmerange}, {Rosen}, {Ruiz-Fuertes}, {Russo}, {Sembay},
  {Serraller Vizcaino}, {Short}, {Siebert}, {Silva}, {Sinachopoulos}, {Slezak},
  {Soffel}, {Sosnowska}, {Strai{\v{z}}ys}, {ter Linden}, {Terrell}, {Theil},
  {Tiede}, {Troisi}, {Tsalmantza}, {Tur}, {Vaccari}, {Vachier}, {Valles}, {Van
  Hamme}, {Veltz}, {Virtanen}, {Wallut}, {Wichmann}, {Wilkinson}, {Ziaeepour},
  \& {Zschocke}}]{2016A&A...595A...2G}
{Gaia Collaboration}, {Brown}, A.~G.~A., {Vallenari}, A., {et~al.} 2016, \aap,
  595, A2

\bibitem[{{Gall} {et~al.}(2017){Gall}, {Hjorth}, {Rosswog}, {Tanvir}, \&
  {Levan}}]{Galletal2017}
{Gall}, C., {Hjorth}, J., {Rosswog}, S., {Tanvir}, N.~R., \& {Levan}, A.~J.
  2017, \apjl, 849, L19

\bibitem[{{Gehrels} {et~al.}(2016){Gehrels}, {Cannizzo}, {Kanner}, {Kasliwal},
  {Nissanke}, \& {Singer}}]{gehrels16}
{Gehrels}, N., {Cannizzo}, J.~K., {Kanner}, J., {et~al.} 2016, \apj, 820, 136

\bibitem[{{Ghirlanda} {et~al.}(2019){Ghirlanda}, {Salafia}, {Paragi},
  {Giroletti}, {Yang}, {Marcote}, {Blanchard}, {Agudo}, {An}, {Bernardini},
  {Beswick}, {Branchesi}, {Campana}, {Casadio}, {Chassande-Mottin}, {Colpi},
  {Covino}, {D'Avanzo}, {D'Elia}, {Frey}, {Gawronski}, {Ghisellini}, {Gurvits},
  {Jonker}, {van Langevelde}, {Melandri}, {Moldon}, {Nava}, {Perego},
  {Perez-Torres}, {Reynolds}, {Salvaterra}, {Tagliaferri}, {Venturi},
  {Vergani}, \& {Zhang}}]{ghirlanda19}
{Ghirlanda}, G., {Salafia}, O.~S., {Paragi}, Z., {et~al.} 2019, Science, 363,
  968

\bibitem[{{Giacobbo} \& {Mapelli}(2018)}]{GiacobboMapelli:2018}
{Giacobbo}, N. \& {Mapelli}, M. 2018, \mnras, 480, 2011

\bibitem[{{Goldstein} {et~al.}(2016){Goldstein}, {Burns}, {Hamburg},
  {Connaughton}, {Veres}, {Briggs}, {Hui}, \& {The GBM-LIGO
  Collaboration}}]{goldstein2016}
{Goldstein}, A., {Burns}, E., {Hamburg}, R., {et~al.} 2016, arXiv e-prints,
  arXiv:1612.02395

\bibitem[{{Goldstein} {et~al.}(2017){Goldstein}, {Veres}, {Burns}, {Briggs},
  {Hamburg}, {Kocevski}, {Wilson-Hodge}, {Preece}, {Poolakkil}, {Roberts},
  {Hui}, {Connaughton}, {Racusin}, {von Kienlin}, {Dal Canton}, {Christensen},
  {Littenberg}, {Siellez}, {Blackburn}, {Broida}, {Bissaldi}, {Cleveland},
  {Gibby}, {Giles}, {Kippen}, {McBreen}, {McEnery}, {Meegan}, {Paciesas}, \&
  {Stanbro}}]{Goldsteinetal2017}
{Goldstein}, A., {Veres}, P., {Burns}, E., {et~al.} 2017, \apjl, 848, L14

\bibitem[{{Goldstein} {et~al.}(2019{\natexlab{a}}){Goldstein}, {Perley},
  {Andreoni}, \& {Kasliwal}}]{2019GCN.25355....1G}
{Goldstein}, D., {Perley}, D.~A., {Andreoni}, I., \& {Kasliwal}, M.~M.
  2019{\natexlab{a}}, GRB Coordinates Network, 25355, 1

\bibitem[{{Goldstein} \& {Anand}(2019)}]{2019GCN.25394....1G}
{Goldstein}, D.~A. \& {Anand}, S. 2019, GRB Coordinates Network, 25394, 1

\bibitem[{{Goldstein} {et~al.}(2019{\natexlab{b}}){Goldstein}, {Andreoni},
  {Hankins}, {Sollerman}, {Anand}, {Ahumada}, {Newman}, {Dobie}, {Gatkine}, \&
  {Kong}}]{2019GCN.25393....1G}
{Goldstein}, D.~A., {Andreoni}, I., {Hankins}, M., {et~al.} 2019{\natexlab{b}},
  GRB Coordinates Network, 25393, 1

\bibitem[{{Goldstein} {et~al.}(2019{\natexlab{c}}){Goldstein}, {Andreoni},
  {Zhou}, {Newman}, {Ahumada}, {Anand}, {Bulla}, {Dahiwale}, {De}, \&
  {Dhawan}}]{2019GCN.25391....1G}
{Goldstein}, D.~A., {Andreoni}, I., {Zhou}, R., {et~al.} 2019{\natexlab{c}},
  GRB Coordinates Network, 25391, 1

\bibitem[{{Gomez} {et~al.}(2019{\natexlab{a}}){Gomez}, {Hosseinzadeh},
  {Berger}, {Blanchard}, {Eftekhari}, {Gill}, {Patton}, {Villar}, {Williams},
  \& {Cowperthwaite}}]{2019GCN.25483....1G}
{Gomez}, S., {Hosseinzadeh}, G., {Berger}, E., {et~al.} 2019{\natexlab{a}}, GRB
  Coordinates Network, 25483, 1

\bibitem[{{Gomez} {et~al.}(2019{\natexlab{b}}){Gomez}, {Hosseinzadeh},
  {Cowperthwaite}, {Villar}, {Berger}, {Gardner}, {Alexand er}, {Blanchard},
  {Chornock}, {Drout}, {Eftekhari}, {Fong}, {Gill}, {Margutti}, {Nicholl},
  {Paterson}, \& {Williams}}]{Gomezetal2019}
{Gomez}, S., {Hosseinzadeh}, G., {Cowperthwaite}, P.~S., {et~al.}
  2019{\natexlab{b}}, \apjl, 884, L55

\bibitem[{{Gompertz} {et~al.}(2020{\natexlab{a}}){Gompertz}, {Cutter},
  {Steeghs}, {Galloway}, {Lyman}, {Ulaczyk}, {Dyer}, {Ackley}, {Dhillon},
  {O'Brien}, {Ramsay}, {Poshyachinda}, {Kotak}, {Nuttall}, {Breton},
  {Pall{\'e}}, {Pollacco}, {Thrane}, {Aukkaravittayapun}, {Awiphan}, {Brown},
  {Burhanudin}, {Chote}, {Chrimes}, {Daw}, {Duffy}, {Eyles-Ferris},
  {Heikkil{\"a}}, {Irawati}, {Kennedy}, {Killestein}, {Levan}, {Littlefair},
  {Makrygianni}, {Marsh}, {Mata S{\'a}nchez}, {Mattila}, {Maund}, {McCormac},
  {Mkrtichian}, {Mong}, {Mullaney}, {M{\"u}ller}, {Obradovic}, {Rol},
  {Sawangwit}, {Stanway}, {Starling}, {Str{\o}m}, {Tooke}, {West}, \&
  {Wiersema}}]{Gompertz20b}
{Gompertz}, B.~P., {Cutter}, R., {Steeghs}, D., {et~al.} 2020{\natexlab{a}},
  arXiv e-prints, arXiv:2004.00025

\bibitem[{{Gompertz} {et~al.}(2020{\natexlab{b}}){Gompertz}, {Levan}, \&
  {Tanvir}}]{Gompertz20}
{Gompertz}, B.~P., {Levan}, A.~J., \& {Tanvir}, N.~R. 2020{\natexlab{b}}, arXiv
  e-prints, arXiv:2001.08706

\bibitem[{{Gompertz} {et~al.}(2018){Gompertz}, {Levan}, {Tanvir}, {Hjorth},
  {Covino}, {Evans}, {Fruchter}, {Gonz{\'a}lez-Fern{\'a}ndez}, {Jin}, {Lyman},
  {Oates}, {O'Brien}, \& {Wiersema}}]{gompertz18}
{Gompertz}, B.~P., {Levan}, A.~J., {Tanvir}, N.~R., {et~al.} 2018, \apj, 860,
  62

\bibitem[{{Gompertz} {et~al.}(2015){Gompertz}, {van der Horst}, {O'Brien},
  {Wynn}, \& {Wiersema}}]{Gompertz15}
{Gompertz}, B.~P., {van der Horst}, A.~J., {O'Brien}, P.~T., {Wynn}, G.~A., \&
  {Wiersema}, K. 2015, \mnras, 448, 629

\bibitem[{{Gonz{\'a}lez-Fern{\'a}ndez}
  {et~al.}(2018){Gonz{\'a}lez-Fern{\'a}ndez}, {Hodgkin}, {Irwin},
  {Gonz{\'a}lez-Solares}, {Koposov}, {Lewis}, {Emerson}, {Hewett},
  {Yolda{\textcommabelow s}}, \& {Riello}}]{GonzalezFernandez18}
{Gonz{\'a}lez-Fern{\'a}ndez}, C., {Hodgkin}, S.~T., {Irwin}, M.~J., {et~al.}
  2018, \mnras, 474, 5459

\bibitem[{{Grado} {et~al.}(2019{\natexlab{a}}){Grado}, {Cappelaro}, {Getman},
  {Yang}, {Brocato}, {D'Avanzo}, {Covino}, {Greco}, {Rossi}, \&
  {Teresa}}]{2019GCN.25371....1G}
{Grado}, A., {Cappelaro}, E., {Getman}, F., {et~al.} 2019{\natexlab{a}}, GRB
  Coordinates Network, 25371, 1

\bibitem[{{Grado} {et~al.}(2019{\natexlab{b}}){Grado}, {Cappellaro}, {Brocato},
  {Covino}, {Getman}, {Greco}, {D'Avanzo}, {Rossi}, {Palazzi}, \&
  {Yang}}]{2019GCN.25669....1G}
{Grado}, A., {Cappellaro}, E., {Brocato}, E., {et~al.} 2019{\natexlab{b}}, GRB
  Coordinates Network, 25669, 1

\bibitem[{{Grado} {et~al.}(2020){Grado}, {Cappellaro}, {Covino}, {Getman},
  {Greco}, {Limatola}, {Yang}, {Amati}, {Benetti}, {Branchesi}, {Brocato},
  {Botticella}, {Campana}, {Cantiello}, {Dadina}, {D'Ammando}, {De Cesare},
  {D'Elia}, {Della Valle}, {Iodice}, {Longo}, {Mapelli}, {Masetti}, {Nicastro},
  {Palazzi}, {Possenti}, {Radovich}, {Rossi}, {Salvaterra}, {Stella},
  {Stratta}, {Testa}, \& {Tomasella}}]{Grado19}
{Grado}, A., {Cappellaro}, E., {Covino}, S., {et~al.} 2020, \mnras, 492, 1731

\bibitem[{{Greiner} {et~al.}(2008){Greiner}, {Bornemann}, {Clemens}, {Deuter},
  {Hasinger}, {Honsberg}, {Huber}, {Huber}, {Krauss}, {Kr{\"u}hler},
  {K{\"u}pc{\"u} Yolda{\textcommabelow s}}, {Mayer-Hasselwand er}, {Mican},
  {Primak}, {Schrey}, {Steiner}, {Szokoly}, {Th{\"o}ne}, {Yolda{\textcommabelow
  s}}, {Klose}, {Laux}, \& {Winkler}}]{greiner08}
{Greiner}, J., {Bornemann}, W., {Clemens}, C., {et~al.} 2008, \pasp, 120, 405

\bibitem[{{Hanna} {et~al.}(2014){Hanna}, {Mandel}, \& {Vousden}}]{Hanna2014}
{Hanna}, C., {Mandel}, I., \& {Vousden}, W. 2014, \apj, 784, 8

\bibitem[{{Hempel} {et~al.}(2012){Hempel}, {Fischer}, {Schaffner-Bielich}, \&
  {Liebend{\"o}rfer}}]{Hempel.etal:2012}
{Hempel}, M., {Fischer}, T., {Schaffner-Bielich}, J., \& {Liebend{\"o}rfer}, M.
  2012, \apj, 748, 70

\bibitem[{{Herner} {et~al.}(2019{\natexlab{a}}){Herner}, {Palmese},
  {Soares-Santes}, {Tucker}, {Allam}, {Annis}, {Garcia}, {Morgan}, {Bachmann},
  \& {Brout}}]{2019GCN.25398....1H}
{Herner}, K., {Palmese}, A., {Soares-Santes}, M., {et~al.} 2019{\natexlab{a}},
  GRB Coordinates Network, 25398, 1

\bibitem[{{Herner} {et~al.}(2019{\natexlab{b}}){Herner}, {Palmese},
  {Soares-Santos}, {Tucker}, {Allam}, {Annis}, {Garcia}, {Morgan}, {Bachmann},
  \& {Brout}}]{2019GCN.25373....1H}
{Herner}, K., {Palmese}, A., {Soares-Santos}, M., {et~al.} 2019{\natexlab{b}},
  GRB Coordinates Network, 25373, 1

\bibitem[{{Hill} {et~al.}(2010){Hill}, {Driver}, {Cameron}, {Cross}, {Liske},
  \& {Robotham}}]{Hill2010}
{Hill}, D.~T., {Driver}, S.~P., {Cameron}, E., {et~al.} 2010, \mnras, 404, 1215

\bibitem[{{Hinderer} {et~al.}(2018){Hinderer}, {Nissanke}, {Foucart},
  {Hotokezaka}, {Vincent}, {Kasliwal}, {Schmidt}, {Williamson}, {Nichols},
  {Duez}, {Kidder}, {Pfeiffer}, \& {Scheel}}]{hinderer18}
{Hinderer}, T., {Nissanke}, S., {Foucart}, F., {et~al.} 2018, arXiv e-prints
  [\eprint[arXiv]{1808.03836}]

\bibitem[{{Hjorth} {et~al.}(2017){Hjorth}, {Levan}, {Tanvir}, {Lyman},
  {Wojtak}, {Schr{\o}der}, {Mandel}, {Gall}, \& {Bruun}}]{hjorth17}
{Hjorth}, J., {Levan}, A.~J., {Tanvir}, N.~R., {et~al.} 2017, \apjl, 848, L31

\bibitem[{{Hu} {et~al.}(2019){Hu}, {Castro-Tirado}, {Valeev}, {Sokolov},
  {Fernandez-Garcia}, {Carrasco}, {Castellon}, {Tucker}, {Soares-Santos},
  {Garcia Alvarez}, \& {Rivero}}]{2019GCN.25588....1H}
{Hu}, Y.-D., {Castro-Tirado}, A.~J., {Valeev}, A.~F., {et~al.} 2019, GRB
  Coordinates Network, 25588, 1

\bibitem[{{Huber} {et~al.}(2019){Huber}, {Smith}, {Chambers}, {Schulz},
  {Smartt}, {Young}, {McBrien}, {Gillanders}, {Strivastav}, \&
  {O'Neil}}]{2019GCN.25356....1H}
{Huber}, M., {Smith}, K.~W., {Chambers}, K., {et~al.} 2019, GRB Coordinates
  Network, 25356, 1

\bibitem[{{IceCube Collaboration}(2019)}]{2019GCN.25321....1I}
{IceCube Collaboration}. 2019, GRB Coordinates Network, 25321, 1

\bibitem[{{Ioka} \& {Nakamura}(2019)}]{ioka19}
{Ioka}, K. \& {Nakamura}, T. 2019, \mnras, 487, 4884

\bibitem[{{Japelj} {et~al.}(2019{\natexlab{a}}){Japelj}, {Jin}, {Kankare},
  {Kool}, {Levan}, {Maguire}, {Malesani}, {Mattlia}, {Milvang-Jensen}, \&
  {Reynolds}}]{2019GCN.25447....1J}
{Japelj}, J., {Jin}, Z., {Kankare}, E., {et~al.} 2019{\natexlab{a}}, GRB
  Coordinates Network, 25447, 1

\bibitem[{{Japelj} {et~al.}(2019{\natexlab{b}}){Japelj}, {Kankare}, {Kool},
  {Levan}, {Maguire}, {Malesani}, {Mattila}, {Milvang-Jensen}, {Reynolds},
  {Sabha}, \& {Vergani}}]{2019GCN.25526....1J}
{Japelj}, J., {Kankare}, E., {Kool}, E., {et~al.} 2019{\natexlab{b}}, GRB
  Coordinates Network, 25526, 1

\bibitem[{{Jin} {et~al.}(2020){Jin}, {Covino}, {Liao}, {Li}, {D'Avanzo}, {Fan},
  \& {Wei}}]{jin20}
{Jin}, Z.-P., {Covino}, S., {Liao}, N.-H., {et~al.} 2020, Nature Astronomy, 4,
  77

\bibitem[{{Jin} {et~al.}(2016){Jin}, {Hotokezaka}, {Li}, {Tanaka}, {D'Avanzo},
  {Fan}, {Covino}, {Wei}, \& {Piran}}]{Jin16}
{Jin}, Z.-P., {Hotokezaka}, K., {Li}, X., {et~al.} 2016, Nature Communications,
  7, 12898

\bibitem[{{Jonker} {et~al.}(2019){Jonker}, {Maguire}, {Fraser}, {Torres},
  {Levan}, \& {Shlentsova}}]{2019GCN.25454....1J}
{Jonker}, P., {Maguire}, K., {Fraser}, M., {et~al.} 2019, GRB Coordinates
  Network, 25454, 1

\bibitem[{{Just} {et~al.}(2015){Just}, {Bauswein}, {Ardevol Pulpillo},
  {Goriely}, \& {Janka}}]{Just15}
{Just}, O., {Bauswein}, A., {Ardevol Pulpillo}, R., {Goriely}, S., \& {Janka},
  H.-T. 2015, \mnras, 448, 541

\bibitem[{{Kasen} {et~al.}(2017){Kasen}, {Metzger}, {Barnes}, {Quataert}, \&
  {Ramirez-Ruiz}}]{kasen17}
{Kasen}, D., {Metzger}, B., {Barnes}, J., {Quataert}, E., \& {Ramirez-Ruiz}, E.
  2017, \nat, 551, 80

\bibitem[{{Kasliwal} {et~al.}(2017){Kasliwal}, {Nakar}, {Singer}, {Kaplan},
  {Cook}, {Van Sistine}, {Lau}, {Fremling}, {Gottlieb}, {Jencson}, {Adams},
  {Feindt}, {Hotokezaka}, {Ghosh}, {Perley}, {Yu}, {Piran}, {Allison},
  {Anupama}, {Balasubramanian}, {Bannister}, {Bally}, {Barnes}, {Barway},
  {Bellm}, {Bhalerao}, {Bhattacharya}, {Blagorodnova}, {Bloom}, {Brady},
  {Cannella}, {Chatterjee}, {Cenko}, {Cobb}, {Copperwheat}, {Corsi}, {De},
  {Dobie}, {Emery}, {Evans}, {Fox}, {Frail}, {Frohmaier}, {Goobar}, {Hallinan},
  {Harrison}, {Helou}, {Hinderer}, {Ho}, {Horesh}, {Ip}, {Itoh}, {Kasen},
  {Kim}, {Kuin}, {Kupfer}, {Lynch}, {Madsen}, {Mazzali}, {Miller}, {Mooley},
  {Murphy}, {Ngeow}, {Nichols}, {Nissanke}, {Nugent}, {Ofek}, {Qi}, {Quimby},
  {Rosswog}, {Rusu}, {Sadler}, {Schmidt}, {Sollerman}, {Steele}, {Williamson},
  {Xu}, {Yan}, {Yatsu}, {Zhang}, \& {Zhao}}]{kasliwal17}
{Kasliwal}, M.~M., {Nakar}, E., {Singer}, L.~P., {et~al.} 2017, Science, 358,
  1559

\bibitem[{Kawaguchi {et~al.}(2016)Kawaguchi, Kyutoku, Shibata, \&
  Tanaka}]{Kawaguchi2016}
Kawaguchi, K., Kyutoku, K., Shibata, M., \& Tanaka, M. 2016, The Astrophysical
  Journal, Volume 825, Issue 1, article id. 52, 12 pp. (2016)., 825
  [\eprint[arXiv]{1601.07711}]

\bibitem[{{Kissler-Patig} {et~al.}(2008){Kissler-Patig}, {Pirard}, {Casali},
  {Moorwood}, {Ageorges}, {Alves de Oliveira}, {Baksai}, {Bedin}, {Bendek},
  {Biereichel}, {Delabre}, {Dorn}, {Esteves}, {Finger}, {Gojak}, {Huster},
  {Jung}, {Kiekebush}, {Klein}, {Koch}, {Lizon}, {Mehrgan}, {Petr-Gotzens},
  {Pritchard}, {Selman}, \& {Stegmeier}}]{kissler-patig08}
{Kissler-Patig}, M., {Pirard}, J.-F., {Casali}, M., {et~al.} 2008, \aap, 491,
  941

\bibitem[{{Kiuchi} {et~al.}(2015){Kiuchi}, {Sekiguchi}, {Kyutoku}, {Shibata},
  {Taniguchi}, \& {Wada}}]{Kiuchi15}
{Kiuchi}, K., {Sekiguchi}, Y., {Kyutoku}, K., {et~al.} 2015, \prd, 92, 064034

\bibitem[{{Kocevksi}(2019)}]{2019GCN.25326....1K}
{Kocevksi}, D. 2019, GRB Coordinates Network, 25326, 1

\bibitem[{{Kruckow} {et~al.}(2018){Kruckow}, {Tauris}, {Langer}, {Kramer}, \&
  {Izzard}}]{Kruckow:2018}
{Kruckow}, M.~U., {Tauris}, T.~M., {Langer}, N., {Kramer}, M., \& {Izzard},
  R.~G. 2018, \mnras, 481, 1908

\bibitem[{{Kr{\"u}hler} {et~al.}(2008){Kr{\"u}hler}, {K{\"u}pc{\"u}
  Yolda{\textcommabelow s}}, {Greiner}, {Clemens}, {McBreen}, {Primak},
  {Savaglio}, {Yolda{\textcommabelow s}}, {Szokoly}, \&
  {Klose}}]{2008ApJ...685..376K}
{Kr{\"u}hler}, T., {K{\"u}pc{\"u} Yolda{\textcommabelow s}}, A., {Greiner}, J.,
  {et~al.} 2008, \apj, 685, 376

\bibitem[{{Kuijken}(2011)}]{Kuijken11}
{Kuijken}, K. 2011, The Messenger, 146, 8

\bibitem[{{Kulkarni} {et~al.}(2018){Kulkarni}, {Perley}, \&
  {Miller}}]{kulkarni18b}
{Kulkarni}, S.~R., {Perley}, D.~A., \& {Miller}, A.~A. 2018, \apj, 860, 22

\bibitem[{{Kyutoku} {et~al.}(2013){Kyutoku}, {Ioka}, \& {Shibata}}]{Kyutoku13}
{Kyutoku}, K., {Ioka}, K., \& {Shibata}, M. 2013, \prd, 88, 041503

\bibitem[{{Kyutoku} {et~al.}(2018){Kyutoku}, {Kiuchi}, {Sekiguchi}, {Shibata},
  \& {Taniguchi}}]{Kyutoku18}
{Kyutoku}, K., {Kiuchi}, K., {Sekiguchi}, Y., {Shibata}, M., \& {Taniguchi}, K.
  2018, \prd, 97, 023009

\bibitem[{{Kyutoku} {et~al.}(2011){Kyutoku}, {Okawa}, {Shibata}, \&
  {Taniguchi}}]{Kyutoku11}
{Kyutoku}, K., {Okawa}, H., {Shibata}, M., \& {Taniguchi}, K. 2011, \prd, 84,
  064018

\bibitem[{{Lamb} \& {Kobayashi}(2016)}]{lamb16}
{Lamb}, G.~P. \& {Kobayashi}, S. 2016, \apj, 829, 112

\bibitem[{{Lamb} \& {Kobayashi}(2017)}]{lamb17}
{Lamb}, G.~P. \& {Kobayashi}, S. 2017, \mnras, 472, 4953

\bibitem[{{Lamb} {et~al.}(2019{\natexlab{a}}){Lamb}, {Lyman}, {Levan},
  {Tanvir}, {Kangas}, {Fruchter}, {Gompertz}, {Hjorth}, {Mandel}, {Oates},
  {Steeghs}, \& {Wiersema}}]{lamb19}
{Lamb}, G.~P., {Lyman}, J.~D., {Levan}, A.~J., {et~al.} 2019{\natexlab{a}},
  \apjl, 870, L15

\bibitem[{{Lamb} {et~al.}(2019{\natexlab{b}}){Lamb}, {Tanvir}, {Levan}, {de
  Ugarte Postigo}, {Kawaguchi}, {Corsi}, {Evans}, {Gompertz}, {Malesani},
  {Page}, {Wiersema}, {Rosswog}, {Shibata}, {Tanaka}, {van der Horst}, {Cano},
  {Fynbo}, {Fruchter}, {Greiner}, {Heintz}, {Higgins}, {Hjorth}, {Izzo},
  {Jakobsson}, {Kann}, {O'Brien}, {Perley}, {Pian}, {Pugliese}, {Starling},
  {Th{\"o}ne}, {Watson}, {Wijers}, \& {Xu}}]{lamb19b}
{Lamb}, G.~P., {Tanvir}, N.~R., {Levan}, A.~J., {et~al.} 2019{\natexlab{b}},
  \apj, 883, 48

\bibitem[{{Lattimer} {et~al.}(1977){Lattimer}, {Mackie}, {Ravenhall}, \&
  {Schramm}}]{lattimer77}
{Lattimer}, J.~M., {Mackie}, F., {Ravenhall}, D.~G., \& {Schramm}, D.~N. 1977,
  \apj, 213, 225

\bibitem[{{Lattimer} \& {Schramm}(1976)}]{Lattimer1976}
{Lattimer}, J.~M. \& {Schramm}, D.~N. 1976, \apj, 210, 549

\bibitem[{{Lazzati} {et~al.}(2018){Lazzati}, {Perna}, {Morsony},
  {Lopez-Camara}, {Cantiello}, {Ciolfi}, {Giacomazzo}, \&
  {Workman}}]{lazzati18}
{Lazzati}, D., {Perna}, R., {Morsony}, B.~J., {et~al.} 2018, Physical Review
  Letters, 120, 241103

\bibitem[{{Levan} {et~al.}(2017){Levan}, {Lyman}, {Tanvir}, {Hjorth}, {Mandel},
  {Stanway}, {Steeghs}, {Fruchter}, {Troja}, {Schr{\o}der}, {Wiersema},
  {Bruun}, {Cano}, {Cenko}, {de Ugarte Postigo}, {Evans}, {Fairhurst}, {Fox},
  {Fynbo}, {Gompertz}, {Greiner}, {Im}, {Izzo}, {Jakobsson}, {Kangas},
  {Khandrika}, {Lien}, {Malesani}, {O'Brien}, {Osborne}, {Palazzi}, {Pian},
  {Perley}, {Rosswog}, {Ryan}, {Schulze}, {Sutton}, {Th{\"o}ne}, {Watson}, \&
  {Wijers}}]{levan17}
{Levan}, A.~J., {Lyman}, J.~D., {Tanvir}, N.~R., {et~al.} 2017, \apjl, 848, L28

\bibitem[{{Li} \& {Paczy{\'n}ski}(1998)}]{li98}
{Li}, L.-X. \& {Paczy{\'n}ski}, B. 1998, \apjl, 507, L59

\bibitem[{{Lipunov} {et~al.}(2017){Lipunov}, {Gorbovskoy}, {Kornilov},
  {.~Tyurina}, {Balanutsa}, {Kuznetsov}, {Vlasenko}, {Kuvshinov}, {Gorbunov},
  {Buckley}, {Krylov}, {Podesta}, {Lopez}, {Podesta}, {Levato}, {Saffe},
  {Mallamachi}, {Potter}, {Budnev}, {Gress}, {Ishmuhametova}, {Vladimirov},
  {Zimnukhov}, {Yurkov}, {Sergienko}, {Gabovich}, {Rebolo}, {Serra-Ricart},
  {Israelyan}, {Chazov}, {Wang}, {Tlatov}, \& {Panchenko}}]{lipunov17}
{Lipunov}, V.~M., {Gorbovskoy}, E., {Kornilov}, V.~G., {et~al.} 2017, \apjl,
  850, L1

\bibitem[{{Lopez-Cruz} {et~al.}(2019{\natexlab{a}}){Lopez-Cruz},
  {Castro-Tirado}, {Macri}, {Valeev}, {Rios-Lopez}, {Hu}, {Diaz},
  {Fernandez-Garcia}, {Chabushyan}, \& {Castellon}}]{2019GCN.25419....1L}
{Lopez-Cruz}, O., {Castro-Tirado}, A.~J., {Macri}, L., {et~al.}
  2019{\natexlab{a}}, GRB Coordinates Network, 25419, 1

\bibitem[{{Lopez-Cruz} {et~al.}(2019{\natexlab{b}}){Lopez-Cruz},
  {Castro-Tirado}, {Macri}, {Valeev}, {Rios-Lopez}, {Hu}, {Diaz},
  {Fernandez-Garcia}, {Troja}, \& {Castellon}}]{2019GCN.25571....1L}
{Lopez-Cruz}, O., {Castro-Tirado}, A.~J., {Macri}, L., {et~al.}
  2019{\natexlab{b}}, GRB Coordinates Network, 25571, 1

\bibitem[{{Lyman} {et~al.}(2018){Lyman}, {Lamb}, {Levan}, {Mandel}, {Tanvir},
  {Kobayashi}, {Gompertz}, {Hjorth}, {Fruchter}, {Kangas}, {Steeghs}, {Steele},
  {Cano}, {Copperwheat}, {Evans}, {Fynbo}, {Gall}, {Im}, {Izzo}, {Jakobsson},
  {Milvang-Jensen}, {O'Brien}, {Osborne}, {Palazzi}, {Perley}, {Pian},
  {Rosswog}, {Rowlinson}, {Schulze}, {Stanway}, {Sutton}, {Th{\"o}ne}, {de
  Ugarte Postigo}, {Watson}, {Wiersema}, \& {Wijers}}]{lyman18}
{Lyman}, J.~D., {Lamb}, G.~P., {Levan}, A.~J., {et~al.} 2018, Nature Astronomy,
  2, 751

\bibitem[{{Magnier} {et~al.}(2016){Magnier}, {Chambers}, {Flewelling},
  {Hoblitt}, {Huber}, {Price}, {Sweeney}, {Waters}, {Denneau}, {Draper},
  {Hodapp}, {Jedicke}, {Kaiser}, {Kudritzki}, {Metcalfe}, {Stubbs}, \&
  {Wainscoast}}]{magnier16a}
{Magnier}, E.~A., {Chambers}, K.~C., {Flewelling}, H.~A., {et~al.} 2016, arXiv
  e-prints, arXiv:1612.05240

\bibitem[{{Margutti} {et~al.}(2018){Margutti}, {Alexander}, {Xie}, {Sironi},
  {Metzger}, {Kathirgamaraju}, {Fong}, {Blanchard}, {Berger}, {MacFadyen},
  {Giannios}, {Guidorzi}, {Hajela}, {Chornock}, {Cowperthwaite}, {Eftekhari},
  {Nicholl}, {Villar}, {Williams}, \& {Zrake}}]{Margutti2018ApJL}
{Margutti}, R., {Alexander}, K.~D., {Xie}, X., {et~al.} 2018, \apjl, 856, L18

\bibitem[{{Matsumoto} {et~al.}(2019){Matsumoto}, {Nakar}, \&
  {Piran}}]{matsumoto19}
{Matsumoto}, T., {Nakar}, E., \& {Piran}, T. 2019, \mnras, 486, 1563

\bibitem[{{McCully} {et~al.}(2017){McCully}, {Hiramatsu}, {Howell},
  {Hosseinzadeh}, {Arcavi}, {Kasen}, {Barnes}, {Shara}, {Williams},
  {V{\"a}is{\"a}nen}, {Potter}, {Romero-Colmenero}, {Crawford}, {Buckley},
  {Cooke}, {Andreoni}, {Pritchard}, {Mao}, {Gromadzki}, \& {Burke}}]{mccully17}
{McCully}, C., {Hiramatsu}, D., {Howell}, D.~A., {et~al.} 2017, \apjl, 848, L32

\bibitem[{{Metzger} {et~al.}(2010){Metzger}, {Mart{\'{\i}}nez-Pinedo},
  {Darbha}, {Quataert}, {Arcones}, {Kasen}, {Thomas}, {Nugent}, {Panov}, \&
  {Zinner}}]{metzger10}
{Metzger}, B.~D., {Mart{\'{\i}}nez-Pinedo}, G., {Darbha}, S., {et~al.} 2010,
  \mnras, 406, 2650

\bibitem[{{Molinari} {et~al.}(1999){Molinari}, {Conconi}, {Pucillo}, \&
  {Monai}}]{Molinari1999}
{Molinari}, E., {Conconi}, P., {Pucillo}, M., \& {Monai}, S. 1999, in Looking
  Deep in the Southern Sky, ed. R.~{Morganti} \& W.~J. {Couch}, 157

\bibitem[{{Molkov} {et~al.}(2019){Molkov}, {Mereghetti}, {Savchenko},
  {Ferrigno}, {Rodi}, \& {Coleiro}}]{2019GCN.25323....1M}
{Molkov}, S., {Mereghetti}, S., {Savchenko}, V., {et~al.} 2019, GRB Coordinates
  Network, 25323, 1

\bibitem[{{Mooley} {et~al.}(2018){Mooley}, {Frail}, {Dobie}, {Lenc}, {Corsi},
  {De}, {Nayana}, {Makhathini}, {Heywood}, {Murphy}, {Kaplan}, {Chandra},
  {Smirnov}, {Nakar}, {Hallinan}, {Camilo}, {Fender}, {Goedhart}, {Groot},
  {Kasliwal}, {Kulkarni}, \& {Woudt}}]{mooley18b}
{Mooley}, K.~P., {Frail}, D.~A., {Dobie}, D., {et~al.} 2018, \apjl, 868, L11

\bibitem[{{Neijssel} {et~al.}(2019){Neijssel}, {Vigna-G{\'o}mez}, {Stevenson},
  {Barrett}, {Gaebel}, {Broekgaarden}, {de Mink}, {Sz{\'e}csi}, {Vinciguerra},
  \& {Mandel}}]{Neijssel:2019}
{Neijssel}, C.~J., {Vigna-G{\'o}mez}, A., {Stevenson}, S., {et~al.} 2019, arXiv
  e-prints [\eprint[arXiv]{1906.08136}]

\bibitem[{{Nicholl} {et~al.}(2017){Nicholl}, {Berger}, {Kasen}, {Metzger},
  {Elias}, {Brice{\~n}o}, {Alexander}, {Blanchard}, {Chornock},
  {Cowperthwaite}, {Eftekhari}, {Fong}, {Margutti}, {Villar}, {Williams},
  {Brown}, {Annis}, {Bahramian}, {Brout}, {Brown}, {Chen}, {Clemens},
  {Dennihy}, {Dunlap}, {Holz}, {Marchesini}, {Massaro}, {Moskowitz},
  {Pelisoli}, {Rest}, {Ricci}, {Sako}, {Soares-Santos}, \&
  {Strader}}]{nicholl17}
{Nicholl}, M., {Berger}, E., {Kasen}, D., {et~al.} 2017, \apjl, 848, L18

\bibitem[{{Nissanke} {et~al.}(2010){Nissanke}, {Holz}, {Hughes}, {Dalal}, \&
  {Sievers}}]{nissanke10}
{Nissanke}, S., {Holz}, D.~E., {Hughes}, S.~A., {Dalal}, N., \& {Sievers},
  J.~L. 2010, \apj, 725, 496

\bibitem[{{Nissanke} {et~al.}(2013){Nissanke}, {Kasliwal}, \&
  {Georgieva}}]{nissanke13}
{Nissanke}, S., {Kasliwal}, M., \& {Georgieva}, A. 2013, \apj, 767, 124

\bibitem[{{Nynka} {et~al.}(2018){Nynka}, {Ruan}, {Haggard}, \&
  {Evans}}]{Nynka2018ApJL}
{Nynka}, M., {Ruan}, J.~J., {Haggard}, D., \& {Evans}, P.~A. 2018, \apjl, 862,
  L19

\bibitem[{{Oertel} {et~al.}(2017){Oertel}, {Hempel}, {Kl{\"a}hn}, \&
  {Typel}}]{Oertel.etal:2017}
{Oertel}, M., {Hempel}, M., {Kl{\"a}hn}, T., \& {Typel}, S. 2017, Reviews of
  Modern Physics, 89, 015007

\bibitem[{{Ohno} {et~al.}(2019){Ohno}, {Axelsson}, {Longo}, {Kocevski},
  {Omodei}, \& {Dirirsa}}]{2019GCN.25385....1O}
{Ohno}, M., {Axelsson}, M., {Longo}, F., {et~al.} 2019, GRB Coordinates
  Network, 25386, 1

\bibitem[{{Oliva} {et~al.}(2001){Oliva}, {Marconi}, {Maiolino}, {Testi},
  {Mannucci}, {Ghinassi}, {Licandro}, {Origlia}, {Baffa}, {Checcucci},
  {Comoretto}, {Gavryussev}, {Gennari}, {Giani}, {Hunt}, {Lisi}, {Lorenzetti},
  {Marcucci}, {Miglietta}, {Sozzi}, {Stefanini}, \& {Vitali}}]{Oliva2001}
{Oliva}, E., {Marconi}, A., {Maiolino}, R., {et~al.} 2001, \aap, 369, L5

\bibitem[{{Palmer} {et~al.}(2019){Palmer}, {Barthelmy}, {Lien}, {Sakamoto},
  {Beardmore}, {Bernardini}, {Breeveld}, {Burrows}, {Campana}, \&
  {Cenko}}]{2019GCN.25341....1P}
{Palmer}, D.~M., {Barthelmy}, S.~D., {Lien}, A.~Y., {et~al.} 2019, GRB
  Coordinates Network, 25341, 1

\bibitem[{{Pannarale} \& {Ohme}(2014)}]{Pannarale2014}
{Pannarale}, F. \& {Ohme}, F. 2014, \apjl, 791, L7

\bibitem[{{Perego} {et~al.}(2017){Perego}, {Radice}, \& {Bernuzzi}}]{perego17}
{Perego}, A., {Radice}, D., \& {Bernuzzi}, S. 2017, \apjl, 850, L37

\bibitem[{{Pian} {et~al.}(2017){Pian}, {D'Avanzo}, {Benetti}, {Branchesi},
  {Brocato}, {Campana}, {Cappellaro}, {Covino}, {D'Elia}, {Fynbo}, {Getman},
  {Ghirlanda}, {Ghisellini}, {Grado}, {Greco}, {Hjorth}, {Kouveliotou},
  {Levan}, {Limatola}, {Malesani}, {Mazzali}, {Melandri}, {M{\o}ller},
  {Nicastro}, {Palazzi}, {Piranomonte}, {Rossi}, {Salafia}, {Selsing},
  {Stratta}, {Tanaka}, {Tanvir}, {Tomasella}, {Watson}, {Yang}, {Amati},
  {Antonelli}, {Ascenzi}, {Bernardini}, {Bo{\"e}r}, {Bufano}, {Bulgarelli},
  {Capaccioli}, {Casella}, {Castro-Tirado}, {Chassande-Mottin}, {Ciolfi},
  {Copperwheat}, {Dadina}, {De Cesare}, {di Paola}, {Fan}, {Gendre},
  {Giuffrida}, {Giunta}, {Hunt}, {Israel}, {Jin}, {Kasliwal}, {Klose}, {Lisi},
  {Longo}, {Maiorano}, {Mapelli}, {Masetti}, {Nava}, {Patricelli}, {Perley},
  {Pescalli}, {Piran}, {Possenti}, {Pulone}, {Razzano}, {Salvaterra},
  {Schipani}, {Spera}, {Stamerra}, {Stella}, {Tagliaferri}, {Testa}, {Troja},
  {Turatto}, {Vergani}, \& {Vergani}}]{pian17}
{Pian}, E., {D'Avanzo}, P., {Benetti}, S., {et~al.} 2017, \nat, 551, 67

\bibitem[{{Pilia} {et~al.}(2019{\natexlab{a}}){Pilia}, {Pittori}, {Tavani},
  {Cardillo}, {Casentini}, {Piano}, {Ursi}, {Lucarelli}, {Verrecchia}, \&
  {Bulgarelli}}]{2019GCN.25327....1P}
{Pilia}, M., {Pittori}, C., {Tavani}, M., {et~al.} 2019{\natexlab{a}}, GRB
  Coordinates Network, 25327, 1

\bibitem[{{Pilia} {et~al.}(2019{\natexlab{b}}){Pilia}, {Pittori}, {Tavani},
  {Lucarelli}, {Verrecchia}, {Longo}, {Cardillo}, {Casentini}, {Ursi}, \&
  {Bulgarelli}}]{2019GCN.25335....1P}
{Pilia}, M., {Pittori}, C., {Tavani}, M., {et~al.} 2019{\natexlab{b}}, GRB
  Coordinates Network, 25335, 1

\bibitem[{{Pirard} {et~al.}(2004){Pirard}, {Kissler-Patig}, {Moorwood},
  {Biereichel}, {Delabre}, {Dorn}, {Finger}, {Gojak}, {Huster}, {Jung}, {Koch},
  {Le Louarn}, {Lizon}, {Mehrgan}, {Pozna}, {Silber}, {Sokar}, \&
  {Stegmeier}}]{pirard04}
{Pirard}, J.-F., {Kissler-Patig}, M., {Moorwood}, A., {et~al.} 2004, in
  \procspie, Vol. 5492, Ground-based Instrumentation for Astronomy, ed.
  A.~F.~M. {Moorwood} \& M.~{Iye}, 1763--1772

\bibitem[{{Poretti}(2018)}]{Poretti2018}
{Poretti}, E. 2018, in Protoplanetary Disks Seen through the Eyes of
  New-Generation High-Resolution Instruments, 28

\bibitem[{{Radice} {et~al.}(2018){Radice}, {Perego}, {Hotokezaka}, {Fromm},
  {Bernuzzi}, \& {Roberts}}]{radice18d}
{Radice}, D., {Perego}, A., {Hotokezaka}, K., {et~al.} 2018, \apj, 869, 130

\bibitem[{{Rantsiou} {et~al.}(2008){Rantsiou}, {Kobayashi}, {Laguna}, \&
  {Rasio}}]{Rantsiou08}
{Rantsiou}, E., {Kobayashi}, S., {Laguna}, P., \& {Rasio}, F.~A. 2008, \apj,
  680, 1326

\bibitem[{{Resmi} {et~al.}(2018){Resmi}, {Schulze}, {Ishwara-Chandra}, {Misra},
  {Buchner}, {De Pasquale}, {Sanchez-Ramirez}, {Klose}, {Kim}, {Tanvir}, \&
  {O'Brien}}]{resmi18}
{Resmi}, L., {Schulze}, S., {Ishwara-Chandra}, C.~H., {et~al.} 2018, \apj, 867,
  57

\bibitem[{{Roberts} {et~al.}(2017{\natexlab{a}}){Roberts}, {Lippuner}, {Duez},
  {Faber}, {Foucart}, {Lombardi}, {Ning}, {Ott}, \& {Ponce}}]{Robertsetal2017}
{Roberts}, L.~F., {Lippuner}, J., {Duez}, M.~D., {et~al.} 2017{\natexlab{a}},
  \mnras, 464, 3907

\bibitem[{{Roberts} {et~al.}(2017{\natexlab{b}}){Roberts}, {Lippuner}, {Duez},
  {Faber}, {Foucart}, {Lombardi}, {Ning}, {Ott}, \& {Ponce}}]{Roberts17}
{Roberts}, L.~F., {Lippuner}, J., {Duez}, M.~D., {et~al.} 2017{\natexlab{b}},
  \mnras, 464, 3907

\bibitem[{{Rodriguez} {et~al.}(2019){Rodriguez}, {Meza-Retamal}, {Quirola},
  {Olivares}, {Cartier}, {Tucker}, {Soares-Santos}, {Martinez-Vasquez},
  {Garcia}, \& {Herner}}]{2019GCN.25423....1R}
{Rodriguez}, A., {Meza-Retamal}, N., {Quirola}, J., {et~al.} 2019, GRB
  Coordinates Network, 25423, 1

\bibitem[{{Rossi} {et~al.}(2019){Rossi}, {D'Avanzo}, {Cappellaro},
  {Botticella}, {Giunta}, {Greco}, {Piranomonte}, {Harutyunyan}, {Magazzu}, \&
  {Mainella}}]{2019GCN.25383....1R}
{Rossi}, A., {D'Avanzo}, P., {Cappellaro}, E., {et~al.} 2019, GRB Coordinates
  Network, 25383, 1

\bibitem[{{Rosswog}(2005)}]{rosswog05}
{Rosswog}, S. 2005, \apj, 634, 1202

\bibitem[{{Rosswog}(2017)}]{rosswog17}
{Rosswog}, S. 2017, Physics Online Journal, 10, 131

\bibitem[{{Rosswog} {et~al.}(2017){Rosswog}, {Feindt}, {Korobkin}, {Wu},
  {Sollerman}, {Goobar}, \& {Martinez-Pinedo}}]{Rosswogetal2017}
{Rosswog}, S., {Feindt}, U., {Korobkin}, O., {et~al.} 2017, Classical and
  Quantum Gravity, 34, 104001

\bibitem[{{Ruan} {et~al.}(2019){Ruan}, {Vieira}, {Haggard}, {Drout}, {Asfari},
  {Carlberg}, {Doyon}, {Fernandez}, {Gaensler}, \&
  {Kaspi}}]{2019GCN.25443....1R}
{Ruan}, J., {Vieira}, N., {Haggard}, D., {et~al.} 2019, GRB Coordinates
  Network, 25443, 1

\bibitem[{{Sagu{\'e}s Carracedo} {et~al.}(2020){Sagu{\'e}s Carracedo}, {Bulla},
  {Feindt}, \& {Goobar}}]{SaguesCarracedo2020}
{Sagu{\'e}s Carracedo}, A., {Bulla}, M., {Feindt}, U., \& {Goobar}, A. 2020,
  arXiv e-prints, arXiv:2004.06137

\bibitem[{{Salafia} {et~al.}(2019){Salafia}, {Ghirlanda}, {Ascenzi}, \&
  {Ghisellini}}]{salafia19}
{Salafia}, O.~S., {Ghirlanda}, G., {Ascenzi}, S., \& {Ghisellini}, G. 2019,
  \aap, 628, A18

\bibitem[{{Salafia} {et~al.}(2015){Salafia}, {Ghisellini}, {Pescalli},
  {Ghirlanda}, \& {Nappo}}]{salafia15}
{Salafia}, O.~S., {Ghisellini}, G., {Pescalli}, A., {Ghirlanda}, G., \&
  {Nappo}, F. 2015, \mnras, 450, 3549

\bibitem[{{Saleem} {et~al.}(2019){Saleem}, {Resmi}, {Arun}, \&
  {Mohan}}]{saleem2019}
{Saleem}, M., {Resmi}, L., {Arun}, K.~G., \& {Mohan}, S. 2019, arXiv e-prints,
  arXiv:1905.00337

\bibitem[{{Salmon} {et~al.}(2019){Salmon}, {Hanlon}, {Jeffrey}, \&
  {Martin-Carrillo}}]{Salmonetal2019}
{Salmon}, L., {Hanlon}, L., {Jeffrey}, R.~M., \& {Martin-Carrillo}, A. 2019,
  arXiv e-prints, arXiv:1912.07304

\bibitem[{{Savchenko} {et~al.}(2017){Savchenko}, {Ferrigno}, {Kuulkers},
  {Bazzano}, {Bozzo}, {Brandt}, {Chenevez}, {Courvoisier}, {Diehl}, {Domingo},
  {Hanlon}, {Jourdain}, {von Kienlin}, {Laurent}, {Lebrun}, {Lutovinov},
  {Martin-Carrillo}, {Mereghetti}, {Natalucci}, {Rodi}, {Roques}, {Sunyaev}, \&
  {Ubertini}}]{savchenko17}
{Savchenko}, V., {Ferrigno}, C., {Kuulkers}, E., {et~al.} 2017, \apjl, 848, L15

\bibitem[{{Schechter}(1976)}]{Schechter1976}
{Schechter}, P. 1976, \apj, 203, 297

\bibitem[{{Schutz}(1986)}]{schutz86}
{Schutz}, B.~F. 1986, \nat, 323, 310

\bibitem[{{Schutz}(2011)}]{Schutz2011}
{Schutz}, B.~F. 2011, Classical and Quantum Gravity, 28, 125023

\bibitem[{{Shappee} {et~al.}(2017){Shappee}, {Simon}, {Drout}, {Piro},
  {Morrell}, {Prieto}, {Kasen}, {Holoien}, {Kollmeier}, {Kelson}, {Coulter},
  {Foley}, {Kilpatrick}, {Siebert}, {Madore}, {Murguia-Berthier}, {Pan},
  {Prochaska}, {Ramirez-Ruiz}, {Rest}, {Adams}, {Alatalo}, {Ba{\~n}ados},
  {Baughman}, {Bernstein}, {Bitsakis}, {Boutsia}, {Bravo}, {Di Mille}, {Higgs},
  {Ji}, {Maravelias}, {Marshall}, {Placco}, {Prieto}, \& {Wan}}]{shappee17}
{Shappee}, B.~J., {Simon}, J.~D., {Drout}, M.~R., {et~al.} 2017, Science, 358,
  1574

\bibitem[{{Shibata} \& {Hotokezaka}(2019)}]{Shibata&Hotokezaka2019}
{Shibata}, M. \& {Hotokezaka}, K. 2019, Annual Review of Nuclear and Particle
  Science, 69, 41

\bibitem[{{Shibata} \& {Taniguchi}(2011)}]{Shibata2011}
{Shibata}, M. \& {Taniguchi}, K. 2011, Living Reviews in Relativity, 14, 6

\bibitem[{{Siebenmorgen} {et~al.}(2011){Siebenmorgen}, {Carraro}, {Valenti},
  {Petr-Gotzens}, {Brammer}, {Garcia}, \& {Casali}}]{siebenmorgen11}
{Siebenmorgen}, R., {Carraro}, G., {Valenti}, E., {et~al.} 2011, The Messenger,
  144, 9

\bibitem[{{Singer} {et~al.}(2016){Singer}, {Chen}, {Holz}, {Farr}, {Price},
  {Raymond}, {Cenko}, {Gehrels}, {Cannizzo}, {Kasliwal}, {Nissanke},
  {Coughlin}, {Farr}, {Urban}, {Vitale}, {Veitch}, {Graff}, {Berry},
  {Mohapatra}, \& {Mandel}}]{singer16b}
{Singer}, L.~P., {Chen}, H.-Y., {Holz}, D.~E., {et~al.} 2016, \apjl, 829, L15

\bibitem[{{Skrutskie} {et~al.}(2006){Skrutskie}, {Cutri}, {Stiening},
  {Weinberg}, {Schneider}, {Carpenter}, {Beichman}, {Capps}, {Chester},
  {Elias}, {Huchra}, {Liebert}, {Lonsdale}, {Monet}, {Price}, {Seitzer},
  {Jarrett}, {Kirkpatrick}, {Gizis}, {Howard}, {Evans}, {Fowler}, {Fullmer},
  {Hurt}, {Light}, {Kopan}, {Marsh}, {McCallon}, {Tam}, {Van Dyk}, \&
  {Wheelock}}]{2006AJ....131.1163S}
{Skrutskie}, M.~F., {Cutri}, R.~M., {Stiening}, R., {et~al.} 2006, \aj, 131,
  1163

\bibitem[{{Smartt} {et~al.}(2016{\natexlab{a}}){Smartt}, {Chambers}, {Smith},
  {Huber}, {Young}, {Cappellaro}, {Wright}, {Coughlin}, {Schultz}, {Denneau},
  {Flewelling}, {Heinze}, {Magnier}, {Primak}, {Rest}, {Sherstyuk}, {Stalder},
  {Stubbs}, {Tonry}, {Waters}, {Willman}, {Anderson}, {Baltay}, {Botticella},
  {Campbell}, {Dennefeld}, {Chen}, {Della Valle}, {Elias-Rosa}, {Fraser},
  {Inserra}, {Kankare}, {Kotak}, {Kupfer}, {Harmanen}, {Galbany}, {Gal-Yam},
  {Le Guillou}, {Lyman}, {Maguire}, {Mitra}, {Nicholl}, {Olivares E},
  {Rabinowitz}, {Razza}, {Sollerman}, {Smith}, {Terreran}, {Valenti}, {Gibson},
  \& {Goggia}}]{smartt16a}
{Smartt}, S.~J., {Chambers}, K.~C., {Smith}, K.~W., {et~al.}
  2016{\natexlab{a}}, \mnras, 462, 4094

\bibitem[{{Smartt} {et~al.}(2016{\natexlab{b}}){Smartt}, {Chambers}, {Smith},
  {Huber}, {Young}, {Chen}, {Inserra}, {Wright}, {Coughlin}, {Denneau},
  {Flewelling}, {Heinze}, {Jerkstrand}, {Magnier}, {Maguire}, {Mueller},
  {Rest}, {Sherstyuk}, {Stalder}, {Schultz}, {Stubbs}, {Tonry}, {Waters},
  {Wainscoat}, {Della Valle}, {Dennefeld}, {Dimitriadis}, {Firth}, {Fraser},
  {Frohmaier}, {Gal-Yam}, {Harmanen}, {Kankare}, {Kotak}, {Kromer}, {Mandel},
  {Sollerman}, {Gibson}, {Primak}, \& {Willman}}]{smartt16b}
{Smartt}, S.~J., {Chambers}, K.~C., {Smith}, K.~W., {et~al.}
  2016{\natexlab{b}}, \apjl, 827, L40

\bibitem[{{Smartt} {et~al.}(2017){Smartt}, {Chen}, {Jerkstrand}, {Coughlin},
  {Kankare}, {Sim}, {Fraser}, {Inserra}, {Maguire}, {Chambers}, {Huber},
  {Kr{\"u}hler}, {Leloudas}, {Magee}, {Shingles}, {Smith}, {Young}, {Tonry},
  {Kotak}, {Gal-Yam}, {Lyman}, {Homan}, {Agliozzo}, {Anderson}, {Angus},
  {Ashall}, {Barbarino}, {Bauer}, {Berton}, {Botticella}, {Bulla}, {Bulger},
  {Cannizzaro}, {Cano}, {Cartier}, {Cikota}, {Clark}, {De Cia}, {Della Valle},
  {Denneau}, {Dennefeld}, {Dessart}, {Dimitriadis}, {Elias-Rosa}, {Firth},
  {Flewelling}, {Fl{\"o}rs}, {Franckowiak}, {Frohmaier}, {Galbany},
  {Gonz{\'a}lez-Gait{\'a}n}, {Greiner}, {Gromadzki}, {Guelbenzu},
  {Guti{\'e}rrez}, {Hamanowicz}, {Hanlon}, {Harmanen}, {Heintz}, {Heinze},
  {Hernandez}, {Hodgkin}, {Hook}, {Izzo}, {James}, {Jonker}, {Kerzendorf},
  {Klose}, {Kostrzewa-Rutkowska}, {Kowalski}, {Kromer}, {Kuncarayakti},
  {Lawrence}, {Lowe}, {Magnier}, {Manulis}, {Martin-Carrillo}, {Mattila},
  {McBrien}, {M{\"u}ller}, {Nordin}, {O'Neill}, {Onori}, {Palmerio},
  {Pastorello}, {Patat}, {Pignata}, {Podsiadlowski}, {Pumo}, {Prentice}, {Rau},
  {Razza}, {Rest}, {Reynolds}, {Roy}, {Ruiter}, {Rybicki}, {Salmon}, {Schady},
  {Schultz}, {Schweyer}, {Seitenzahl}, {Smith}, {Sollerman}, {Stalder},
  {Stubbs}, {Sullivan}, {Szegedi}, {Taddia}, {Taubenberger}, {Terreran}, {van
  Soelen}, {Vos}, {Wainscoat}, {Walton}, {Waters}, {Weiland}, {Willman},
  {Wiseman}, {Wright}, {Wyrzykowski}, \& {Yaron}}]{smartt17}
{Smartt}, S.~J., {Chen}, T.-W., {Jerkstrand}, A., {et~al.} 2017, \nat, 551, 75

\bibitem[{{Smartt} {et~al.}(2019{\natexlab{a}}){Smartt}, {Malesani}, {Smith},
  {Huber}, {Chambers}, {Schulz}, {Young}, {de Boer}, {Bulger}, \&
  {Fairlamb}}]{2019GCN.25386....1S}
{Smartt}, S.~J., {Malesani}, D.~B., {Smith}, K.~W., {et~al.}
  2019{\natexlab{a}}, GRB Coordinates Network, 25386, 1

\bibitem[{{Smartt} {et~al.}(2019{\natexlab{b}}){Smartt}, {Smith}, {Srivastav},
  {Chen}, {Huber}, {Chambers}, {Young}, {McBrien}, {Gillanders}, \&
  {O'Neill}}]{2019GCN.25455....1S}
{Smartt}, S.~J., {Smith}, K.~W., {Srivastav}, S., {et~al.} 2019{\natexlab{b}},
  GRB Coordinates Network, 25455, 1

\bibitem[{{Soares-Santos} {et~al.}(2019{\natexlab{a}}){Soares-Santos}, {Annis},
  {Garcia}, {Herner}, {Palmese}, {Tucker}, {Allam}, {Morgan}, {Bachmann}, \&
  {Sherman}}]{2019GCN.25425....1S}
{Soares-Santos}, M., {Annis}, J., {Garcia}, A., {et~al.} 2019{\natexlab{a}},
  GRB Coordinates Network, 25425, 1

\bibitem[{{Soares-Santos} {et~al.}(2019{\natexlab{b}}){Soares-Santos}, {Annis},
  {Garcia}, {Herner}, {Palmese}, {Tucker}, {Allam}, {Morgan}, {Bachmann}, \&
  {Sherman}}]{2019GCN.25438....1S}
{Soares-Santos}, M., {Annis}, J., {Garcia}, A., {et~al.} 2019{\natexlab{b}},
  GRB Coordinates Network, 25438, 1

\bibitem[{{Soares-Santos} {et~al.}(2019{\natexlab{c}}){Soares-Santos}, {Annis},
  {Herner}, {Palmese}, {Garcia}, {Tucker}, {Allam}, {Morgan}, {Bachmann}, \&
  {Sherman}}]{2019GCN.25486....1S}
{Soares-Santos}, M., {Annis}, J., {Herner}, K., {et~al.} 2019{\natexlab{c}},
  GRB Coordinates Network, 25486, 1

\bibitem[{{Soares-Santos} {et~al.}(2017){Soares-Santos}, {Holz}, {Annis},
  {Chornock}, {Herner}, {Berger}, {Brout}, {Chen}, {Kessler}, {Sako}, {Allam},
  {Tucker}, {Butler}, {Palmese}, {Doctor}, {Diehl}, {Frieman}, {Yanny}, {Lin},
  {Scolnic}, {Cowperthwaite}, {Neilsen}, {Marriner}, {Kuropatkin}, {Hartley},
  {Paz-Chinch{\'o}n}, {Alexander}, {Balbinot}, {Blanchard}, {Brown}, {Carlin},
  {Conselice}, {Cook}, {Drlica-Wagner}, {Drout}, {Durret}, {Eftekhari}, {Farr},
  {Finley}, {Foley}, {Fong}, {Fryer}, {Garc{\'{\i}}a-Bellido}, {Gill},
  {Gruendl}, {Hanna}, {Kasen}, {Li}, {Lopes}, {Louren{\c c}o}, {Margutti},
  {Marshall}, {Matheson}, {Medina}, {Metzger}, {Mu{\~n}oz}, {Muir}, {Nicholl},
  {Quataert}, {Rest}, {Sauseda}, {Schlegel}, {Secco}, {Sobreira}, {Stebbins},
  {Villar}, {Vivas}, {Walker}, {Wester}, {Williams}, {Zenteno}, {Zhang},
  {Abbott}, {Abdalla}, {Banerji}, {Bechtol}, {Benoit-L{\'e}vy}, {Bertin},
  {Brooks}, {Buckley-Geer}, {Burke}, {Carnero Rosell}, {Carrasco Kind},
  {Carretero}, {Castander}, {Crocce}, {Cunha}, {D{'}Andrea}, {da Costa},
  {Davis}, {Desai}, {Dietrich}, {Doel}, {Eifler}, {Fernandez}, {Flaugher},
  {Fosalba}, {Gaztanaga}, {Gerdes}, {Giannantonio}, {Goldstein}, {Gruen},
  {Gschwend}, {Gutierrez}, {Honscheid}, {Jain}, {James}, {Jeltema}, {Johnson},
  {Johnson}, {Kent}, {Krause}, {Kron}, {Kuehn}, {Kuhlmann}, {Lahav}, {Lima},
  {Maia}, {March}, {McMahon}, {Menanteau}, {Miquel}, {Mohr}, {Nichol}, {Nord},
  {Ogando}, {Petravick}, {Plazas}, {Romer}, {Roodman}, {Rykoff}, {Sanchez},
  {Scarpine}, {Schubnell}, {Sevilla-Noarbe}, {Smith}, {Smith}, {Suchyta},
  {Swanson}, {Tarle}, {Thomas}, {Thomas}, {Troxel}, {Vikram}, {Wechsler},
  {Weller}, {Dark Energy Survey}, \& {Dark Energy Camera GW-EM
  Collaboration}}]{soares-santo17}
{Soares-Santos}, M., {Holz}, D.~E., {Annis}, J., {et~al.} 2017, \apjl, 848, L16

\bibitem[{{Soares-Santos} {et~al.}(2019{\natexlab{d}}){Soares-Santos},
  {Tucker}, {Allam}, {Annis}, {Garcia}, {Herner}, {Davis}, {Sherman}, {Morgan},
  \& {Vivas}}]{2019GCN.25336....1S}
{Soares-Santos}, M., {Tucker}, D., {Allam}, S., {et~al.} 2019{\natexlab{d}},
  GRB Coordinates Network, 25336, 1

\bibitem[{{Song} {et~al.}(2019){Song}, {Ai}, {Wang}, {Xing}, {Gao}, \&
  {Zhang}}]{song19}
{Song}, H.-R., {Ai}, S.-K., {Wang}, M.-H., {et~al.} 2019, \apjl, 881, L40

\bibitem[{{Srivastas} {et~al.}(2019){Srivastas}, {Huber}, {Smartt}, {Smith},
  {Chambers}, {Young}, {McBrien}, {Gillanders}, {O'Neill}, \&
  {Clark}}]{2019GCN.25417....1S}
{Srivastas}, S., {Huber}, M., {Smartt}, S.~J., {et~al.} 2019, GRB Coordinates
  Network, 25417, 1

\bibitem[{{Stalder} {et~al.}(2017){Stalder}, {Tonry}, {Smartt}, {Coughlin},
  {Chambers}, {Stubbs}, {Chen}, {Kankare}, {Smith}, {Denneau}, {Sherstyuk},
  {Heinze}, {Weiland}, {Rest}, {Young}, {Huber}, {Flewelling}, {Lowe},
  {Magnier}, {Schultz}, {Waters}, {Wainscoat}, {Willman}, {Wright}, {Chu},
  {Sanders}, {Inserra}, {Maguire}, \& {Kotak}}]{stalder17}
{Stalder}, B., {Tonry}, J., {Smartt}, S.~J., {et~al.} 2017, \apj, 850, 149

\bibitem[{{Steele} {et~al.}(2004){Steele}, {Smith}, {Rees}, {Baker}, {Bates},
  {Bode}, {Bowman}, {Carter}, {Etherton}, {Ford}, {Fraser}, {Gomboc}, {Lett},
  {Mansfield}, {Marchant}, {Medrano-Cerda}, {Mottram}, {Raback}, {Scott},
  {Tomlinson}, \& {Zamanov}}]{2004SPIE.5489..679S}
{Steele}, I.~A., {Smith}, R.~J., {Rees}, P.~C., {et~al.} 2004, {The Liverpool
  Telescope: performance and first results}, Vol. 5489 (Society of
  Photo-Optical Instrumentation Engineers (SPIE) Conference Series), 679--692

\bibitem[{{Steiner} {et~al.}(2013){Steiner}, {Hempel}, \&
  {Fischer}}]{Steiner.etal:2013}
{Steiner}, A.~W., {Hempel}, M., \& {Fischer}, T. 2013, \apj, 774, 17

\bibitem[{{Stewart} {et~al.}(2019){Stewart}, {Dobie}, {Murphy}, {Lenc}, {Wang},
  {Kaplan}, {Hotan}, \& {Moss}}]{2019GCN.25487....1S}
{Stewart}, A., {Dobie}, D., {Murphy}, T., {et~al.} 2019, GRB Coordinates
  Network, 25487, 1

\bibitem[{{Stone} {et~al.}(2017){Stone}, {Metzger}, \&
  {Haiman}}]{Stoneetal2017}
{Stone}, N.~C., {Metzger}, B.~D., \& {Haiman}, Z. 2017, \mnras, 464, 946

\bibitem[{{Sugizaki} {et~al.}(2019){Sugizaki}, {Kawai}, {Negoro}, {Sugita},
  {Serino}, {Nakajima}, {Maruyama}, {Aoki}, {Kobayashi}, \&
  {Mihara}}]{2019GCN.25329....1S}
{Sugizaki}, M., {Kawai}, N., {Negoro}, H., {et~al.} 2019, GRB Coordinates
  Network, 25329, 1

\bibitem[{{Sutherland} {et~al.}(2015){Sutherland}, {Emerson}, {Dalton},
  {Atad-Ettedgui}, {Beard}, {Bennett}, {Bezawada}, {Born}, {Caldwell}, {Clark},
  {Craig}, {Henry}, {Jeffers}, {Little}, {McPherson}, {Murray}, {Stewart},
  {Stobie}, {Terrett}, {Ward}, {Whalley}, \& {Woodhouse}}]{sutherland2015}
{Sutherland}, W., {Emerson}, J., {Dalton}, G., {et~al.} 2015, \aap, 575, A25

\bibitem[{{Svinkin} {et~al.}(2019){Svinkin}, {Golenetskii}, {Aptekar},
  {Frederiks}, {Ulanov}, {Tsvetkova}, {Lysenko}, {Kozlova}, \&
  {Cline}}]{2019GCN.25369....1S}
{Svinkin}, D., {Golenetskii}, S., {Aptekar}, R., {et~al.} 2019, GRB Coordinates
  Network, 25369, 1

\bibitem[{{Tanaka} {et~al.}(2014){Tanaka}, {Hotokezaka}, {Kyutoku}, {Wanajo},
  {Kiuchi}, {Sekiguchi}, \& {Shibata}}]{Tanakaetal2014}
{Tanaka}, M., {Hotokezaka}, K., {Kyutoku}, K., {et~al.} 2014, \apj, 780, 31

\bibitem[{{Tanaka} {et~al.}(2019){Tanaka}, {Kato}, {Gaigalas}, \&
  {Kawaguchi}}]{Tanaka2019arXiv}
{Tanaka}, M., {Kato}, D., {Gaigalas}, G., \& {Kawaguchi}, K. 2019, arXiv
  e-prints, arXiv:1906.08914

\bibitem[{{Tanvir} {et~al.}(2013){Tanvir}, {Levan}, {Fruchter}, {Hjorth},
  {Hounsell}, {Wiersema}, \& {Tunnicliffe}}]{tanvir13}
{Tanvir}, N.~R., {Levan}, A.~J., {Fruchter}, A.~S., {et~al.} 2013, \nat, 500,
  547

\bibitem[{{Tanvir} {et~al.}(2017){Tanvir}, {Levan},
  {Gonz{\'a}lez-Fern{\'a}ndez}, {Korobkin}, {Mandel}, {Rosswog}, {Hjorth},
  {D'Avanzo}, {Fruchter}, {Fryer}, {Kangas}, {Milvang-Jensen}, {Rosetti},
  {Steeghs}, {Wollaeger}, {Cano}, {Copperwheat}, {Covino}, {D'Elia}, {de Ugarte
  Postigo}, {Evans}, {Even}, {Fairhurst}, {Figuera Jaimes}, {Fontes}, {Fujii},
  {Fynbo}, {Gompertz}, {Greiner}, {Hodosan}, {Irwin}, {Jakobsson},
  {J{\o}rgensen}, {Kann}, {Lyman}, {Malesani}, {McMahon}, {Melandri},
  {O'Brien}, {Osborne}, {Palazzi}, {Perley}, {Pian}, {Piranomonte}, {Rabus},
  {Rol}, {Rowlinson}, {Schulze}, {Sutton}, {Th{\"o}ne}, {Ulaczyk}, {Watson},
  {Wiersema}, \& {Wijers}}]{tanvir17}
{Tanvir}, N.~R., {Levan}, A.~J., {Gonz{\'a}lez-Fern{\'a}ndez}, C., {et~al.}
  2017, \apjl, 848, L27

\bibitem[{{The LIGO Scientific Collaboration and the Virgo
  Collaboration}(2019{\natexlab{a}})}]{2019GCN.25324....1L}
{The LIGO Scientific Collaboration and the Virgo Collaboration}.
  2019{\natexlab{a}}, GRB Coordinates Network, 25324, 1

\bibitem[{{The LIGO Scientific Collaboration and the Virgo
  Collaboration}(2019{\natexlab{b}})}]{2019GCN.25333....1T}
{The LIGO Scientific Collaboration and the Virgo Collaboration}.
  2019{\natexlab{b}}, GRB Coordinates Network, 25333, 1

\bibitem[{{Tonry} {et~al.}(2018{\natexlab{a}}){Tonry}, {Denneau}, {Flewelling},
  {Heinze}, {Onken}, {Smartt}, {Stalder}, {Weiland}, \& {Wolf}}]{refcat2}
{Tonry}, J.~L., {Denneau}, L., {Flewelling}, H., {et~al.} 2018{\natexlab{a}},
  \apj, 867, 105

\bibitem[{{Tonry} {et~al.}(2018{\natexlab{b}}){Tonry}, {Denneau}, {Heinze},
  {Stalder}, {Smith}, {Smartt}, {Stubbs}, {Weiland}, \& {Rest}}]{tonry18}
{Tonry}, J.~L., {Denneau}, L., {Heinze}, A.~N., {et~al.} 2018{\natexlab{b}},
  \pasp, 130, 064505

\bibitem[{{Tonry} {et~al.}(2012){Tonry}, {Stubbs}, {Lykke}, {Doherty},
  {Shivvers}, {Burgett}, {Chambers}, {Hodapp}, {Kaiser}, {Kudritzki},
  {Magnier}, {Morgan}, {Price}, \& {Wainscoat}}]{tonry12}
{Tonry}, J.~L., {Stubbs}, C.~W., {Lykke}, K.~R., {et~al.} 2012, \apj, 750, 99

\bibitem[{{Troja} {et~al.}(2019){Troja}, {Castro-Tirado}, {Becerra
  Gonz{\'a}lez}, {Hu}, {Ryan}, {Cenko}, {Ricci}, {Novara},
  {S{\'a}nchez-R{\'a}mirez}, {Acosta-Pulido}, {Ackley}, {Caballero
  Garc{\'\i}a}, {Eikenberry}, {Guziy}, {Jeong}, {Lien}, {M{\'a}rquez}, {Pand
  ey}, {Park}, {Sakamoto}, {Tello}, {Sokolov}, {Sokolov}, {Tiengo}, {Valeev},
  {Zhang}, \& {Veilleux}}]{Troja2019}
{Troja}, E., {Castro-Tirado}, A.~J., {Becerra Gonz{\'a}lez}, J., {et~al.} 2019,
  \mnras, 489, 2104

\bibitem[{{Troja} {et~al.}(2018{\natexlab{a}}){Troja}, {Piro}, {Ryan}, {van
  Eerten}, {Ricci}, {Wieringa}, {Lotti}, {Sakamoto}, \&
  {Cenko}}]{Troja2018MNRAS}
{Troja}, E., {Piro}, L., {Ryan}, G., {et~al.} 2018{\natexlab{a}}, \mnras, 478,
  L18

\bibitem[{{Troja} {et~al.}(2017){Troja}, {Piro}, {van Eerten}, {Wollaeger},
  {Im}, {Fox}, {Butler}, {Cenko}, {Sakamoto}, {Fryer}, {Ricci}, {Lien}, {Ryan},
  {Korobkin}, {Lee}, {Burgess}, {Lee}, {Watson}, {Choi}, {Covino}, {D'Avanzo},
  {Fontes}, {Gonz{\'a}lez}, {Khandrika}, {Kim}, {Kim}, {Lee}, {Lee}, {Kutyrev},
  {Lim}, {S{\'a}nchez-Ram{\'{\i}}rez}, {Veilleux}, {Wieringa}, \&
  {Yoon}}]{troja17}
{Troja}, E., {Piro}, L., {van Eerten}, H., {et~al.} 2017, \nat, 551, 71

\bibitem[{{Troja} {et~al.}(2018{\natexlab{b}}){Troja}, {Ryan}, {Piro}, {van
  Eerten}, {Cenko}, {Yoon}, {Lee}, {Im}, {Sakamoto}, {Gatkine}, {Kutyrev}, \&
  {Veilleux}}]{troja18b}
{Troja}, E., {Ryan}, G., {Piro}, L., {et~al.} 2018{\natexlab{b}}, Nature
  Communications, 9, 4089

\bibitem[{{Tucker} {et~al.}(2019{\natexlab{a}}){Tucker}, {Allam}, {Wiesner},
  {Cartier}, {Garcia}, {Palmese}, {Zenteno}, {Herner}, {Soares-Santos}, \&
  {Annis}}]{2019GCN.25379....1T}
{Tucker}, D., {Allam}, S., {Wiesner}, M., {et~al.} 2019{\natexlab{a}}, GRB
  Coordinates Network, 25379, 1

\bibitem[{{Tucker} {et~al.}(2019{\natexlab{b}}){Tucker}, {Butner}, {Wiesner},
  {Rodriguez}, {Meza-Retamal}, {Quirola}, {Olivares}, {Points}, {Cartier}, \&
  {Allam}}]{2019GCN.25484....1T}
{Tucker}, D., {Butner}, M., {Wiesner}, M., {et~al.} 2019{\natexlab{b}}, GRB
  Coordinates Network, 25484, 1

\bibitem[{{Tunnicliffe} {et~al.}(2014){Tunnicliffe}, {Levan}, {Tanvir},
  {Rowlinson}, {Perley}, {Bloom}, {Cenko}, {O'Brien}, {Cobb}, {Wiersema},
  {Malesani}, {de Ugarte Postigo}, {Hjorth}, {Fynbo}, \&
  {Jakobsson}}]{tunnicliffe14}
{Tunnicliffe}, R.~L., {Levan}, A.~J., {Tanvir}, N.~R., {et~al.} 2014, \mnras,
  437, 1495

\bibitem[{{Typel} {et~al.}(2010){Typel}, {R{\"o}pke}, {Kl{\"a}hn}, {Blaschke},
  \& {Wolter}}]{Typel.etal:2010}
{Typel}, S., {R{\"o}pke}, G., {Kl{\"a}hn}, T., {Blaschke}, D., \& {Wolter},
  H.~H. 2010, \prc, 81, 015803

\bibitem[{{Utsumi} {et~al.}(2017){Utsumi}, {Tanaka}, {Tominaga}, {Yoshida},
  {Barway}, {Nagayama}, {Zenko}, {Aoki}, {Fujiyoshi}, {Furusawa}, {Kawabata},
  {Koshida}, {Lee}, {Morokuma}, {Motohara}, {Nakata}, {Ohsawa}, {Ohta},
  {Okita}, {Tajitsu}, {Tanaka}, {Terai}, {Yasuda}, {Abe}, {Asakura}, {Bond},
  {Miyazaki}, {Sumi}, {Tristram}, {Honda}, {Itoh}, {Itoh}, {Kawabata},
  {Morihana}, {Nagashima}, {Nakaoka}, {Ohshima}, {Takahashi}, {Takayama},
  {Aoki}, {Baar}, {Doi}, {Finet}, {Kanda}, {Kawai}, {Kim}, {Kuroda}, {Liu},
  {Matsubayashi}, {Murata}, {Nagai}, {Saito}, {Saito}, {Sako}, {Sekiguchi},
  {Tamura}, {Tanaka}, {Uemura}, \& {Yamaguchi}}]{utsumi17}
{Utsumi}, Y., {Tanaka}, M., {Tominaga}, N., {et~al.} 2017, \pasj, 69, 101

\bibitem[{{Valenti} {et~al.}(2017){Valenti}, {David}, {Sand}, {Yang},
  {Cappellaro}, {Tartaglia}, {Corsi}, {Jha}, {Reichart}, {Haislip}, \&
  {Kouprianov}}]{valenti17}
{Valenti}, S., {David}, {Sand}, J., {et~al.} 2017, \apjl, 848, L24

\bibitem[{{Veitch} {et~al.}(2015){Veitch}, {Raymond}, {Farr}, {Farr}, {Graff},
  {Vitale}, {Aylott}, {Blackburn}, {Christensen}, {Coughlin}, {Del Pozzo},
  {Feroz}, {Gair}, {Haster}, {Kalogera}, {Littenberg}, {Mandel},
  {O'Shaughnessy}, {Pitkin}, {Rodriguez}, {R{\"o}ver}, {Sidery}, {Smith}, {Van
  Der Sluys}, {Vecchio}, {Vousden}, \& {Wade}}]{veitch15}
{Veitch}, J., {Raymond}, V., {Farr}, B., {et~al.} 2015, \prd, 91, 042003

\bibitem[{{V{\'e}ron-Cetty} \& {V{\'e}ron}(2001)}]{veron01}
{V{\'e}ron-Cetty}, M.~P. \& {V{\'e}ron}, P. 2001, \aap, 374, 92

\bibitem[{{Vieira} {et~al.}(2020){Vieira}, {Ruan}, {Haggard}, {Drout}, {Nynka},
  {Boyce}, {Spekkens}, {Safi-Harb}, {Carlberg}, {Fern{\'a}ndez}, {Piro},
  {Afsariardchi}, \& {Moon}}]{Vieiraetal2020}
{Vieira}, N., {Ruan}, J.~J., {Haggard}, D., {et~al.} 2020, arXiv e-prints,
  arXiv:2003.09437

\bibitem[{{Waters} {et~al.}(2016){Waters}, {Magnier}, {Price}, {Chambers},
  {Burgett}, {Draper}, {Flewelling}, {Hodapp}, {Huber}, {Jedicke}, {Kaiser},
  {Kudritzki}, {Lupton}, {Metcalfe}, {Rest}, {Sweeney}, {Tonry}, {Wainscoat},
  {Wood-Vasey}, \& {Builders}}]{waters16}
{Waters}, C.~Z., {Magnier}, E.~A., {Price}, P.~A., {et~al.} 2016, arXiv
  e-prints, arXiv:1612.05245

\bibitem[{{Watson} {et~al.}(2020){Watson}, {Butler}, {Lee}, {Becerra},
  {Pereyra}, {Angeles}, {Farah}, {Figueroa}, {Gonz{\'a}lez-Buitrago},
  {Quir{\'o}s}, {Ru{\'\i}z-D{\'\i}az-Soto}, {Tejada de Vergas}, {Tinoco}, \&
  {Wolfram}}]{Watson2020arXiv}
{Watson}, A.~M., {Butler}, N.~R., {Lee}, W.~H., {et~al.} 2020, \mnras, in press
  (arXiv:2001.05436), arXiv:2001.05436

\bibitem[{{Watson} {et~al.}(2019){Watson}, {Hansen}, {Selsing}, {Koch},
  {Malesani}, {Andersen}, {Fynbo}, {Arcones}, {Bauswein}, {Covino}, {Grado},
  {Heintz}, {Hunt}, {Kouveliotou}, {Leloudas}, {Levan}, {Mazzali}, \&
  {Pian}}]{watson19}
{Watson}, D., {Hansen}, C.~J., {Selsing}, J., {et~al.} 2019, \nat, 574, 497

\bibitem[{{Wiesner} {et~al.}(2019{\natexlab{a}}){Wiesner}, {Butner}, {Allam},
  {Tucker}, {Wood}, {Mann}, {Rodriguez}, {Meza-Retamal}, {Quirola}, \&
  {Olivares}}]{2019GCN.25596....1W}
{Wiesner}, M., {Butner}, M., {Allam}, S., {et~al.} 2019{\natexlab{a}}, GRB
  Coordinates Network, 25596, 1

\bibitem[{{Wiesner} {et~al.}(2019{\natexlab{b}}){Wiesner}, {Butner}, {Tucker},
  {Rodriguez}, {Meza-Retamal}, {Quirola}, {Olivares}, {Cartier}, {Points}, \&
  {Allam}}]{2019GCN.25540....1W}
{Wiesner}, M., {Butner}, M., {Tucker}, D., {et~al.} 2019{\natexlab{b}}, GRB
  Coordinates Network, 25540, 1

\bibitem[{{Willmer}(2018)}]{Willmer2018}
{Willmer}, C. N.~A. 2018, \apjs, 236, 47

\bibitem[{{Wollaeger} {et~al.}(2018){Wollaeger}, {Korobkin}, {Fontes},
  {Rosswog}, {Even}, {Fryer}, {Sollerman}, {Hungerford}, {van Rossum}, \&
  {Wollaber}}]{Wollaegeretal2018}
{Wollaeger}, R.~T., {Korobkin}, O., {Fontes}, C.~J., {et~al.} 2018, \mnras,
  478, 3298

\bibitem[{{Yang} {et~al.}(2015){Yang}, {Jin}, {Li}, {Covino}, {Zheng},
  {Hotokezaka}, {Fan}, {Piran}, \& {Wei}}]{yang15}
{Yang}, B., {Jin}, Z.-P., {Li}, X., {et~al.} 2015, Nature Communications, 6,
  7323

\bibitem[{{Yang}(2018)}]{grawitaml}
{Yang}, S. 2018, Ph.D. thesis, http://paduaresearch.cab.unipd.it/11854/1/phd

\bibitem[{{Yang} {et~al.}(2019){Yang}, {Cappellaro}, {Grado}, {Brocato},
  {Covino}, {Getman}, {Greco}, {Rossi}, {Palazzi}, \&
  {Melandri}}]{2019GCN.25748....1Y}
{Yang}, S., {Cappellaro}, E., {Grado}, A., {et~al.} 2019, GRB Coordinates
  Network, 25748, 1

\bibitem[{{Zappa} {et~al.}(2019){Zappa}, {Bernuzzi}, {Pannarale}, {Mapelli}, \&
  {Giacobbo}}]{Zappaetal2019}
{Zappa}, F., {Bernuzzi}, S., {Pannarale}, F., {Mapelli}, M., \& {Giacobbo}, N.
  2019, \prl, 123, 041102

\end{thebibliography}

$^{1}$ School of Physics and Astronomy, Monash University, Clayton, Victoria 3800, Australia\\
$^{2}$ INAF - Osservatorio di Astrofisica e Scienza dello Spazio di Bologna, via Piero Gobetti 93/3, 40129 Bologna, Italy\\
$^{3}$ INAF / Brera Astronomical Observatory, via Bianchi 46, 23807, Merate (LC), Italy\\
$^{4}$ INFN - Sezione di Milano-Bicocca, Piazza della Scienza 3, I-20126 Milano (MI), Italy\\
$^{5}$ University of Milano-Bicocca, Department of Physics "G. Occhialini", Piazza della Scienza 3, I-20126 Milano, Italy\\
$^{6}$ Instituto de Astrof{\'{\i}}sica and Centro de Astroingenier{\'{\i}}a, Facultad de F{\'{i}}sica, Pontificia Uni versidad Cat{\'{o}}lica de Chile, Casilla 306, Santiago 22, Chile\\
$^{7}$ Millennium Institute of Astrophysics (MAS), Nuncio Monse{\~{n}}or S{\'{o}}tero Sanz 100, Providencia, Santiago , Chile\\
$^{8}$ Space Science Institute, 4750 Walnut Street, Suite 205, Boulder, Colorado 80301, USA\\
$^{9}$ INAF, Osservatorio Astronomico di Padova, I-35122 Padova, Italy\\
$^{10}$ Space Telescope Science Institute, 3700 San Martin Drive, Baltimore, MD 21218, USA\\
$^{11}$ INAF Osservatorio Astronomico di Capodimonte, Via Moiariello 16, 80131 Napoli, Italy\\
$^{12}$ Gran Sasso Science Institute, Viale F. Crispi 7, I-67100, L'Aquila (AQ), Italy\\
$^{13}$ INFN - Laboratori Nazionali del Gran Sasso, I-67100, L'Aquila (AQ), Italy\\
$^{14}$ INAF, Osservatorio Astronomico di Roma, Via di Frascati 33, 00078 Monteporzio Catone (RM), Italy\\
$^{15}$ INAF - Osservatorio Astronomico d'Abruzzo, Via M. Maggini s.n.c.~, I-64100 Teramo, Italy\\
$^{16}$ DARK, Niels Bohr Institute, University of Copenhagen, Lyngbyvej 2, 2100 Copenhagen \O, Denmark\\
$^{17}$ Nordita, KTH Royal Institute of Technology and Stockholm University, Roslagstullsbacken 23, SE-106 91 Stockholm, Sweden\\
$^{18}$ Instituto de Astrof\'isica de Andaluc\'ia (IAA-CSIC), Glorieta de la Astronom\'ia s/n, 18008 Granada, Spain\\
$^{19}$ Institute for Astronomy, University of Hawai'i, 2680 Woodlawn Drive, Honolulu, HI 96822, USA\\
$^{20}$ APC, Univ Paris Diderot, CNRS/IN2P3, CEA/Irfu, Obs de Paris, Sorbonne Paris Cit\'e, France\\
$^{21}$ AIM, CEA, CNRS, Universit\'e Paris-Saclay, Universit\'e Paris Diderot, Sorbonne Paris Cit\'e, F-91191 Gif-sur-Yvette, France\\
$^{22}$ The Oskar Klein Centre, Department of Astronomy, Stockholm University, AlbaNova, SE-10691 Stockholm, Sweden\\
$^{23}$ Max-Planck-Institut f{\"u}r Extraterrestrische Physik, Giessenbachstra\ss e 1, 85748, Garching, Germany\\
$^{24}$ INFN, Sezione di Padova, I-35131 Padova, Italy\\
$^{25}$ Astrophysics Research Institute, IC2 building, Liverpool Science Park, 146 Brownlow Hill, Liverpool L3 5RF, UK\\
$^{26}$ Department of Physics, University of Warwick, Coventry, CV4 7AL, UK\\
$^{27}$ INAF - Istituto di radioastronomia Bologna, Italy\\
$^{28}$ ASI Science Data Centre, Via del Politecnico snc, 00133 Rome, Italy\\
$^{29}$ Faculty of Science, Department of Astronomy and Space Sciences, Istanbul University, Beyaz\i t, 34119, Istanbul, Turkey\\
$^{30}$ Department of Physics and Astronomy, University of Sheffield, Sheffield, S3 7RH, UK\\
$^{31}$ Instituto de Astrof\'{i}sica de Canarias, E-38205 La Laguna, Tenerife, Spain\\
$^{32}$ Institute of Space Sciences (ICE, CSIC), Campus UAB, Carrer de Can Magrans s/n, 08193 Barcelona, Spain\\
$^{33}$ School of Physics and Astronomy, University of Leicester, University Road, LE1 7RH, UK\\
$^{34}$ Physics and Astronomy Department Galileo Galilei, University of Padova, Italy\\
$^{35}$ School of Physics, O'Brien Centre for Science North, University College Dublin, Belfield, Dublin 4, Ireland\\
$^{36}$ Cosmic Dawn Center (DAWN)\\
$^{37}$ Niels Bohr Institute, University of Copenhagen, Lyngbyvej 2, 2100 Copenhagen \O, Denmark\\
$^{38}$ Departamento de F\'isica Te\'orica y del Cosmos, Universidad de Granada, E-18071 Granada, Spain\\
$^{39}$ Astrophysics Research Centre, School of Mathematics and Physics, Queen's University Belfast, BT7 1NN, UK\\
$^{40}$ University of Nova Gorica, Center for Astrophysics and Cosmology, Vipavska 13, 5000 Nova Gorica, Slovenia\\
$^{41}$ Institute of Astronomy, University of Cambridge, Madingley Road, Cambridge, CB3 0HA, UK\\
$^{42}$ CENTRA-Centro de Astrof\'{\i}isica e Gravita\c{c}\~ao and Departamento de F\'{\i}sica, Instituto Superior T\'ecnico, Universidade de Lisboa, Avenida Rovisco Pais, 1049-001 Lisboa, Portugal\\
$^{43}$ Universit\`a degli Studi di Urbino `Carlo Bo', I-61029 Urbino, Italy\\
$^{44}$ INFN, Sezione di Firenze, I-50019 Sesto Fiorentino, Firenze, Italy\\
$^{45}$ Astronomical Observatory, University of Warsaw, Al. Ujazdowskie 4, 00-478 Warszawa, Poland\\
$^{46}$ Department of Astrophysics/IMAPP, Radboud University, P.O.~Box 9010, 6500 GL, Nijmegen, The Netherlands\\
$^{47}$ Department of Astronomy, University of Cape Town, Private Bag X3, Rondebosch, 7701, South Africa\\
$^{48}$ South African Astronomical Observatory, P.O. Box 9, Observatory, 7935, South Africa\\
$^{49}$ The Inter-University Institute for Data Intensive Astronomy, University of Cape Town, Private Bag X3, Rondebosch, 7701, South Africa\\
$^{50}$ Department of Physics and Astronomy, University of Southampton, Southampton, SO17 1BJ, UK\\
$^{51}$ Department of Physics and Astronomy, University of Turku, Vesilinnantie 5, Turku, FI-20014, Finland\\
$^{52}$ Centre for Astrophysics and Cosmology, Science Institute, University of Iceland, Dunhagi 5, 107 Reykjav\'ik, Iceland\\
$^{53}$ School of Physics \& Astronomy, Cardiff University, Queens Buildings, The Parade, Cardiff, CF24 3AA, UK\\
$^{54}$ Anton Pannekoek Institute for Astronomy, University of Amsterdam, Science Park 904, 1098 XH Amsterdam, The Netherlands\\
$^{55}$ Key Laboratory of Dark Matter and Space Astronomy, Purple Mountain Observatory, Chinese Academy of Sciences, Nanjing 210008, China\\
$^{56}$ SRON, Netherlands Institute for Space Research, Sorbonnelaan 2, 3584~CA, Utrecht, The Netherlands\\
$^{57}$ School of Physics and Astronomy, The University of Manchester, Manchester, M13 9PL, UK\\
$^{58}$ Max-Planck-Institut f{\"u}r Astronomie K{\"o}nigstuhl 17, D-69117, Heidelberg, Germany\\
$^{59}$ Th\"uringer Landessternwarte Tautenburg, Sternwarte 5, 07778 Tautenburg, Germany\\
$^{60}$ Finnish Centre for Astronomy with ESO (FINCA), FI-20014 University of Turku, Finland\\
$^{61}$ DTU Space, National Space Institute, Technical University of Denmark, Elektrovej 327, 2800 Kongens Lyngby, Denmark\\
$^{62}$ Universit\`{a} degli Studi di Trieste and INFN, sezione di Trieste, I-34127 Trieste, Italy\\
$^{63}$ School of Physics, Trinity College Dublin, University of Dublin, College Green, Dublin 2, Ireland\\
$^{64}$ Institute for Gravitational Wave Astronomy and School of Physics and Astronomy, University of Birmingham, Edgbaston, Birmingham B15 2TT, UK\\
$^{65}$ Astronomical Observatory Institute, Faculty of Physics, Adam Mickiewicz University, ul.~S{\l}oneczna 36, 60-286 Pozna{\'n}, Poland\\
$^{66}$ Institute of Cosmology and Gravitation, University of Portsmouth, Portsmouth, PO1 3FX, UK\\
$^{67}$ Istituto di Astrofisica e Planetologia Spaziali (INAF), via del Fosso del Cavaliere 100, Roma, I-00133, Italy\\
$^{68}$ Universit\'a di Pisa, Largo B. Pontecorvo 3, I-56127 Pisa, Italy\\
$^{69}$ INFN - Sezione di Pisa, Largo B. Pontecorvo 3, I-56127 Pisa, Italy\\
$^{70}$ Department of Physics, Trento University, Via Sommarive 14, 38123 Povo, Trento, Italy\\
$^{71}$ Departamento de Astrof\'\i{}sica, Universidad de La Laguna, E-38206 La Laguna, Tenerife, Spain\\
$^{72}$ Departamento de Ciencias Fisicas, Universidad Andres Bello, Avda. Republica 252, Santiago, Chile\\
$^{73}$ National Astronomical Research Institute of Thailand, Ministry of Science and Technology, Chiang Mai 50180, Thailand\\
$^{74}$ INAF - Osservatorio Astronomico di Cagliari, via della Scienza 5, 09047 Selargius (CA), Italy\\
$^{75}$ University of Cagliari, Dept of Physics, S.P. Monserrato-Sestu Km 0,700 - 09042 Monserrato (CA), Italy\\
$^{76}$ Dipartimento di Fisica e Astronomia ``E. Majorana'', Universit\`a degli studi di Catania, Via Santa Sofia 64, I-95123 Catania, Italy\\
$^{77}$ INFN - Laboratori Nazionali del Sud, Via Santa Sofia 62, I-95123 Catania, Italy\\
$^{78}$ INFN, Sezione di Napoli, Complesso Universitario di Monte S. Angelo, Via Cintia Edificio 6, 80126 Napoli, Italy\\
$^{79}$ Department of Physics, Via Cinthia, I-80126 Fuorigrotta, Naples, Italy\\
$^{80}$ Armagh Observatory and Planetarium, Armagh, BT61 9DG, UK\\
$^{81}$ Department of Physics and Astronomy, Johns Hopkins University, Baltimore, MD 21218, USA\\
$^{82}$ Institut f\"{u}r Astro- und Teilchenphysik, Universit\"{a}t Innsbruck, Technikerstrasse 25/8, 6020 Innsbruck, Austria\\
$^{83}$ The Oskar Klein Centre, Department of Physics, Stockholm University, AlbaNova, SE-10691 Stockholm, Sweden\\
$^{84}$ INAF - Istituto di Astrofisica Spaziale e Fisica Cosmica di Milano, via A. Corti 12, I-20133 Milano, Italy\\
$^{85}$ Physics Department, University of Calabria, via P. Bucci, 87036 Rende, Italy\\
$^{86}$ European Southern Observatory, Alonso de C\'ordova, 3107, Vitacura, Santiago 763-0355, Chile\\
$^{87}$ Department of Physics, University of Bath, Bath BA2 7AY, UK\\
$^{88}$ Department of Physics, Harvard University, Cambridge, MA 02138, USA\\
$^{89}$ Department of Physics, The George Washington University, 725 21st Street NW, Washington, DC 20052, USA\\
$^{90}$ Astronomy, Physics, and Statistics Institute of Sciences (APSIS), The George Washington University, Washington, DC 20052, USA\\
$^{91}$ GEPI, Observatoire de Paris, PSL University, CNRS, 5 Place Jules Janssen, 92190 Meudon, France\\

\appendix

\section{From limiting magnitudes to limits in the model parameter space}\label{sec:from_maglims_to_param_lims}

For an EM counterpart model defined by an intrinsic light curve $dL/d\nu(\nu,t)$ (specific luminosity at a given rest-frame frequency $\nu$, as a function of rest-frame post-merger time $t$), one can in principle use the galaxy-targeted and wide-field observations to exclude the presence of such emission in a given galaxy. We define here a framework that allows us to combine the results of different searches with a heterogeneous range of telescopes. We work under the simplifying assumption that each observation has a well-defined limiting flux, above which we can exclude a detection with high confidence.

A galaxy-targeted search consists of a set of observations of $N_\mathrm{gal}$ galaxies, each observed $N_\mathrm{obs,i}$ times ($i$ here runs on the galaxies). Each observation takes place at a time $t_{i,j}$ post-merger, and reaches a limiting flux $F_\mathrm{lim,i,j}$ in a band whose central frequency is $\nu_{i,j}$. We can exclude that galaxy $i$ hosted the putative EM counterpart as long as 
\begin{equation}
F_{\mathrm{lim},i,j} < F_{\mathrm{model},i,j} = \frac{(1+z_i)}{4\pi d_{\mathrm{L},i}^2}\frac{dL}{d\nu}\left((1+z_i)\nu_{i,j},\frac{t_{i,j}}{(1+z_i)}\right),
\label{eq:galaxy_host_exclusion_condition}
\end{equation}
for any index $j$ running over the $N_{\mathrm{obs},i}$ observations of that galaxy. In that case, the observations contribute a total of $P_{\mathrm{gal},i}$ to the confidence at which the particular EM counterpart model can be excluded. Formally, calling $\xi$ the set of parameters and assumptions that define a particular EM counterpart model, we can exclude $\xi$ with a confidence defined by
\begin{equation}
    1-P(\xi) = \sum_{i=1}^{N_\mathrm{gal}}P_{\mathrm{gal},i}\mathcal{E}_{i}(\xi)
    \label{eq:galaxy-targeted_exclusion_confidence_level},
\end{equation}
with 
\begin{equation}
    \mathcal{E}_{i}(\xi)=1-\Pi_{j=1}^{N_{\mathrm{obs},i}}H\left(F_{\mathrm{lim},i,j}-F_{\mathrm{model}}(\xi,\nu_{i,j},t_{i,j})\right),
\end{equation}
where $H$ is the Heaviside step function. The quantity $\mathcal{E}_{i,j}(\xi)$ is $1$ if at least one observation of galaxy $i$ is constraining (i.e.~it satisfies inequality reported in Eq.~\ref{eq:galaxy_host_exclusion_condition}), and $0$ otherwise. In the absence of a detection, the quantity defined by Eq.~\ref{eq:galaxy-targeted_exclusion_confidence_level} is most commonly referred to as the ``covered probability'' with respect to a particular source model.

In the case of a wide-field search, one can define the corresponding exclusion confidence as\footnote{For wide-field observations we assume the localisation probability density to follow the GW 3D skymap, i.e.~we do not distribute such probability to galaxies as in the galaxy-targeted search case. This relies on the assumption that the galaxy density averages out on scales as large as those probed by wide-field observations. Figure~\ref{fig:Integrated_Pgal_P3D_comparison} shows that this is a good approximation in our case, and this removes the uncertainty on the catalogue completeness.}
\begin{equation}
    1-P(\xi)=\sum_{i=1}^{N_\mathrm{tiles}}P_{\mathrm{tile},i}\int_0^\infty dr \frac{dP}{dr}\mathcal{E}_{i}(r,\xi)
    \label{eq:wide-field_exclusion_confidence_level},
\end{equation}
with
\begin{equation}
    \mathcal{E}_{i}(r,\xi) = 1-\Pi_{j=1}^{N_{\mathrm{obs},i}}H\left(F_{\mathrm{lim},i,j}-F_{\mathrm{model}}(r,\xi,\nu_{i,j},t_{i,j})\right).
\end{equation}
Here the sum runs over a number $N_\mathrm{tiles}$ of non-overlapping sky tiles (e.g.~a healpix tessellation), each observed $N_{\mathrm{obs},i}$ times, at epochs $t_{i,j}$, with limiting fluxes $F_\mathrm{lim,i,j}$ in bands whose central frequencies are $\nu_{i,j}$. $P_{\mathrm{tile},i}$ is the (2D) skymap probability density integrated over the tile, while $dP/dr$ defines how this probability density is distributed over luminosity distance $r\equiv d_\mathrm{L}$ (i.e.~sky-position-conditional distance probability density of the tile in the 3D skymap -- \citealt{singer16b}). Here $\mathcal{E}_{i}(r,\xi)$ equals $1$ up to the distance beyond which the putative EM counterpart becomes too faint to be detected by the wide-field observations of tile $i$, and $0$ for longer distances. This general framework allows for combining constraints from different wide-field searches. 

The results from wide-field and galaxy-targeted searches can be combined conservatively by taking the most constraining between the two for each particular EM counterpart model.

\section{Data tables}
\begin{figure}
    \centering
    \includegraphics[width=\columnwidth]{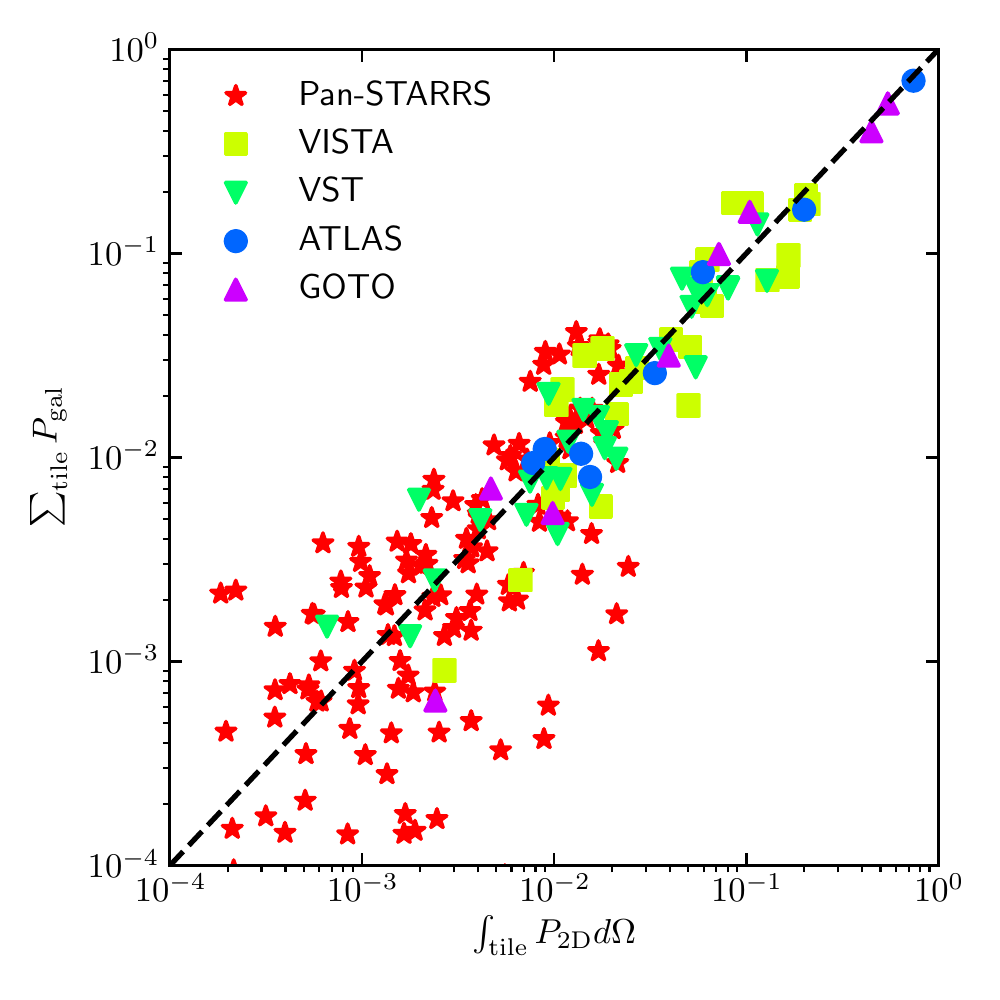}
    \caption{GW probability \textit{versus} galaxy-weighted probability in wide-field observations. Comparison between the \texttt{LALInference} GW sky localisation probability density ($P_\mathrm{2D}$) integrated over the tiles of our wide-field observations and the sum of individual galaxy probabilities $P_\mathrm{gal}$ (\S\ref{sec:probability}) over galaxies that fall in the same tile. The data are reported in Table~\ref{tab:widefield-obs-list}. Due to the wide field of view of these facilities, the galaxy density averages out, and the probabilities computed in the two ways are very similar. Note that the larger scatter for Pan-STARRS observations is due to the fact that observations are divided into smaller `sky cells'. Grouping sky cells in the same observation would reduce the scatter.}
    \label{fig:Integrated_Pgal_P3D_comparison}
\end{figure}

In this appendix, we provide tables which list exhaustively both our galaxy-targeted (Tab.~\ref{tab:targeted-galaxies-LK-Pgal}) and wide-field (Tab.~\ref{tab:widefield-obs-list}) observations. Figure~\ref{fig:Integrated_Pgal_P3D_comparison} compares two different possible definitions of the probability contained in a wide-field observation, namely the \texttt{LALInference} sky localisation probability integrated over the observed tile, and the sum of the individual galaxy probabilities contained in the same tile, showing that the two are essentially equivalent in the case of wide-field observations.

\longtab[1]{
\begin{landscape}
\begin{longtable}{rrrrrrrr|rrrr}
\caption{\label{tab:targeted-galaxies-LK-Pgal} List of galaxies targeted in our search, in order of decreasing, combined, probability ($P_{3D}\times P_{\rm gal}$). For each galaxy we report: HyperLEDA name (except for those followed by 2MX in parentheses, which are from 2MASSX), right ascension, declination and luminosity distance as listed in the GLADE catalogue$^*$, Ks-band magnitude as measured in our own VISTA/VINROUGE data, $K_{\rm s}$-band luminosity, 3D skymap probability density per unit volume at the galaxy position, and localisation probability (Eq.~\ref{eq:single_gal_prob}), followed by a list of observations of that galaxy, for which we specify the telescope, the post-merger time at which the observation was performed, the band, and the limiting magnitude.\\ \tiny{$^*$the distances in GLADE are computed assuming a flat cosmology with $H_0=70\,\mathrm{km\,s^{-1}Mpc^{-1}}$ and $\Omega_\mathrm{m}=0.27$ \citep{dalya18}. The difference in luminosity distance when compared to our assumed cosmology ($H_0=70\,\mathrm{km\,s^{-1}Mpc^{-1}}$ and $\Omega_\mathrm{m}=0.30$) is $<1\,\mathrm{Mpc}$ within $400\,\mathrm{Mpc}$, but we anyway report the distance computed with our assumed cosmology.}}\\
\toprule
          \multicolumn{8}{c}{{Targeted galaxies}} & \multicolumn{4}{c}{{Observations of each galaxy}}\\
\midrule
          Galaxy &      RA &  Dec & $d_\mathrm{L}$ & $K_{\rm s}$ & $\log(L_\mathrm{K})$ & $P_\mathrm{3D}$ &  $P_\mathrm{gal}$ & Telescope/ & $\Delta T$ & Band & $m_\mathrm{lim}$ \\
          ~ &  (Deg) & (Deg) & $(\mathrm{Mpc})$ &  (Vega) & $(\mathrm{L_\odot})$ & $(10^{-5}\mathrm{Mpc^{-3}})$ &  ~ & Instrument & (days) & ~ & (AB) \\
\midrule
\endfirsthead

\toprule
          \multicolumn{8}{c}{{Targeted galaxies}} & \multicolumn{4}{c}{{Observations of each galaxy}}\\
\midrule
          Galaxy &      RA &  Dec & $d_\mathrm{L}$ & $K_{\rm s}$ & $\log(L_\mathrm{K})$ & $P_\mathrm{3D}$ &  $P_\mathrm{gal}$ & Telescope/ & $\Delta T$ & Band & $m_\mathrm{lim}$ \\
          ~ &  (Deg) & (Deg) & $(\mathrm{Mpc})$ &  (Vega) & $(\mathrm{L_\odot})$ & $(10^{-5}\mathrm{Mpc^{-3}})$ &  ~ & Instrument & (days) & ~ & (AB) \\
\midrule
\endhead
\midrule
\multicolumn{12}{r}{{Continued on next page}} \\
\midrule
\endfoot

\bottomrule
\endlastfoot

198197  & 12.091079 & -25.126814 & 297.2 & 12.4 & 11.30 & 24.341 & 0.02607 & LT/IO:O & 0.20 & r & 20.3 \\
& & & & & & & & LT/IO:O & 0.21 & i & 20.3 \\
& & & & & & & & VLT/HAWKI & 1.37 & K & 22.3 \\
& & & & & & & & VLT/HAWKI & 4.51 & K & 22.0 \\
ESO474-035  & 13.173257 & -25.733852 & 271.0 & 12.5 & 11.15 & 31.691 & 0.02402 & LT/IO:O & 0.26 & r & 20.3 \\
& & & & & & & & VLT/HAWKI & 1.39 & K & 22.5 \\
& & & & & & & & VLT/HAWKI & 4.54 & K & 21.5 \\
797191  & 12.178233 & -23.773075 & 253.9 & 12.3 & 11.22 & 24.334 & 0.02150 & LT/IO:O & 0.33 & r & 20.3 \\
\end{longtable}
\end{landscape}
}

\longtab[2]{
\begin{longtable}{lllllllll}
\caption{\label{tab:widefield-obs-list} List of wide-field observations. For each facility we report a list of unique tiles, identified by a Tile ID and by the RA and Dec of the tile centre, along with the \texttt{LALInference} sky location probability density ($P_\mathrm{2D}$) integrated over the solid angle subtended by the tile, and the sum of the galaxy probabilities $P_\mathrm{gal}$ over galaxies that fall within the tile (see Figure~\ref{fig:Integrated_Pgal_P3D_comparison} for a comparison plot). For each tile, we then report the MJD, Filter and Limiting magnitude of the observations carried out of these tiles.}\\
\toprule
Facility & Tile ID & $\int_\mathrm{tile} P_\mathrm{2D}\,d\Omega$ & $\sum_\mathrm{tile} P_\mathrm{gal}$ & RA & Dec & MJD & Filter & Limiting mag \\
~ & ~ & [\%] & [\%] & [deg] & [deg] & ~ & ~ & [AB]\\
\midrule 

\endfirsthead

\toprule
Facility & Tile ID & $\int_\mathrm{tile} P_\mathrm{2D}\,d\Omega$ & $\sum_\mathrm{tile} P_\mathrm{gal}$ & RA & Dec & MJD & Filter & Limiting mag \\
~ & ~ & [\%] & [\%] & [deg] & [deg] & ~ & ~ & [AB]\\

\midrule

\endhead
\midrule
\multicolumn{9}{r}{{Continued on next page}} \\
\midrule
\endfoot

\bottomrule
\endlastfoot

VST & 0 & 1.692 & 1.578 & 10.573625 & -23.734472 & 58711.36 & r & 21.90\\
~ & ~ & ~ & ~ & ~ & ~ & 58715.36 & r & 21.60\\
~ & ~ & ~ & ~ & ~ & ~ & 58717.36 & r & 22.00\\
~ & ~ & ~ & ~ & ~ & ~ & 58724.36 & r & 22.10\\
VST & 1 & 1.421 & 1.710 & 10.558375 & -24.729083 & 58711.36 & r & 20.90\\
~ & ~ & ~ & ~ & ~ & ~ & 58715.36 & r & 21.50\\
~ & ~ & ~ & ~ & ~ & ~ & 58717.36 & r & 22.10\\
~ & ~ & ~ & ~ & ~ & ~ & 58724.36 & r & 22.20\\
\end{longtable}
}

\section{Candidate counterparts}
\label{sect:candidates}
\longtab[1]{
\begin{landscape}
\begin{longtable}{lllllllll}
\caption{\label{tab:candidates} List of candidate counterparts reported to the GW 190814bv. We only list candidates that were first reported to the TNS after 7 Aug (one week prior to the GW event), and that were discovered up to two weeks after the GW event. The ``Comments'' column lists the reason why candidates were discounted: Non.Det. = no subsequent detection after first report; Ast = asteroid, moving object; Spec.Host-z = spectroscopic redshift of host is inconsistent with GW; Phot.Host-z = photometric redshift of host is inconsistent with GW (note that photometric redshifts are reported to one decimal place in the redshift column); SN = spectroscopically classified as a supernova; SN? = probable supernova; Pre.Det. = historical detection in survey data prior to S190814bv; HPM = high proper motion star; AGN/AGN? = active galactic nucleus flare or other nuclear transient (or probable nuclear transient); Andreoni = candidates which are reported and ruled out in \cite{Andreonietal2019}. Candidates which we are unable to classify are indicated with ``?''. 
Objects with an asterisk in the comment column are discussed in more detail in the text of Appendix\,\ref{sect:candidates}}.\\
\toprule
IAU name & Survey name & Coordinates & Discovery date & Mag  & Source & z &  GCNs & Comments \\
\midrule
\endfirsthead
\toprule
IAU name & Survey name & Coordinates & Discovery date & Mag  & Source & z &  GCNs & Comments \\
\midrule
\endhead
\midrule
\multicolumn{9}{r}{{Continued on next page}} \\
\midrule
\endfoot

\bottomrule
\endlastfoot
AT2019nbp   & PS19ekf                       & 00:46:57.393 -24:21:42.70     & 8/09 13:58:04 & w=20.0    & PS1       & ?         &                           & Pre.Det.      \\   
AT2019nmd   & desgw-190814a                 & 00:51:29.004 -22:28:16.96     & 8/15 06:40:19 & i=20.6    & DESGW     & -         & 25336,25348               & Ast           \\ 
AT2019nme   & desgw-190814b                 & 00:50:32.558 -22:13:33.70     & 8/15 06:40:19 & i=19.3    & DESGW     & -         & 25336,25355               & Ast           \\
AT2019noq   & PS19epf,                      & 00:48:47.882 -25:18:23.46     & 8/15 14:02:24 & i=19.9    & PS1       & 0.07      & 25356,25423,              & SN            \\
            & VSTJ004847.88-251823.5        &                               &                           &           &           & 25669                     &               \\
AT2019nor   & PS19eph                       & 00:49:51.992 -24:16:17.71     & 8/15 14:02:24 & i=19.7    & PS1       & ?         & 25356                     & SN?*          \\ 
AT2019npd   & DG19hqpgc, PS19epw,           & 00:46:56.711 -25:22:36.43     & 8/15 06:43:12 & i=19.1    & GROWTH    & 0.0008    & 25362,25669               & Spec.Host-z*  \\ 
            & VSTJ004656.70-252236          &                               &                           &           &           &                           &               \\
AT2019npe   & DG19prsgc,                    & 00:41:33.330 -23:44:31.94     & 8/15 06:37:26 & z=20.1    & GROWTH    & -         & 25362                     & Non.Det.*     \\
            & VSTJ004133.33-234432.0        &                               &               &           &           &           &                           &               \\
AT2019npj   & DG19frdgc,PS19epv             & 00:46:11.590 -22:07:07.82     & 8/15 07:20:38 & z=19.7    & GROWTH    & ?         & 25362                     & Pre.Det.*     \\ 
AT2019npv   & DG19wxnjc, PS19eqi,           & 00:53:32.309 -23:49:58.64     & 8/16 07:45:07 & i=20.7    & GROWTH    & 0.056     & 25393,25454               & SN            \\ 
            & VSTJ005332.32-234958.5        &                               &               &           &           &           &                           &               \\
AT2019npw   & DG19wgmjc, PS19eqj,           & 00:55:52.405 -25:46:59.80     & 8/16 07:39:21 & i=20.6    & GROWTH    & 0.163     & 25362,25484               & SN            \\  
            & VSTJ005552.40-254659.8        &                               &               &           &           &           &                           &               \\
AT2019npy   & DG19sevhc,PS19epy             & 00:56:12.020 -25:29:24.28     & 8/16 05:48:28 & i=20.4    & GROWTH    & -         & 25362,25383,              & HPM           \\  
            &                               &                               &               &           &           &           & 25374,25386               &               \\
AT2019npz   & DG19tzyhc,                    & 00:53:05.560 -24:21:38.71     & 8/16 05:57:07 & z=20.6    & GROWTH    & ?         & 25362                     & AGN?*         \\
            & VSTJ005305.56-242138.7        &                               &               &           &           &           &                           &               \\
AT2019nqc   & DG19qabkc                     & 01:29:03.673 -32:42:18.50     & 8/16 08:23:59 & i=20.6    & GROWTH    & 0.078     & 25362,25571               & SN            \\
AT2019nqe   & DG19etlkc                     & 01:25:49.625 -34:07:10.49     & 8/16 08:52:48 & i=21.2    & GROWTH    & ?         & 25398                     & ?*            \\ 
AT2019nqp   & PS19epx                       & 00:56:50.403 -24:20:49.95     & 8/15 14:05:16 & z=20.7    & PS1       & ?         & -                         & Pre.Det.*     \\ 
AT2019nqq   & desgw-190814c                 & 01:23:49.217 -33:02:05.00     & 8/15 07:09:07 & i=20.8    & DESGW     & 0.071     & 25373,25419               & SN            \\  
AT2019nqr   & desgw-190814d                 & 01:34:17.649 -32:44:30.41     & 8/16 06:31:40 & i=18.3    & DESGW     & 0.084     & 25373,25379               & SN            \\      
AT2019nqs   & desgw-190814e                 & 01:33:35.164 -31:46:48.48     & 8/16 07:07:40 & i=20.4    & DESGW     & 0.1263    & 25373,25384               & Spec.Host-z   \\
AT2019nqw   & DG19xczjc, PS19eqk,           & 00:56:46.717 -25:09:33.42     & 8/16 08:18:14 & z=20.0    & GROWTH    & ?         & 25748                     & SN?*           \\
            & VSTJ005646.69-250933.3        &                               &               &           &           &           &                           &               \\
AT2019nqz   & DG19ayfjc                     & 00:46:46.416 -24:20:12.06     & 8/16 07:22:04 & z=21.1    & GROWTH    & 0.1076    & 25391,25571               & Spec.Host-z   \\
AT2019nra   & DG19njmjc                     & 00:55:10.705 -25:56:57.12     & 8/16 06:01:26 & i=20.5    & GROWTH    & ?         & 25395                     & ?*        \\
AT2019nsm   & DG19wlpmc,                    & 00:43:30.111 -22:43:29.46     & 8/17 07:20:38 & z=20.4    & GROWTH    & ?         & 25393                     & Pre.Det.*     \\
            & VSTJ004330.16-224329.4        &                               &               &           &           &           &                           &               \\
AT2019nte   & desgw-190814f                 & 01:34:13.766 -31:43:18.12     & 8/16 07:07:40 & i=21.0    & DESGW     & 0.0704    & 25398,25784               & AGN?*         \\
AT2019ntm   & DG19jqzkc,                    & 00:48:04.438 -23:47:50.91     & 8/17 04:58:04 & i=19.7    & GROWTH    & 0.116?    & 25486                     & Spec.Host-z?  \\
            & VSTJ004804.40-234750.9        &                               &               &           &           &           &                           &               \\
AT2019ntn   & DG19rtekc, desgw-190814i      & 01:34:53.352 -31:22:49.74     & 8/16 07:07:40 & i=21.6    & GROWTH    & 0.10      & 25393,25423               & SN            \\
AT2019ntp   & DG19gcwjc, PS19eqq,           & 00:50:12.060 -26:11:52.81     & 8/16 07:35:02 & z=21.1    & GROWTH    & ?         & 25393,25596               & SN*           \\
            & VSTJ005012.07-261152.6        &                               &               &           &           &           &                           &               \\
AT2019ntr   & DG19sbzkc                     & 01:00:01.879 -26:42:51.65     & 8/16 07:53:45 & z=20.6    & GROWTH    & 0.2       & 25393,25540               & SN            \\ 
            & VSTJ010001.84-264251.3        &                               &               &           &           &           &                           &               \\
AT2019nts   & DG19vodmc                     & 00:48:31.378 -23:06:39.24     & 8/17 06:50:23 & z=20.9    & GROWTH    & ?         & 25393                     & SN?*          \\
AT2019nuj   & DG19hqhjc,                    & 00:49:01.738 -23:14:04.93     & 8/16 07:29:16 & z=20.8    & GROWTH    & ?         & 25393                     & Pre.Det*      \\ 
            & VSTJ004901.74-231404.9        &                               &               &           &           &           &                           &               \\
AT2019nuk   & DG19dnjlc                     & 00:54:57.827 -26:08:04.61     & 8/17 05:22:33 & i=21.6    & GROWTH    & 0.076     & 25393,25394               & ?*            \\ 
AT2019nul   & DG19kpykc                     & 00:55:16.443 -26:56:34.57     & 8/16 07:53:45 & z=20.3    & GROWTH    & 0.098     & 25393                     & Spec.Host-z*  \\
AT2019num   & DG19rzhoc,                    & 00:55:31.602 -22:58:08.48     & 8/18 06:41:45 & i=21.1    & GROWTH    & 0.113     & 25393,25484               & SN            \\ 
            & VSTJ005531.60-225808.5        &                               &               &           &           &           &                           &               \\
AT2019nun   & DG19tvtnc                     & 00:56:48.599 -24:54:30.48     & 8/18 06:17:16 & i=21.7    & GROWTH    & 0.131     & 25393                     & Spec.Host-z * \\
AT2019nuq   & DG19kxdnc                     & 01:32:36.497 -33:55:01.88     & 8/17 07:50:52 & z=20.5    & GROWTH    & 0.104     & 25393                     & Spec.Host-z   \\      
AT2019nuw   & PS19epz                       & 00:50:26.340 -25:52:57.83     & 8/18 12:59:02 & z=21.9    & PS1       & ?         & 25417                     & SN?*          \\
AT2019nux   & PS19eqa                       & 00:50:21.019 -23:42:46.79     & 8/18 12:59:02 & z=21.8    & PS1       & ?         & 25417                     & SN?*          \\
AT2019nuy   & PS19eqb                       & 00:50:50.397 -25:29:29.55     & 8/18 12:59:02 & z=21.1    & PS1       & ?         & -                         & ?*            \\ 
AT2019nuz   & PS19eqc                       & 00:49:52.264 -25:31:25.62     & 8/18 12:59:02 & z=21.9    & PS1       & ?         & 25417                     & SN?*          \\
AT2019nva   & PS19eqd,                      & 00:52:43.395 -23:37:54.01     & 8/18 12:59:02 & z=21.5    & PS1       & ?         & 25417                     & SN?*          \\
            & VSTJ005243.34-233753.6        &                               &               &           &           &           &                           &               \\
AT2019nvb   & PS19eqe                       & 00:46:51.168 -25:25:39.31     & 8/18 12:59:02 & z=21.7    & PS1       & 0.0008    & 25417                     & Spec.Host-z*  \\
AT2019nvc   & PS19eqf                       & 00:52:18.326 -26:19:42.07     & 8/18 13:01:55 & z=21.3    & PS1       & 0.07      & 25417                     & {\bf SN?}     \\ 
AT2019nvd   & PS19eqg,                      & 00:55:42.309 -24:41:50.28     & 8/18 12:59:02 & z=21.5    & PS1       & ?         & 25417                     & SN?*          \\
            & VSTJ005542.30-244149.9        &                               &               &           &           &           &                           &               \\
AT2019nve   & PS19eqh,                      & 00:56:05.510 -24:38:26.40     & 8/18 12:59:02 & z=21.3    & PS1       & ?         & 25417,25669               & SN?*          \\
            & VSTJ005605.55-243826.4        &                               &               &           &           &           &                           &               \\
AT2019nvr   & PS19eqo                       & 00:48:16.088 -25:28:14.99     & 8/18 12:56:09 & z=20.9    & PS1       & ?         & 25417                     & SN?*          \\
AT2019nvs   & PS19eqp                       & 00:52:37.759 -26:11:41.40     & 8/18 13:10:33 & z=21.4    & PS1       & ?         & 25417                     & SN?*          \\
AT2019nxd   & desgw-190814i                 & 00:42:44.598 -24:57:20.34     & 8/17 04:52:19 & i=21.8    & DESGW     & 1.3       & 25486                     & Phot.Host-z   \\
AT2019nxe   & desgw-190814j,                & 00:46:16.814 -24:22:21.19     & 8/16 07:29:16 & z=22.1    & DESGW     & 0.0777    & 25425,25543               & SN            \\
            & VSTJ004616.81-242221.2        &                               &               &           &           &           &                           &               \\
AT2019nys   & desgw-190814k,                & 00:57:56.903 -24:34:00.56     & 8/16 05:55:40 & i=21.8    & DESGW     & 0.4       & 25486                     & Phot.Host-z*  \\
            & VSTJ005756.90-243400.5        &                               &               &           &           &           &                           &               \\
AT2019nyv   & DG19mulnc,                    & 00:46:59.451 -23:05:59.50     & 8/18 06:01:26 & z=20.6    & GROWTH    &           & 25669                     & Pre.Det.      \\  
            & VSTJ004659.45-230559.5        &                               &               &           &           &           &                           &               \\
AT2019nyy$^a$   & DG19zoonc                 & 00:48:16.652 -26:38:26.97     & 8/18 06:08:38 & z=21.1    & GROWTH    & ?         & \cite{Andreonietal2019}   & Andreoni      \\
AT2019nzd   & DG19kzvqc,                    & 00:58:06.435 -24:50:14.35     & 8/21 04:35:02 & i=22.2    & GROWTH    & ?         & 25486                     & ?*            \\
            & VSTJ005806.46-245014.3        &                               &                           &           &           &                           &               \\
AT2019nzr   & desgw-190814m                 & 00:47:21.410 -24:34:36.58     & 8/16 05:36:57 & i=21.8    & DESGW     & ?         & 25425,25486               & ?* \\
AT2019oab   & desgw-190814o                 & 00:58:59.398 -25:46:12.66     & 8/16 08:18:14 & z=21.7    & DESGW     & 0.5       & 25425,25486               & Phot.Host-z*  \\
AT2019oac   & DG19zujoc                     & 00:53:02.879 -21:39:04.57     & 8/18 06:43:12 & z=21.4    & GROWTH    & ?         & 25486                     & SN?*          \\
AT2019obc   & desgw-190814q                 & 00:58:16.023 -24:08:23.18     & 8/18 06:17:16 & i=21.7    & DESGW     & 0.22      & 25438,25543               & SN            \\ 
AT2019odc   & desgw-190814r                 & 00:46:01.689 -25:27:32.94     & 8/17 04:52:19 & i=21.6    & DESGW     & 0.0540    & 25486,25588               & ?*            \\
AT2019oer   & CFHT\_cand, desgw-190814s     & 00:47:28.027 -25:26:14.15     & 8/18          & i=22.9    & CFHT      & ?         & 25443,25447               & Pre.Det.      \\      
AT2019ofb   & PS19erd                       & 00:55:19.237 -26:11:50.77     & 8/21 13:00:28 & z=21.4    & PS1       &           & 25455                     & ?             \\
AT2019okr   & desgw-190814t                 & 00:47:23.696 -25:27:30.78     & 8/17 05:06:43 & z=22.7    & DESGW     & ?         & 25486,25526               & Pre.Det.*     \\  
AT2019oks   & desgw-190814u                 & 01:02:08.319 -24:54:21.70     & 8/17 05:47:02 & i=22.8    & DESGW     & 0.193?    & 25486                     & Spec.Host-z?* \\
AT2019omt   & desgw-190814v                 & 00:59:26.742 -25:59:41.28     & 8/18 06:25:55 & z=22.4    & DESGW     & 0.1564    & 25486,25588               & SN            \\ 
AT2019omu   & desgw-190814w                 & 01:33:58.890 -34:20:20.01     & 8/16 06:37:26 & i=21.3    & DESGW     & 0.7       & 25486                     & Phot.Host-z*  \\
AT2019omv   & desgw-190814x                 & 01:39:54.812 -33:23:01.39     & 8/16 06:38:52 & z=22.0    & DESGW     & 0.4       & 25486                     & Phot.Host-z   \\
AT2019omw   & desgw-190814y                 & 00:48:56.255 -23:10:12.49     & 8/18 06:01:26 & z=22.6    & DESGW     & ?         & 25486                     & SN?*          \\
AT2019omx   & desgw-190814z                 & 01:36:44.246 -33:18:09.64     & 8/16 07:13:26 & i=22.6    & DESGW     & 0.275     & 25486,25540               & Spec.Host-z   \\
AT2019onj   & desgw-190814ab                & 00:47:26.006 -25:26:55.13     & 8/17 04:55:12 & z=21.7    & DESGW     & ?         & 25486,25526               & Pre.Det.*     \\      
AT2019opp   & desgw-190814ac                & 00:57:38.254 -25:16:45.00     & 8/17 05:48:28 & z=22.5    & DESGW     & ?         & 25486                     & SN?*          \\
AT2019osy   & ASKAP 005547-270433           & 00:55:47.000 -27:04:33.00     & 8/23 13:44:59 & i=22.3    & ASKAP     &  -        & 25488,25801,              & AGN?*         \\
            &                               &                               &               &           &           &           & 25487                     &               \\
AT2019paw$^a$   & DG19ggesc                 & 00:48:34.270 -25:05:25.81     & 8/21 06:00:00 & i=22.2    & GROWTH    & ?         & \cite{Andreonietal2019}   & Andreoni      \\
AT2019qbu   & VSTJ005109.17-221740.7        & 00:51:09.173 -22:17:40.69     & 8/20 03:47:38 & r=21.1    & GRAWITA   & ?         & -                         & Non.det.*     \\
AT2019qby   & VSTJ004414.33-250744.3        & 00:44:14.334 -25:07:44.32     & 8/16 06:36:00 & r=21.4    & GRAWITA   & ?         & -                         & SN?*           \\
AT2019qbz   & VSTJ005653.99-275921.4        & 00:56:53.980 -27:59:21.37     & 8/16 07:10:33 & r=20.9    & GRAWITA   & ?         &                           & SN?*           \\
AT2019qca   & VSTJ004548.54-264939.0        & 00:45:48.540 -26:49:39.01     & 8/22 04:01:55 & r=21.8    & GRAWITA   & ?         & -                        & SN?*          \\
AT2019qcb   & VSTJ004619.06-260843.2,       & 00:46:19.038 -26:08:43.61     & 8/16 05:31:12 & r=21.6    & GRAWITA   & ?         & -                         & SN?*          \\
            & PS19gaz                       &                               &               &           &           &           &                           &               \\
AT2019qcc   & VSTJ005349.82-244549.6,       & 00:53:49.844 -24:45:49.40     & 8/20 04:16:19 & r=22.0    & GRAWITA   & ?         & -                         & SN?*          \\
            & PS19gba                       &                               &               &           &           &           &                           &               \\
AT2019thm$^a$   & DG19gyvx                  & 00:47:56.625 -26:54:02.80     & 8/15 07:00:28 & ?         & GROWTH    & ?         & \cite{Andreonietal2019}   & Andreoni        \\
AT2019tih$^a$   & DG19ilqnc                 & 00:47:26.675 -27:36:02.82     & 8/18 06:05:45 & i=21.3    & GROWTH    & ?         & \cite{Andreonietal2019}   & Andreoni      \\
AT2019tii$^a$   & DG19tedsc                 & 00:49:35.203 -27:02:09.33     & 8/21 05:52:48 & i=21.9    & GROWTH    & ?         & \cite{Andreonietal2019}   & Andreoni      \\
AT2019tix$^a$   & PS19gjp, DG19bown         & 00:48:45.660 -24:38:50.56     & 8/21 04:11:51 & i=22.9    & GROWTH    & ?         & \cite{Andreonietal2019}   & Andreoni      \\
 AT2019aacd & CFHT0054-2456zau            & 00:54:11.280 -24:56:17.66	& 8/18 12:39:22 & i=20.8    & CFHT      & ?         & \cite{Vieiraetal2020}      & Bogus?*       \\
\end{longtable}
\footnotesize{$^a$Not reported in GCN, candidate announced on TNS and in \cite{Andreonietal2019}}.
\footnotesize{$^b$References for GCNs listed in the table:
\protect\citet{2019GCN.25336....1S,2019GCN.25348....1D,2019GCN.25355....1G,2019GCN.25356....1H,2019GCN.25362....1A,2019GCN.25372....1C,2019GCN.25373....1H,2019GCN.25374....1D,2019GCN.25379....1T,2019GCN.25383....1R,2019GCN.25384....1B,2019GCN.25386....1S,2019GCN.25391....1G,2019GCN.25393....1G,2019GCN.25394....1G,2019GCN.25395....1D,2019GCN.25398....1H,2019GCN.25417....1S,2019GCN.25419....1L,2019GCN.25423....1R,2019GCN.25425....1S,2019GCN.25438....1S,2019GCN.25443....1R,2019GCN.25447....1J,2019GCN.25454....1J,2019GCN.25455....1S,2019GCN.25484....1T,2019GCN.25486....1S,2019GCN.25487....1S,2019GCN.25488....1A,2019GCN.25526....1J,2019GCN.25540....1W,2019GCN.25543....1C,2019GCN.25571....1L,2019GCN.25588....1H,2019GCN.25596....1W,2019GCN.25669....1G,2019GCN.25748....1Y,2019GCN.25784....1C,2019GCN.25801....1B}
}

\end{landscape}
}

The list of publicly reported candidate counterparts for S190814bv is given in Tab. \ref{tab:candidates}. All photometry of the candidates discovered by our search will be made available as online supplementary material. Some detailed comments on specific candidates follow:
\begin{itemize}
    \item {\bf AT2019nor} Pan-STARRS lightcurve lasts for at least 70 days from discovery, with a slow decline of 1.2 mag per 100 days in $w$ consistent with a Type IIP SN on the plateau.
    \item {\bf AT2019npd} Likely associated with the foreground galaxy NGC 253.
    \item {\bf AT2019npe} No detection in Pan-STARRS images taken on 58710.58 (7~hr after the GROWTH detection) to a limiting magnitude of $w=21.2$.
    \item {\bf AT2019npj} Ruled out by Pan-STARRS detection on 2019-08-04, ten days before GW event.
    \item {\bf AT2019npz} Consistent with the nucleus of a compact galaxy, could be AGN or other nuclear transient. Flat lightcurve around $w=21.5$ in Pan-STARRS images taken between 20 and 70 days after the GW.
    \item {\bf AT2019nqe} Reported to the TNS with i=21.2 on 08/16, there are no historic detections in Pan-STARRS, ATLAS or reported by ZTF.
    \item {\bf AT2019nqp} Archival Pan-STARRS detections from 2018.
    \item {\bf AT2019nqw} Still present in GRAWITA images two weeks after GW event.
    \item {\bf AT2019nra} A J-band spectrum taken 3 days after discovery was reported in GCN 25395 to be featureless.
    \item {\bf AT2019nsm} Seen in PS1 images prior to GW event.
    \item {\bf AT2019nte} The transient fades by 1 mag in i-band over two consecutive nights (from 20.95 on 08/16 to 21.92 on 08/17. However, the transient is still visible at i=22.3 (GCN 25598) ten days later on 08/27.The source is consistent with the nucleus of it's host galaxy.
    \item {\bf AT2019ntm} An 1800s spectrum of the host was taken on 2019-09-09 with the William Herschel Telescope (WHT)+ACAM using the V400 grism. The spectrum reveals a single emission feature, that if associated with H$\alpha$ corresponds to a redshift of 0.116.
    \item {\bf AT2019ntp} Spectrum in GCN 25596 is reported to be that of a broad-lined Type Ic SN, no redshift is listed in GCN.
    \item {\bf AT2019nts} z=20.9 on 8/17, and i=20.3 on 8/18 (DECAM photometry via TNS) implies that the transient is either rising or has a blue i-z colour 4 days after the GW event.
    \item {\bf AT2019nuj} A detection at w=22 on MJD 58699 was recovered in Pan-STARRS data. The lightcurve is consistent with a SN.
    \item {\bf AT2019nuk} Transient is not offset from its host galaxy. Spectroscopic redshift is consistent with GW, while DECAM photometry on TNS appears to show a rapid decline (2.6 mag over one day).  However, if associated with S190814bv, then the absolute magnitude three days after the GW event would be $i=-18.1$.
    \item {\bf AT2019nul} An 1800s spectrum of the host was taken on 2019-09-12 with the William Herschel Telescope (WHT)+ACAM using the V400 grism. 
    \item {\bf AT2019nun} An 1800s spectrum of the host was taken on 2019-09-11 with the William Herschel Telescope (WHT)+ACAM using the V400 grism.
    \item {\bf AT2019nuw} Likely SN, flat lightcurve in Pan-STARRS.
    \item {\bf AT2019nux} Likely SN, flat lightcurve in Pan-STARRS.
    \item {\bf AT2019nuy} Offset from faint host in Pan-STARRS.
    \item {\bf AT2019nuz} Likely SN, flat lightcurve in Pan-STARRS.
    \item {\bf AT2019nva} Likely SN, flat lightcurve in Pan-STARRS.
    \item {\bf AT2019nvb} Likely associated with the foreground galaxy NGC 253.
    \item {\bf AT2019nvd} Likely SN, flat lightcurve in Pan-STARRS.
    \item {\bf AT2019nve} Likely SN, flat lightcurve in Pan-STARRS.
    \item {\bf AT2019nvr} Likely SN, flat lightcurve in Pan-STARRS.
    \item {\bf AT2019nvs} Likely SN, flat lightcurve in Pan-STARRS.
    \item {\bf AT2019nys} The DES photometric redshift catalogue reports z=0.41$\pm$0.06 for the host.
    \item {\bf AT2019nzd} Brightens by 0.4 mag between two DECAM i-band images taken 0.7 hr apart on 08/21.
    \item {\bf AT2019nzr} An 1800s spectrum of the host was taken on 2019-09-09 with the William Herschel Telescope (WHT)+ACAM using the V400 grism. The spectrum revealed a featureless continuum with no clear emission features, and we were unable to secure a redshift. While the DES colours for the host are consistent with an AGN (GCN 25486), the spectrum does not show Seyfert features.
    \item {\bf AT2019oab} While the reported photometric redshift appears grossly inconsistent with the distance to the GW event, we note that the reported lightcurve fades by 0.8 mag over two days.
    \item {\bf AT2019oac} Rises by 0.2 mag in z between 08/18 and 08/21
    \item {\bf AT2019odc} The host redshift is consistent with S190814bv. GTC spectroscopy (reported in GCN 25588) with the slit covering the position of the transient reveals no broad features in the spectrum.
    \item {\bf AT2019oer} Detection in VISTA archive imaging \citep{2010Msngr.139....6A}, published in ENGRAVE GCN 25447.
    \item {\bf AT2019okr} Detection in VISTA archival imaging \citep{2010Msngr.139....6A}, published in ENGRAVE GCN 25526
    \item {\bf AT2019oks} No change in i-band magnitude over 4 days, suggests unrelated to GW event. An 1800s spectrum of the host was taken on 2019-09-09 with the William Herschel Telescope (WHT)+ACAM using the V400 grism. The spectrum reveals a single emission feature, that if associated with H$\alpha$ corresponds to a redshift of 0.193.
    \item {\bf AT2019omu} i-band photometry reported by DES reveals a decline of only 0.3 mag over the five days after discovery. Moreover, the photometric redshift of the host from the DES photometric redshift catalogue is reported to be 0.66$\pm$0.03
    \item {\bf AT2019omw} Flat lightcurve in i-band between 08/18 and 08/21 from DECAM photometry reported on TNS.
    \item {\bf AT2019onj} Detection in VISTA archival imaging \citep{2010Msngr.139....6A}, published in ENGRAVE GCN 25526
    \item {\bf AT2019opp} Lightcurve from DECAM reported on TNS shows a rise in magnitude one week after the GW event.
    \item {\bf AT2019osy} Radio transient found by the Australian Square Kilometre Array Pathfinder (ASKAP). Followup with HST and ALMA confirms peculiar transient unrelated to GW event.
    \item {\bf AT2019qbu} Non-detection to r=22.3 on 8/16 (after GW event).
    \item {\bf AT2019qby} 0.5 mag decline in r-band over 6 days.
    \item {\bf AT2019qbz} Fades by 0.2 mag over 6 days. Apparently hostless.
    \item {\bf AT2019qca} Apparently hostless source that has a constant magnitude from discovery until at least 08/30.
    \item {\bf AT2019qcb} Fades by only 0.2 mag over two weeks from discovery.
    \item {\bf AT2019qcc} 0.2 mag decline over 10 days after discovery.
    \item {\bf AT2019aacd} Reported by \cite{Vieiraetal2020} (with the identifier CFHT0054-2345zau) as a potential counterpart after the first submission of this manuscript, and subsequently added to Table \ref{tab:candidates} upon revision. On 08/21 we observed a similar footprint with the Pan-STARRS2 telescope in the \ips\ filter. We did not carry out an independent transient search in these data, since the facility was still in science commissioning mode, but we stored the data to provide additional photometry for any interesting source. We cross-checked these data to confirm the proposed $i$-band transient found by the CFHT search: we have a nearby 5$\sigma$ detection on MJD 58716.573 (5 days after the CFHT discovery), but it is 0\farcs6 from CFHT0054-2345zau, closer to the galaxy’s core (0\farcs48 separation) and, if real, its measured \ips\ magnitude is 22.2$\pm0.2$. However, this is clearly offset from the CFHT object, and no excess flux is visible at the position of the CFHT object to \ips$<$22.3. \citeauthor{Vieiraetal2020} find $i=21.5$ just 1 day later, which is therefore incompatible with our PS2 images. There is no detection in the \zps data of PS1 either, hence we consider both the CFHT object and the excess flux in the PS2 \ips-band to be bogus artefacts from image subtraction.
    \end{itemize}

\end{document}